\documentclass[lettersize,journal]{IEEEtran}
\usepackage{graphicx}
\usepackage{amsmath}
\usepackage{amssymb}
\usepackage{booktabs}
\usepackage{xcolor}
\usepackage{caption}
\usepackage{subcaption}
\usepackage{tikz}
\usepackage{soul}
\usepackage{wrapfig}
\usepackage{multirow}
\usepackage{cite}
\usepackage{xcolor}
\usepackage[switch]{lineno} %
\usepackage{makecell}

\usepackage[pagebackref,breaklinks,colorlinks]{hyperref}

\usepackage[capitalize]{cleveref}
\crefname{section}{Sec.}{Secs.}
\crefname{table}{Tab.}{Tabs.}
\crefname{figure}{Fig.}{Figs.}
\crefname{equation}{Eq.}{Eqs.}

\DeclareMathOperator*{\argmin}{argmin} %

\begin{document}

\title{Learning Lens Blur Fields}
\author{Esther Y. H. Lin,~\IEEEmembership{Student Member,~IEEE,}
Zhecheng Wang,~\IEEEmembership{Student Member,~IEEE,} \\
Rebecca Lin,~\IEEEmembership{Student Member,~IEEE,}
Daniel Miau,~\IEEEmembership{Member,~IEEE,}
Florian Kainz,~\IEEEmembership{Member,~IEEE,}\\
Jiawen Chen,~\IEEEmembership{Member,~IEEE,}
Xuaner Zhang,~\IEEEmembership{Member,~IEEE,}\\
David B. Lindell,~\IEEEmembership{Member,~IEEE,}
Kiriakos N. Kutulakos,~\IEEEmembership{Member,~IEEE,}
}

\maketitle

\definecolor{olivegreen}{HTML}{3C8031}

\newcommand{\PSF}{\text{PSF}}
\renewcommand{\S}{\text{S}}
\newcommand{\T}{\text{T}}

\newcommand{\fignum}[1]{\ref{#1}}
\newcommand{\sectnum}[1]{\ref{#1}}

\newcommand{\edit}[1]{\textcolor{black}{#1}} %

\newcommand{\finaledit}[1]{\textcolor{green}{#1}} %

\begin{figure*}[!ht]
  \centering
  \includegraphics[width=0.90\textwidth]{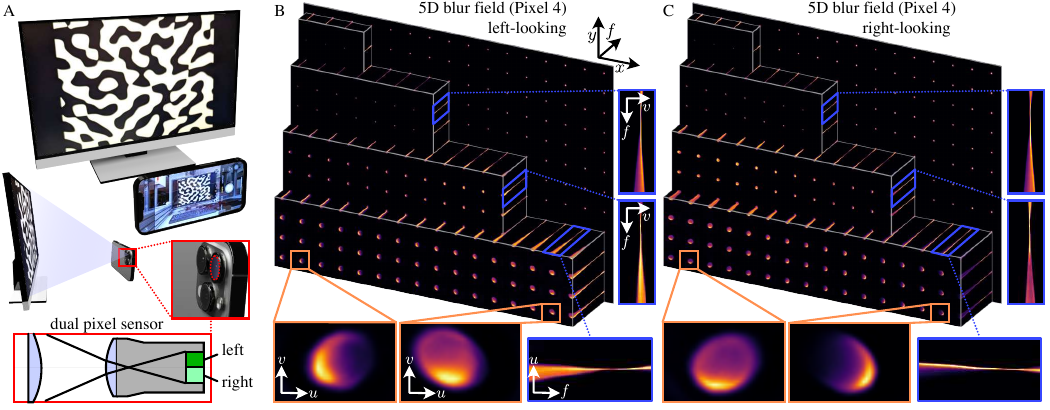}
  \caption{Illustration of learning lens blur fields. We prescribe an easy-to-use method for acquiring a lens blur field---a neural representation that captures high-dimensional variations in the point spread function of an imaging system. (A) We first capture focal stacks of calibration patterns displayed on a monitor, for example, using the Google Pixel 4 as depicted. This particular smartphone camera has a dual-pixel sensor which places two pixels behind a microlens focused on the objective lens. (B-C) We use captured calibration patterns to optimize a multi-layer perceptron that parameterizes variations in the PSF ($u$, $v$) across image plane positions ($x$, $y$) and focus ($f$). We visualize slices of the recovered 5D blur field for the left-looking and right-looking green-channel pixels of the sensor.}
  \label{fig:teaser}
\end{figure*}
  
\begin{abstract}
Optical blur is an inherent property of any lens system and is challenging to model in modern cameras because of their complex optical elements. To tackle this challenge, we introduce a high-dimensional neural representation of blur---\emph{the lens blur field}---and a practical method for acquiring it. The lens blur field is a multilayer perceptron (MLP) designed to (1) accurately capture variations of the lens 2D point spread function over image plane location, focus setting and, optionally, depth and (2) represent these variations parametrically as a single, sensor-specific function. The representation models the combined effects of defocus, diffraction, aberration, and accounts for sensor features such as pixel color filters and pixel-specific micro-lenses. To learn the real-world blur field of a given device, we formulate a generalized non-blind deconvolution problem that directly optimizes the MLP weights using a small set of focal stacks as the only input. We also provide a first-of-its-kind dataset of 5D blur fields---for smartphone cameras, camera bodies equipped with a variety of lenses, \emph{etc.} Lastly, we show that acquired 5D blur fields are expressive and accurate enough to reveal, for the first time, differences in optical behavior of smartphone devices of the same make and model. Code and data can be found at \href{https://blur-fields.github.io/}{blur-fields.github.io}.
\end{abstract}


\section{Introduction}

\IEEEPARstart{A}{ccurately} modeling an optical imaging system requires characterizing the blur induced by optical aberration, defocus, and diffraction.
Such models of optical blur underlie many computational imaging~\cite{cossairt2010spectral, tang2012utilizing, heide2013modern} and sensing ~\cite{levoy2006light, tao2013depth, tao2015depth} techniques and are becoming increasingly important for high-performance computer vision in automotive~\cite{pasqual1998use} and mobile photography applications~\cite{suwajanakorn2015depthff, tang2017depth, punnappurath2020modeling, abuolaim2022multi,xin2021defocus}.

Despite the potentially wide-ranging impact of such models, existing approaches for estimating optical blur in real-world camera systems trade-off representational accuracy against ease of acquisition. \edit{Accurately representing optical blur across multiple dimensions requires densely sampling the parameter space, which increases acquisition time and estimation complexity.} To address this major limitation, we propose a novel method for blur acquisition that enables highly accurate modeling of smartphone and interchangeable-lens camera systems without sacrificing scalability across many potential dimensions (2D pixel location, depth, focus setting, wavelength, etc.) or ease of use. A key property of our approach is that it outputs a model that is \emph{continuous}, making it possible to represent both large scale and minute blur variations across all those dimensions. Although our specific focus in this paper is conventional cameras for which a thin-lens model can serve as a (very crude) approximation, generalization to systems employing coded optics~\cite{levin2007image, asif2016flatcam, antipa2018diffusercam,levin20094d, pavani2008high} may be possible as well. 

\subsection{Background and Related Work} 
Optical blur is typically characterized using a point spread function (PSF), which describes the image formed by a point light source on the camera's sensor plane. Recovering the PSF of a camera is a challenging problem: blur can vary considerably over the sensor plane and is dependent on the distance to the light source, the camera's focus setting, aperture, wavelength, \emph{etc.}

Historically, PSFs have been modeled in three different ways: parametrically, non-parametrically, or via geometric or Fourier optics simulation. Parametric methods~\cite{born1999principles,punnappurath2020modeling,tang2018modeling,mahajan1994aberration, smith2008modern} rely on idealized models of lenses and image formation and thus have limited expressiveness.
For example, models based on Gaussian kernels~\cite{kee2011modeling,jang2016modeling} or hand-crafted formulations~\cite{schuler2012blind,simpkins2014parameterized} are too simple to model real-world camera PSFs; Zernike polynomials assume rotational symmetry and circular pupils~\cite{zernicke1934physica, born1999principles, mahajan1994aberration, song2019full, bezdidko1974use, tango1977thecp, debarnot2021learning}; and Seidel aberrations~\cite{mahajan1994aberration,tang2018modeling,smith2008modern} assume axially-symmetric lens systems and cannot model blur of dual-pixel auto-focus systems, which do not have such symmetry. 

Non-parametric methods sample the space of a camera's PSFs by acquiring a sparse grid of 2D PSFs over the sensor plane~\cite{joshi2008psf,mannan,kee2011modeling,Gwak:20} and optionally over focus settings~\cite{joshi2008psf} and depth~\cite{mannan}. These approaches are fundamentally limited by a density-accuracy tradeoff because they solve an independent, local blind~\cite{joshi2008psf,schuler2012blind,hirsch2015self,gwak2020modeling,hu2011psf,cho2011blur} or non-blind~\cite{rav2005two,yuan2007image,brauers2010direct,schuler2011non,heide2013high} deconvolution problem for each PSF, which becomes degenerate as the image patches used for deconvolution become smaller and/or the PSFs spatial extent becomes larger~\cite{mannan, delbracio2012non}. As a result, these methods sample PSFs sparsely, limiting their ability to model PSFs that vary rapidly along one or more dimensions~\cite{hansen2006deblurrinq}. \edit{Although in theory these limitations can be side-stepped by adhering to the PSF definition and imaging a point light source to capture the PSF directly, in practice it is difficult to construct an ideal point light source~\cite{gkioulekas2015micron}, i.e., a source with a perfectly circular shape that projects to a region smaller than a pixel and has uniform radiant intensity. Moreover, sources with the micron-scale sizes required for the tiny pixels in modern smartphones are subject to diffraction~\cite{yang2021designing, shih2012image} and have very low light efficiency, making it even more difficult to treat their images as that of a ``ground truth PSF". For all these reasons, accurate estimation of spatially varying PSFs has long relied on deconvolution of known patterns~\cite{joshi2008psf} as opposed to direct imaging.}

Optical simulation systems~\cite{tseng2021differentiable,zemax,sitzmann2018end2end, metzler2020deep, baek2021single}, on the other hand, assume that a physical model of the complete camera-plus-lens system is known precisely and that the corresponding real-world devices conform to it exactly. In many cases, however, precise models of widely-used compound lens systems may be proprietary or unknown~\cite{yang2022sub}, and their precision cannot be taken for granted due to device age, mechanical impacts from daily use, manufacturing variabilities, etc. 
While many deep learning techniques tackle the blind~\cite{xu2017motion,li2018learning,ren2020neural,shajkofci2020spatially,sun2015learning, ren_2020_cvpr} or non-blind~\cite{sureau2020deep,rego2021robust,yanny2022deep, zhang2017learning, dong2020deep} deconvolution problems, none of these methods apply to the high-dimensional\edit{, spatially-varying} cases we consider. Moreover, these methods rely on a discrete PSF representation instead of an MLP and thus suffer from the curse of dimensionality as PSF dimensions increase.
Consequently, the recovery of accurate, densely-sampled, high-dimensional PSFs for commodity cameras has remained beyond the reach of the broader computational photography community due to these technical obstacles.

\subsection{Contributions} In this paper, we seek to fill this gap by introducing a novel neural representation of blur---\textit{the lens blur field}---along with a practical method of acquiring it. 
The lens blur field represents a camera's PSF as a multi-layer perceptron (MLP) that is designed to capture blur variations across sensor position, focus setting and, optionally, depth. This representation is particularly suitable for PSF acquisition because
it yields a continuous representation of the PSF across multiple dimensions instead of sparse discrete samples; it can be optimized by leveraging the advanced MLP training schemes already available; \edit{it can learn high-dimensional signals more accurately while using only a fraction of the parameters of non-parametric approaches~\cite{xie2022neuralfields, takikawa2023compact};}
 and it can accurately capture difficult-to-model effects such as defocus, chromatic aberration, and higher-order aberrations with relatively few parameters. \edit{Overall, the lens blur field diminishes the tradeoff between representational accuracy and ease of acquisition by leveraging the interpolation properties of MLPs~\cite{ramasinghe2022regularizingcoordinatemlps, ramasinghi2022frequency}, enabling scalable modeling within the input dimensions while accurately capturing detailed optical lens effects. }

Recovering the lens blur field involves training its MLP on a dataset of focal stacks of a computer screen that can be easily acquired in less than five minutes. Our approach is applicable to a wide array of devices, from smartphone cameras and dual-pixel sensors to single-lens reflex (SLR) cameras and lenses and we show how to reconstruct 5D and 6D blur fields from such data.

Our method offers the following key technical contributions over the state of the art:
\begin{itemize}
    \item We develop a PSF acquisition procedure that enables reliable, fast, lightweight acquisition of continuous, high-dimensional PSFs of commodity cameras. 
    
    \item We reduce the problem of high-dimensional deconvolution to the problem of training a continuous MLP, whose network architecture can remain fixed across all devices we consider (smartphones, SLR lenses). 

    \item We assemble a first-of-its-kind dataset of 5D and 6D lens blur fields.
\end{itemize}

Our work provides new insights and enables new applications. 
For example we show that PSFs vary considerably between smartphone models, including the iPhone 12 Pro and iPhone 14 Pro; that dual-pixel auto-focus systems have PSFs that depart significantly from the analytical models used hitherto; and that lens blur fields are reliable and expressive enough to reveal subtle differences in the optical behavior of smartphone devices of the same make and model, yielding a device-specific ``blur signature''. 
We also show that lens blur fields can be used as space-efficient computational proxies of device optics, allowing us to apply high-dimensional blurs to render images with device-specific PSFs (\cref{fig:siemens-star}). 
Taken together, we provide new ways to both model and simulate the performance of common imaging systems, including smartphone cameras used by billions worldwide.

\begin{figure*}[ht!]
\centering
\includegraphics[width=0.90\textwidth]{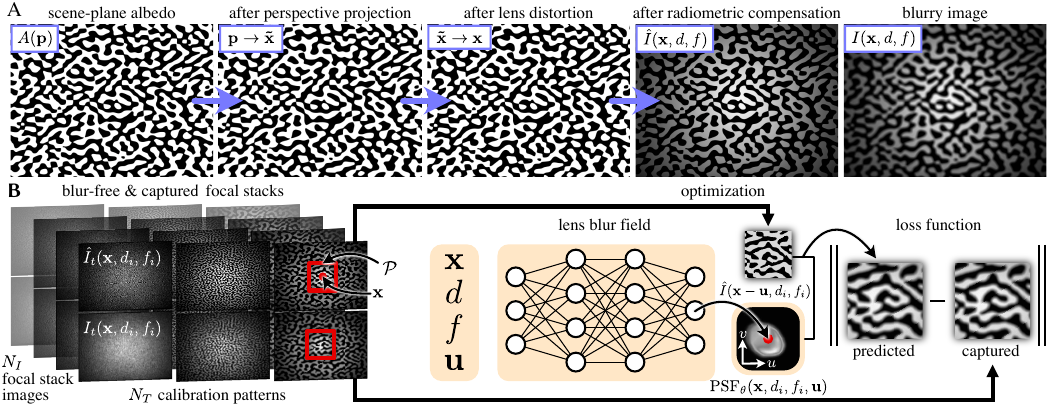}
\caption{Overview of learning lens blur fields. (A) Our approach learns a blur field by solving a non-blind deconvolution problem between blur-free calibration patterns $\hat{I}$ and blurry images $I$ of the patterns. Blur-free images are computed by applying perspective projection, lens distortion, and radiometric compensation to the albedo $A$ at the scene plane (top left). 
(B) An MLP is used to represent the lens blur field, which parameterizes the PSF over sensor position $\mathbf{x}$, target distance $d_i$, focus distance $f_i$, and a displacement $\mathbf{u}$ from $\mathbf{x}$. We train the MLP on a dataset of blur-free and captured focal stack images of \edit{synthetic noise} patterns~\cite{couture2011unstructured} and optimize MLP parameters to minimize the error between predicted and captured blurry image patches. \edit{Choice of noise patterns are discussed in Supplement Sec. S-IV. }
} 
\label{fig:setup}
\end{figure*}

\section{Learning Lens Blur Fields}

Our approach leverages multilayer perceptrons (MLPs) both for modeling the high-dimensional PSF of a camera system and for acquiring it from image data. 

\subsection{Lens Blur Field Representation}
Consider an incoherent point light source $\mathbf{p}\in\mathbb{R}^3$ at distance $d$ from the front pupil plane of a lens (see \cref{fig:setup}).
Let $\mathbf{\tilde{x}} \in \mathbb{R}^2$ be the perspective projection of point $\mathbf{p}$ onto the camera's sensor plane and let $\mathbf{x} \in \mathbb{R}^2$ be the point's actual projection after accounting
for non-linear geometric distortions caused by the lens. Since a camera's focus setting $f$ may affect the overall magnification of captured images~\cite{herrmann2020learning} and since dual-pixel AF sensors induce disparity between left-looking and right-looking pixels,
in general $\mathbf{\tilde{x}}$ and $\mathbf{x}$ depend on both $f$ and the pixel type $c$. In the following, we express the one-to-one relation between geometrically-distorted points on the sensor plane and point sources at distance $d$ using
\begin{align}
	\mathbf{p} ~=~ \mathrm{invproj}(\mathbf{x}, d, f, c). 
    \label{eq:invproj}
\end{align}
The lens blur field, denoted by $\PSF_\theta(\mathbf{x}, d, f, \mathbf{u})$, is an MLP with trainable parameters $\theta$ that describes the blurry 2D image of a light source at $\mathbf{p}$ 
as a function of the displacement $\mathbf{u} \in \mathbb{R}^2$ from $\mathbf{x}$.

For camera sensors with $C>1$ types of pixels we consider the MLP to be a vector-valued function with $C$ output dimensions. 
 For example, $C=4$ for a sensor with a conventional RGB Bayer color filter array, $C=2$ for a monochrome sensor with dual-pixel autofocus~\cite{xin2021defocus} and $C=8$ for a color dual-pixel AF sensor.

\subsection{Continuous Image Formation}
\label{ssec:blurfield}
Now suppose we image a fronto-parallel plane at distance $d$ from the lens.
We express the plane's blur-free image for pixel type $c$ and focus setting $f$ as the product
\begin{align}
	\hat{I}^{(c)}(\mathbf{x}, d, f) ~=~ A^{(c)}(\mathbf{p})~E^{(c)}(\mathbf{p}, f)\edit{,}
\end{align}
where $A^{(c)}(\mathbf{p}) \in [0,1]$ denotes the plane's unit-less albedo at point $\mathbf{p}$ and 
$E^{(c)}(\mathbf{p}, f)$ is the total irradiance incident on the sensor plane 
due to light falling onto (or emitted from) point $\mathbf{p}$. Intuitively, the albedo $A^{(c)}(\mathbf{p})$ represents the
plane's spatially-varying texture (or emission pattern) whereas the factor $E^{(c)}(\mathbf{p}, f)$
accounts for all camera- and plane-specific radiometric quantities such as vignetting, cosine-fourth fall-off~\cite{aggarwal2001cosine}, the plane's reflectance or emission properties, and the solid angle
at $\mathbf{p}$ subtended by the lens pupil~\cite{kolb1995realistic}.

The plane's blurry image is then given by the integral 
\begin{align}
    \label{eq:image_formation}
    I^{(c)}(\mathbf{x}, d, f) = \int_\mathcal{U} \PSF^{(c)}_{\theta}(\mathbf{x}, d, f, \mathbf{u}) \cdot \hat{I}^{(c)}(\mathbf{x} - \mathbf{u}, d, f)   \, \mathrm{d}\mathbf{u},
\end{align}
where $\mathbf{u}$ is a displacement from the projection of each point $\mathbf{x}$ on the sensor plane.
Integration is performed over $\mathcal{U}$, which is the spatial support of the blur field over $\mathbf{u}$.

\paragraph{Learning the Blur Field}
As illustrated in \cref{fig:setup}, we optimize the parameters of the blur field MLP by minimizing a photometric loss between a set of captured blurry  images of known 2D patterns and corresponding rendered images estimated using \cref{eq:image_formation}. 

Specifically, we solve the following optimization problem.
\begin{equation}
    \resizebox{0.96\columnwidth}{!}{%
    $
    \begin{aligned}
        &\argmin_{\theta} \sum_{\substack{t,c,i\\\mathbf{x} \in \mathcal{P}}} \Bigl\lVert I^{(c)}_{t}(\mathbf{x}, d_i,f_i) \nonumber - \sum_{\mathbf{u}} \PSF^{(c)}_\theta(\mathbf{x}, d_i, f_i, \mathbf{u}) \cdot \hat{I}_t^{(c)}(\mathbf{x} - \mathbf{u},d_i,f_i) \Bigr\rVert^2_2, \\
        &\text{subject to } \PSF_\theta \geq \mathbf{0}\ \ .  \nonumber
    \end{aligned}
    $
    }
    \label{eqn:objective}
\end{equation}

The outer summation is performed over 
all pixel channels $c$, all sensor points $\mathbf{x}$, focus and distance settings $f_i$ and $d_i$, $1\leq i \leq N_I$, and for each of $N_T$ known patterns ($1 \leq t \leq N_T$).
The inner summation is an approximation of the integral formulation of \cref{eq:image_formation} with a discrete sum over $\mathbf{u}$. 
We consider two instances of the training procedure depending on whether or not the distance $d_i$ is the same for all input images.
In the former case $d$ is fixed, yielding a 5D blur field; otherwise we optimize for a 6D blur field.

\paragraph{Implementation Details}
We train on raw captured images, with the summation over $\mathbf{x}$ in \cref{eqn:objective} taken only over pixels for which we have image measurements for a given channel. For example, for a sensor with a Bayer color filter array, we supervise the red, green, and blue channels only on coordinates $\mathbf{x}$ that are explicitly measured.  

The MLP uses seven fully-connected layers of 512 channels each with ReLU activation functions.
The final layer has a sigmoid activation to satisfy the non-negativity constraint.
Unlike recent work employing MLPs, we do not apply positional encoding to the MLP's input coordinates~\cite{tancik2020fourier} as we find it unnecessary for accurate PSF acquisition.
Blur fields are trained for between $4\times 10^6$ and $10\times 10^6$ iterations with the Adam optimizer ($\beta_1 = 0.5$, $\beta_2 = 0.999$) on a single NVIDIA A6000 using the \texttt{tiny-cuda-nn} framework~\cite{tiny-cuda-nn}.
Details on other hyperparameters (learning rate, exponential decay, \emph{etc.}) are included in Supplement Sec. S-III. \edit{Ablations on MLP architecture are included in Supplement Sec. S-IV.}

Additionally, to improve computational efficiency, we assume that the blur field is locally stationary and neglect variations over sufficiently small image patches $\mathcal{P}$ centered at $\mathbf{x}$.
This allows us to (1) reduce the number of computationally-expensive MLP forward passes to a single forward pass per patch and (2) formulate the inner summation as a matrix-vector product that can be efficiently parallelized on GPUs. Note that despite employing a patch-based forward model, our approach does not suffer from the density-accuracy tradeoff characteristic of local deconvolution methods~\cite{hirsch2010efficient}. These methods rely
on patches to solve a local \emph{inverse} problem, thereby restricting the pixels contributing to its solution and reducing the solution's signal-to-noise ratio. Here, deconvolution is performed globally and implicitly by the MLP, with all pixels across all images potentially contributing to it. 

During training, we set one batch element to comprise a 2D PSF obtained by first randomly sampling a focus setting $f_i$ and a point $\mathbf{x}$ at a pixel center. 
Then we sample a grid of points in the two dimensions of $\mathbf{u}$ within the patch around $\mathbf{x}$ and query the MLP at these points to retrieve a 2D PSF. 
\cref{eqn:objective} is computed using the output of the convolution of the 2D PSF with $\hat{I}$ within the area of the patch.
In our experiments, we use a batch size of one and allow 2D PSFs up to $120 \times 120$ samples in size. 
The size of a patch $\mathcal{P}$ is set to be 1.5 times the size of the 2D PSF along each dimension, and the convolution is performed within the valid region of the patch to avoid boundary artifacts.

\section{Image Acquisition Procedure}

\label{sec:calibration}
We describe the procedure to acquire the blur-free and captured blurry images of the calibration patterns used to optimize the lens blur field.  
The steps of the procedure are outlined in \cref{fig:setup} (A) and consist of first computing the blur-free calibration image by applying (1) perspective projection via homography, (2) lens distortion, and (3) radiometric compensation to the scene-plane albedo $A$.

\subsection{Homography Estimation}
The perspective projection of the scene-plane albedo $A(\mathbf{p})$ onto the sensor plane (i.e., $\mathbf{p}\rightarrow\mathbf{\tilde{x}}$---see \cref{fig:setup}) is computed via a homography $H(\mathbf{\tilde{x}}, d_i, c)$.
To estimate the homography, we require a pattern that enables robust point localization with sub-pixel accuracy, even in the presence of large defocus. 
To this end, we use two dot-grid calibration patterns consisting of white circles on a black background and black circles on a white background (shown, right).
We compute the albedo dot-grid images and capture in-focus dot-grid images for which we also remove radial distortion using the procedure described in the next \edit{subsection}.
Following Scharstein and Szeliski~\cite{scharstein2003high}, we binarize the dot-grid images and detect the pixel coordinates of the dot centers in each image.
Then we refine the dot center locations in both images using a center-of-mass calculation on the dots in the original albedo and captured images.
Finally, we compute the homography using RANSAC~\cite{fischler1981random} between the dot center point correspondences in the images.
Applying the homography to the scene-plane albedo as $A(H^{-1}(\mathbf{\tilde{x}}, d_i, c))$ gives the sensor-plane albedo.

\subsection{Lens Distortion Estimation}
Now, we wish to map the sensor-plane albedo to an image of albedo that includes the effects of magnification due to changes in focus setting and the effects of lens distortion.
To account for magnification changes, we acquire an undistorted captured focal stack of dot-grid images for each distance $d_i$ and solve for the scale transformations $S(\mathbf{\tilde{x}}, d_i, f_i,  c)$ that align the dot centers from each out-of-focus image to the single in-focus image.
We apply the resulting scale transformations (one for each focus setting and distance) to the sensor-plane albedo image computed after homography estimation.

\sloppy{
    To estimate the lens distortion we use the sixth-order Brown-Conrady model of radial distortion implemented in OpenCV's \texttt{calibrateCamera} function~\cite{opencv_library} and apply it to the dot-grid center positions of the captured in-focus image described in the homography estimation step.
We apply this model of radial distortion $D(\mathbf{\tilde{x}}, c)$ to all scaled sensor-plane albedo images over $f_i$ and $d_i$. 
The composition of homography $H$, scale transformation $S$, and radial distortion $D$ define the \text{invproj} operator introduced in \cref{eq:invproj}, which maps points $\mathbf{x}$ on the sensor plane to points $\mathbf{p}$ on the scene plane. }

\subsection{Radiometric Compensation}
Next, we apply radiometric compensation to the distorted scene-plane albedo computed for a given $d_i$ and $f_i$.
To do this, we capture two images, $I_{\text{min}}$ and $I_\text{max}$ under an all-black and an all-white pattern, corresponding to albedos $A(\mathbf{p}) = A_\text{min}  > 0$ and $A(\mathbf{p}) = A_\text{max}$, respectively.
These images can be thought of as scaled versions of the total irradiance $E(\mathbf{x},d_i,f_i)$ because this factor is nearly constant over the monitor’s plane and thus it is not affected by the lens blur even when the 2D PSF at $\mathbf{x}$ has a very large spatial extent:
\begin{equation}
    \resizebox{0.95\columnwidth}{!}{%
    $
    \begin{aligned}
        I_\text{min}(\mathbf{x})  &= A_\text{min} \int_\mathbf{u}  \text{PSF}_\theta(\mathbf{x},d_i,f_i,\mathbf{u}) \cdot E(\text{invproj}(\mathbf{x},d_i,f_i)-\mathbf{u})  \mathrm d\mathbf{u} \\
                                 &\approx A_\text{min} \cdot E(\text{invproj}(\mathbf{x},d_i,f_i)).
    \end{aligned}
    $
    }
    \label{eq:radiometric}
\end{equation}

By combining \cref{eq:radiometric} and its analogue for $I_\text{max}$ which follows trivially, the blur-free image $\hat{I}$ can now be expressed as an affine combination of $I_\text{min}$ and $I_\text{max}$.
\begin{align}
    \hat{I}(\mathbf{x}) &= \left[1-A(\text{invproj}(\mathbf{x},d_i,f_i))\right]\cdot I_\text{min}(\mathbf{x}) \\&+ A(\text{invproj}(\mathbf{x},d_i,f_i)) \cdot I_\text{max}(\mathbf{x})\nonumber.
\end{align}

\subsection{Image Capture}
Finally, we describe the procedure for capturing focal stack images displayed on a monitor.
We use a 32-inch 4K LED monitor to display the calibration patterns for both smartphone and SLR captures. We mount each camera on a tripod and position it fronto-parallel to the monitor.
Depending on the distance $d$ to the scene plane, the image of the calibration pattern may span the entire sensor plane or just a portion of it. 
In the former case, a single focal stack of images is captured for lens distortion estimation and radiometric compensation. 
In the latter, the camera and/or monitor are physically moved to maintain $d$ while also adjusting the patterns to project to different parts of the sensor plane (see result in \cref{sec:results}).
For smartphone experiments, focal stacks are automatically captured using software, and we sample a range of focus positions denoted as $N_I \in \{7, 24\}$, which depends on hardware and device limitations.
Finally, we set the ISO to the lowest possible value and capture bursts with exposure times of 100~ms (for iPhones) or 250~ms (for Pixel and SLR) to avoid capturing screen refresh, PWM, and flicker. Supplement Sec. S-II contains additional details on the acquisition procedure for all devices in our dataset.

\begin{figure*}[!t]
\centering
\includegraphics[width=0.95\linewidth]{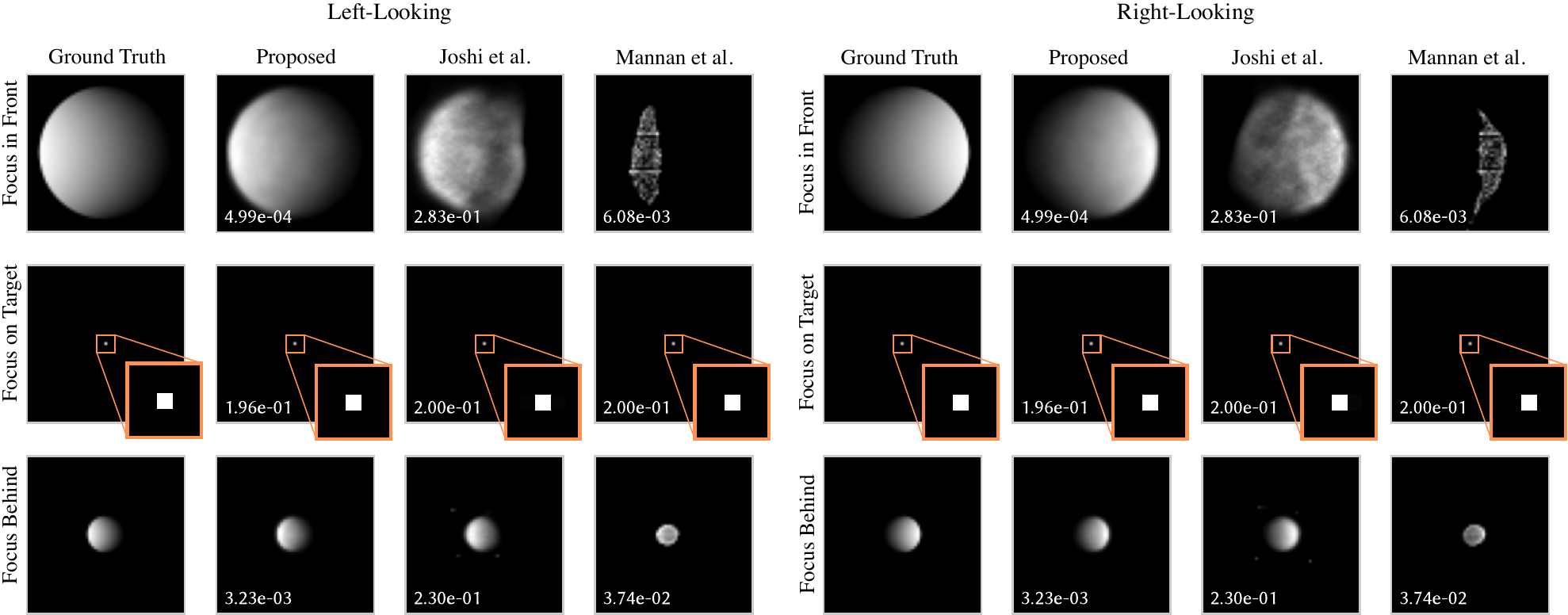}
\caption{Simulated recovery of spatially invariant dual-pixel blur kernels created with the parametric models proposed by Punnappurath et al.~\cite{punnappurath2020modeling}. Each sub-panel gives a side-by-side comparison of the ground truth (left), our recovered PSF (middle left), the result of Joshi et al.~\cite{joshi2008psf} (middle right), and Mannan \& Langer~\cite{mannan} (right).
Root-mean-square error (RMSE) is estimated on the inset regions shown in  \cref{fig:simulations} to mitigate bias from large regions of zero-valued pixels. The values are shown in the bottom left corner of each plot. \edit{Supplement Sec. S-IV contains more extensive comparisons and cross-sectional plots.}
}
\label{fig:simulations}
\end{figure*}

\section{Results}
\label{sec:results}
We evaluate our method in simulation and on captured data, and perform quantitative and qualitative comparisons to other non-blind PSF estimation methods.
We also describe recovered blur fields for a wide range of lenses and devices, ranging from iPhone and Google Pixel cameras to varying SLR lenses. 

\subsection{Simulated Evaluation}
To evaluate our method in simulation, we use calibration patterns blurred with a known PSF.
Specifically, we use the synthetic noise patterns~\cite{couture2011unstructured} \edit{shown in~\cref{fig:setup}} and blur them with a spatially-invariant parametric model of lens blur for dual-pixel cameras~\cite{punnappurath2020modeling}.
The focus is adjusted by varying the radius of the PSFs using the thin lens model~\cite{malacara2003handbook}, with a 30 mm focal length, 40 cm calibration target distance, and linear sampling in diopter space.
The maximum and minimum defocus blur radii are set to 24 and 0.5 pixels respectively. The resulting focal stack comprises $N_I=21$ images for each synthetic noise pattern. Gaussian read noise with a standard deviation of 0.01 is added to the blurred images. Additionally, we simulate a 16-bit capture with Poisson noise. Overall, our noise model incorporates read noise (Gaussian), shot noise (Poisson), and ADC (quantization) noise.

We train on the paired blur-free and blurry images, holding out every other image in the focal stack for validation. 
The proposed approach is compared to PSFs reconstructed using the methods of Joshi et al.~\cite{joshi2008psf} and Mannan et al.~\cite{mannan} across focus settings.
Both baseline methods employ constrained least squares optimization, penalizing deviations between patches in predicted and captured data.
Mannan et al.\ include an additional penalty over filtered versions of the predicted and captured data (\emph{e.g.,} to facilitate reconstructing edges). ~\cref{fig:simulations} compares the ground truth PSFs to those recovered by our method and the two baselines.

Qualitatively, our lens blur field recovers PSFs closer to the ground truth: they have smooth variations with far fewer of the high-frequency artifacts that appear in other methods as shown in ~\cref{fig:simulations}.
We find that the proposed method also significantly outperforms the recovered PSFs in terms of quantitative accuracy, as evaluated using root-mean-square error (RMSE), indicated for each PSF displayed in ~\cref{fig:simulations}. 

We also evaluate how well the predicted PSFs estimate the blurry calibration patterns and report the peak signal-to-noise ratio, RMSE, and structural similarity between the predicted and ground truth blurry patterns. To assess performance at withheld focal stack positions, we query our MLP at the corresponding focus setting. For the other methods, we perform linear interpolation between the nearest optimized PSFs. We find that our method outperforms the baselines when evaluating on the validation lens positions (i.e., positions not seen during training or optimized explicitly by the baselines) as shown in \cref{tab:image_recon}. The image reconstruction metric is less sensitive to the high-frequency artifacts in the PSF obseved in Joshi et al.'s method~\cite{joshi2008psf}, and so their approach performs roughly on par with our method in terms of overfitting to the blurry image. Supplement Sec. S-IV contains additional \edit{comparisons with neural deconvolution baselines}. \edit{Results in~\cref{tab:image_recon} also show that directly obtaining PSF predictions at validation lens positions from the MLP (using the MLP's interpolation) outperforms a baseline that obtains PSFs predictions only at training lens positions and then uses linear interpolation to estimate the validation lens positions.}

\begin{table*}
  \centering
  \resizebox{0.99\textwidth}{!}{
  \renewcommand{\arraystretch}{1.25}
  \begin{tabular}{lc@{\qquad}ccc@{\qquad}cccc}
    \toprule
    \multirow{2}{*}{\raisebox{-\heavyrulewidth}{}} && \multicolumn{3}{c}{Training Lens Positions} & \multicolumn{3}{c}{Validation Lens Positions} \\
    \cmidrule{3-9}
    && Proposed & Joshi et al.~\cite{joshi2008psf} & Mannan \& Langer~\cite{mannan} 
    & Proposed & Joshi et al.~\cite{joshi2008psf} & Mannan \& Langer~\cite{mannan} & \edit{Lin. Interpolation + Proposed}\\
    \midrule
    \multirow{3}{*}{\raisebox{-\heavyrulewidth}{\makecell[l]{Training \\ Patterns}}}
                & PSNR $\uparrow$ & $ 32.109 \pm 1.201$ & $ 31.829 \pm 1.016$ & $ 15.420 \pm 0.381$ 
                                  & $ 30.462 \pm 3.850$ & $ 30.797 \pm 3.172$ & $ 15.291 \pm 0.513$ & \edit{$29.394 \pm 3.607$}\\
                & SSIM $\uparrow$   & $ 0.929 \pm 0.072$ & $ 0.950 \pm 0.071$ & $ 0.601 \pm 0.021$ 
                                    & $ 0.924 \pm 0.066$ & $ 0.944 \pm 0.066$ & $ 0.600 \pm 0.024$ & \edit{$0.898 \pm 0.077$}\\
                & RMSE $\downarrow$ & $ 0.022 \pm 0.004$ & $ 0.025 \pm 0.003$ & $ 0.166 \pm 0.010$ 
                                    & $ 0.031 \pm 0.023$ & $ 0.031 \pm 0.013$ & $ 0.169 \pm 0.013$ & \edit{$0.034 \pm 0.022$}\\
    \midrule
    \multirow{3}{*}{\raisebox{-\heavyrulewidth}{\makecell[l]{Validation \\ Patterns}}}
                & PSNR $\uparrow$ & $ 32.051 \pm 1.179$ & $ 31.981 \pm 1.277$ & $ 16.753 \pm 3.189$ 
                                  & $ 30.396 \pm 3.840$ & $ 28.964 \pm 4.493$ & $ 16.400 \pm 3.086$ & \edit{$29.351 \pm 3.651$}\\

                & SSIM $\uparrow$   & $ 0.924 \pm 0.078$ & $ 0.916 \pm 0.103$ & $ 0.461 \pm 0.290$ 
                                    & $ 0.919 \pm 0.072$ & $ 0.874 \pm 0.100$ & $ 0.452 \pm 0.278$ & \edit{$0.887 \pm 0.091$}\\
                & RMSE $\downarrow$ & $ 0.022 \pm 0.004$ & $ 0.022 \pm 0.003$ & $ 0.139 \pm 0.079$ 
                    & $ 0.031 \pm 0.023$ & $ 0.038 \pm 0.029$ & $ 0.145 \pm 0.077$ & \edit{$0.034 \pm 0.023$}\\
    \bottomrule
  \end{tabular}}
  \caption{Evaluation of reconstruction accuracy on synthetic noise patterns. The proposed approach quantitatively outperforms non-parametric, optimization-based methods in terms of peak signal-to-noise ratio (PSNR) and root-mean-square error (RMSE). Validation patterns have the same frequencies as the training set. \edit{MLP interpolation outperforms linear interpolation (denoted as ``Lin. Interpolation + Proposed"), where PSFs at validation lens positions are obtained by linearly interpolating the MLP predictions at the training lens positions. The difference is most pronounced near the in-focus position because uniform diopter sampling yields non-uniform metric distance sampling, and PSF size changes become more significant due to a derivative sign change near focus. MLP interpolation effectively handles the resulting complex, nonlinear defocus blur radius model derived from the thin lens equation.} Additional evaluations are included in Supplement Sec. S-IV.}
\label{tab:image_recon}
\end{table*}

Finally, we also validate our approach using experimental data, as presented in \cref{tab:image_valid}. We blur a ground-truth calibration image with our learned blur fields and compare the output to real-world blurry captures. Metrics are computed on both training and a held-out validation dataset, the latter containing calibration patterns not seen during training. Our method also generalizes well to the validation dataset.

\begin{table}[!t]
  \centering
  \resizebox{0.45\textwidth}{!}{
  \renewcommand{\arraystretch}{1.25}
  \begin{tabular}{l@{\qquad}l@{\qquad}l@{\qquad}l}
    \toprule
    & &\multicolumn{1}{c}{PSNR $\uparrow$}  &\multicolumn{1}{c}{SSIM $\uparrow$} \\
    \midrule
    \multirow{9}{*}{\rotatebox{90}{Training Dataset}} 
    & iPhone 14 Pro Device 0 & $ 36.969 \pm 1.926$ & $ 0.976 \pm 0.0158$ \\
    & iPhone 14 Pro Device 1 & $ 38.715 \pm 1.806$ & $ 0.977 \pm 0.0157$ \\
    & iPhone 12 Pro Device 0 & $ 36.668 \pm 1.587$ & $ 0.972 \pm 0.0192$ \\
    & iPhone 12 Pro Device 1 & $ 36.789 \pm 1.593$ & $ 0.959 \pm 0.0369$ \\
    & Pixel 4 & $ 29.940 \pm 1.635$ & $ 0.964 \pm 0.0201$ \\
    & Canon EF 14mm f/2.8L & $ 37.276 \pm 1.937$ & $ 0.996 \pm 0.0028$ \\
    & Canon EF 24-70mm f/2.8L & $ 32.293 \pm 0.725$ & $ 0.985 \pm 0.0476$ \\
    & Canon EF 50mm f/1.2L & $ 31.068 \pm 0.411$ & $ 0.987 \pm 0.0225$ \\
    & Canon EF 50mm f/1.4 & $ 33.286 \pm 0.301$ & $ 0.991 \pm 0.0231$ \\
    \midrule
    \midrule
    \multirow{9}{*}{\rotatebox{90}{Validation Dataset}} 
    & iPhone 14 Pro Device 0 & $ 36.736 \pm 1.949$ & $ 0.973 \pm 0.0196$ \\
    & iPhone 14 Pro Device 1 & $ 38.232 \pm 1.814$ & $ 0.976 \pm 0.0175$ \\
    & iPhone 12 Pro Device 0 & $ 32.605 \pm 0.650$ & $ 0.930 \pm 0.0303$ \\
    & iPhone 12 Pro Device 1 & $ 31.046 \pm 0.804$ & $ 0.904 \pm 0.0370$ \\
    & Pixel 4 & $ 24.559 \pm 1.974$ & $ 0.916 \pm 0.0260$ \\
    & Canon EF 14mm f/2.8L & $ 37.273 \pm 1.919$ & $ 0.996 \pm 0.0028$ \\
    & Canon EF 24-70mm f/2.8L & $ 31.531 \pm 0.633$ & $ 0.982 \pm 0.0484$ \\
    & Canon EF 50mm f/1.2L & $ 31.134 \pm 0.399$ & $ 0.987 \pm 0.0216$ \\
    & Canon EF 50mm f/1.4 & $ 33.160 \pm 0.302$ & $ 0.991 \pm 0.0225$ \\
    \bottomrule
  \end{tabular}}
  \caption{Models for each device are validated by comparing artificially blurred images to blurry captured images, for both training and unseen validation datasets. In general, lenses across both the training and validation sets collectively exhibit comparable performance. Our method also produces consistent performance across training and validation sets for most devices.}
  \label{tab:image_valid}
\end{table}

\subsection{Comparison to Prior Methods} The proposed method outperforms prior methods for dense stacks as shown in \cref{tab:comparisons}. While a direct comparison of runtime and memory for a single PSF favors prior work, our method scales better with the number of images---a requirement for accurate calibration. For the example shown in \cref{tab:comparisons}, consider optimizing a 5D PSF for a 12 MP (4032x3024) iPhone 12 Pro; if we optimize blur kernels at 75x100 sensor positions (i.e., assume each blur kernel is constant within a 40 pixel area), 15 focus positions, 73x73 blur kernel samples, and 4 RGGB color channels, then Joshi et al. (2008) would require 34 days for optimization, Mannan and Langer (2016) would require 1945 days, and our approach requires 14 hours. Moreover storing this discretely sampled 5D PSF requires 9.3 GB whereas our MLP requires 19.1 MB. Prior methods were also observed to be more noise-sensitive, often produce estimations with high-frequency artifacts, and deteriorates for large PSFs (see \cref{fig:simulations} and Supplement Sec. S-IV).

\begin{table}
  \centering
  \resizebox{0.48\textwidth}{!}{
  \renewcommand{\arraystretch}{1.25}
    \begin{tabular}{llcc}
      \toprule
       &  & Optimization Time & Memory Usage \\
      \midrule
      \multirow{2}{*}{Proposed} & Single PSF & N/A & N/A \\
                                & 5D PSF  & 14 hours & 19.1MB \\
      \midrule
      \multirow{2}{*}{Joshi et al.~\cite{joshi2008psf}}& Single PSF & 6.5 sec & 39KB \\
                                                       & 5D PSF  & 34 days & 9.3GB \\
      \midrule
      \multirow{2}{*}{Mannan \& Langer~\cite{mannan}}  & Single PSF & 373.4 sec & 39KB \\
                                                       & 5D PSF  & 1945 days & 9.3GB \\
      \bottomrule
    \end{tabular}
  }
  \caption{Optimization time and memory requirements for a single (73x73) blur kernel and (75x100x15x73x73) 5D PSF, where we consider optimizing a 5D PSF for a 12 MP (4032x3024) iPhone 12 Pro. If we optimize blur kernels at 75x100 sensor positions (i.e., assume each blur kernel is constant within a 40 pixel area), 15 focus positions, 73x73 blur kernel samples, and 4 RGGB color channels.}
  \label{tab:comparisons}
\end{table}

\subsection{Comparing iPhone 12 Pro and 14 Pro Lens Blur Fields}
\begin{figure*}[!ht]
\centering
\includegraphics[width=0.98\textwidth]{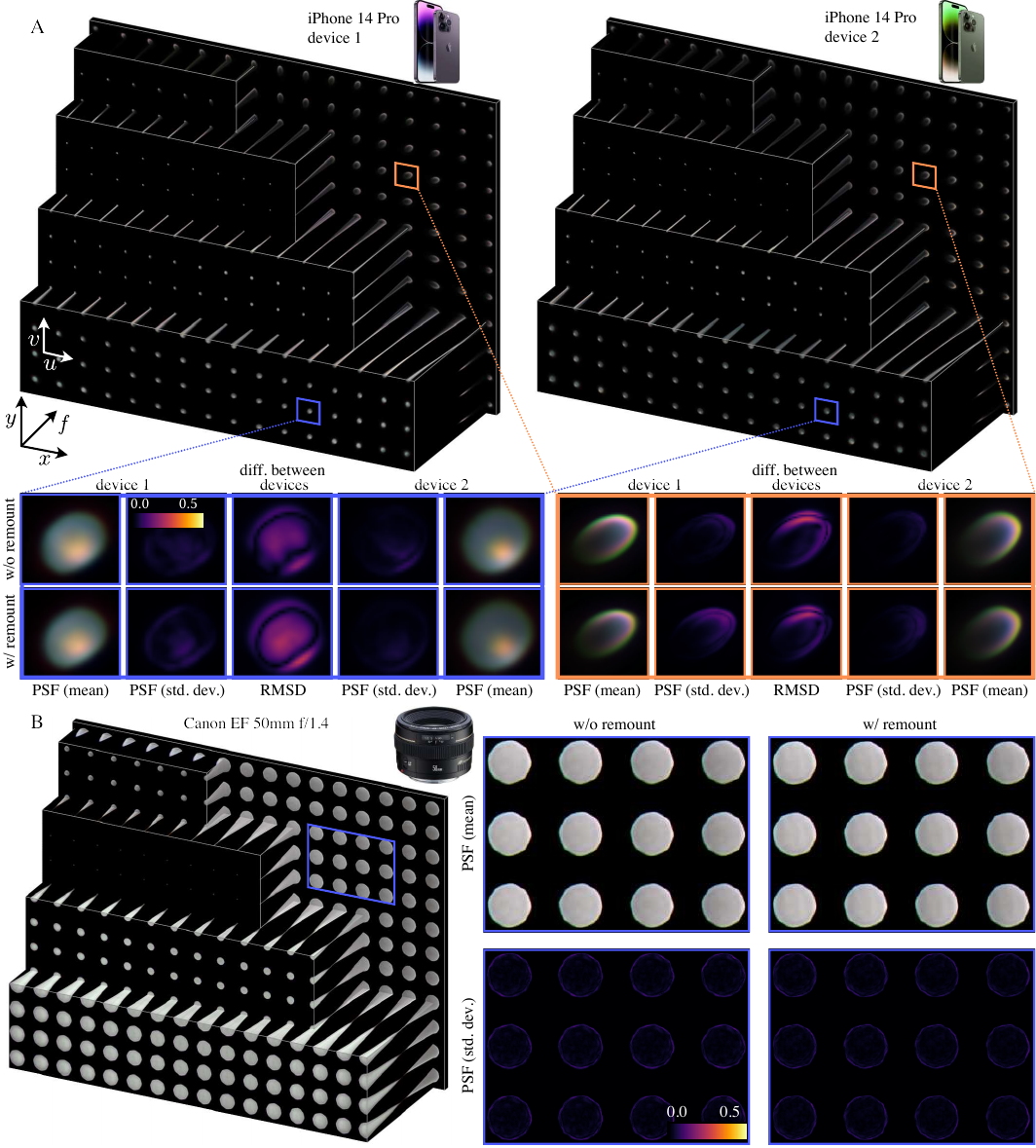}
\caption{(A) Comparison between calibrated PSFs from the wide cameras of two iPhone 14 Pro devices. We estimate PSFs using four separate sets of calibration measurements captured independently for each device, for each repeatability investigation. Visualizing slices of the mean estimated 5D PSF function for each device (top) reveals broad agreement in the PSF structure; however, significant variations can be seen upon close inspection (insets, bottom). While imperfections in the calibration and capture process result in differences between PSF estimates for the same device (see standard deviation of PSFs, bottom left and right), this effect is generally weaker than PSF differences between devices (bottom center, shown as root-mean-square deviation), suggesting that each device has a specific blur signature. (B) Comparison between static and dynamic experiments for a Canon EF 50mm f/1.4 lens. We capture five separate sets of measurements for each repeatbility experiment (with and without remount) and visualize the mean and standard deviations for each experiment with the same color scale. We observe little differences between experiment, indicating that remounting the lens does not alter the blur field of the lens.}
\label{fig:device-id}
\end{figure*}

Our method can resolve the differences in optical behavior between the wide-angle cameras on two separate iPhone 12 Pro devices (Supplement Sec. S-II) and two separate iPhone 14 Pro devices (\cref{fig:device-id}). To demonstrate this, we conduct two types of repeatability experiments. 

First, we assesses the variation in estimated blur fields for a single camera with successive captures, where the only difference is the closing and re-launching the acquisition app, while camera placement remains constant. 
This first type of evaluation measures the sensitivity of our method to device-centric parameters, such as focus-position repeatability or repeatability of camera software settings.
Second, we evaluate the variation in estimated blur fields after closing the acquisition app, unmounting, and remounting the phone, which assesses repeatability of the entire acquisition procedure. 

For both experiments, we use each device to capture 4 separate focal stack datasets and train a separate model for each repeatablity investigation (\emph{i.e.,} a total of 8 datasets per iPhone).
\cref{fig:device-id} shows recovered 5D lens blur fields and 2D PSF slices recovered for each device, averaged across the four trials.
We find that the standard deviation of estimated PSF values across trials is significantly smaller than the RMS difference between the mean PSFs of each device. Moreover, we see clear differences between the PSFs in the Supplement, especially near the PSF boundaries. Our method is sufficiently sensitive to uncover such cross-device differences without specialized hardware. Supplement Sec. S-II contains additional details on these repeatability experiments. 

\subsection{Visualizing xy-slices of the iPhone 12 Pro Blur Field}
In the supplementary video, we visualize xy-slices of the blur fields from two different iPhone 12 Pro devices of the same make and model, showcasing the high spatial frequency content and variations across the sensor plane. This analysis highlights the challenges of capturing and analyzing such complexity using local deconvolution on a sparse patch grid.

\subsection{Comparing SLR Lens Blur Fields}
Our methodology demonstrates robustness against dismounting and remounting of SLR lenses, indicating negligible optical axis misalignment between lens removals.
Repeatability experiments were conducted using a Canon EF 50mm f/1.4 lens for separate trials with and without removing and re-attaching the lenses (see \cref{fig:device-id}). 
In each experiment, we captured five focal stack datasets and trained individual models for each, resulting in a total of ten datasets. 
Comparing the blur fields between remount and without remount experiments showed insignificant differences.

\subsection{Reconstructing Dual-Pixel Lens Blur Fields}
Our proposed method enables reconstructing lens blur fields for left- and right-looking pixels on dual-pixel cameras, such as the Google Pixel 4 (shown in ~\cref{fig:teaser}).
We observe a complex dependency of the PSF on image plane location and focus setting, with the observed blur patterns departing radically from simpler parametric models for dual-pixel PSFs~\cite{punnappurath2020modeling}. Notably, we find that PSF centers of mass do not always follow the expected left-right orientations, and PSFs are significantly skewed at locations near the periphery of the image sensor.

\subsection{Additional SLR Lens Blur Fields}
\begin{figure*}[!ht]
    \centering
    \includegraphics[width=0.90\textwidth]{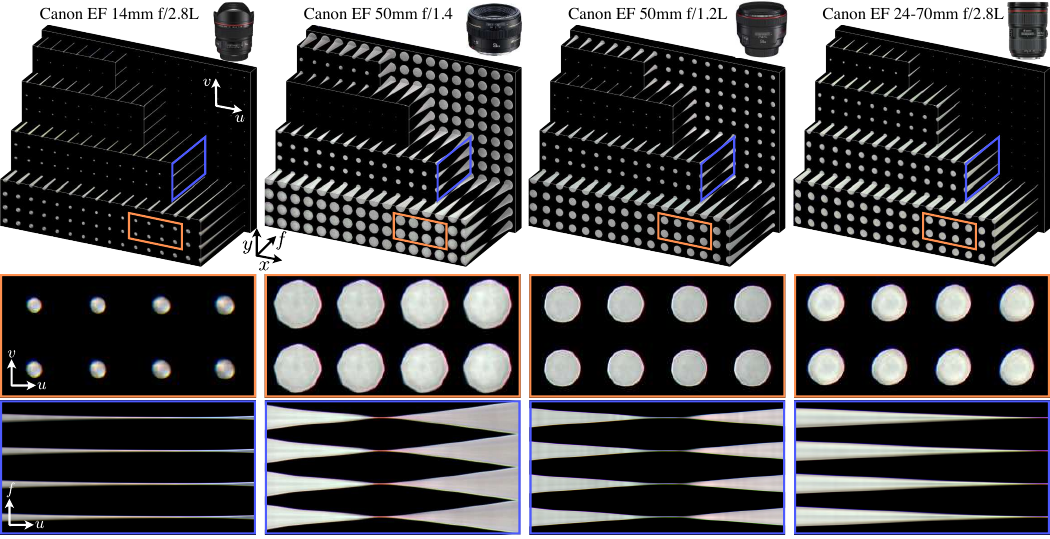}
    \caption{Reconstructed lens blur fields for a collection of Canon lenses. Our method recovers interesting features in the PSFs of the Canon EF 50 mm f/1.4 lens and the Canon EF 24--70 mm zoom lens (evaluated at the 50 mm focal position). For example, the f/1.4 lens has octagonal PSFs from the aperture blades, and the PSF of the zoom lens is distorted, perhaps due vignetting effects. The prime 14 mm f/2.8 and 50 mm f/1.2 lenses have round, smooth PSFs.}
    \label{fig:dslr-models}
    \end{figure*}

We recover lens blur fields for a variety of Canon SLR lenses of quality and type, shown in \cref{fig:dslr-models}. 
Our method is sensitive enough to recover various properties of the PSFs, including the octagonal shape of the 50 mm Canon f/1.4 lens (due to the aperture blades), and the irregular shape of the Canon 24--70 mm f/2.8L zoom lens (evaluated at 50 mm focal setting), which is most likely due to vignetting.
The Canon EF 14 mm f/2.8L and Canon EF 50 mm f/1.2L prime lenses show smooth, round defocus blur with minimal spatial variation.

\subsection{6D Lens Blur Fields}
\begin{figure}[!h]
  \centering
  \includegraphics[width=0.42\textwidth]{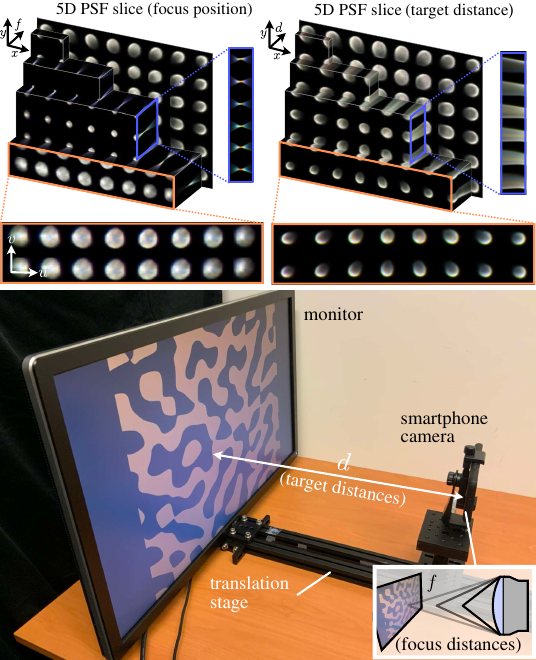}
  \caption{6D lens blur field calibrated for an iPhone 12 Pro wide lens and the capture setup for 6D blur fields. We show 5D slices of the 6D function, which vary over focus distance (top left) and target distance (top right), ranging from 20.5 cm to 41.5 cm. While we observe small changes in the PSF over target distance, the most dramatic changes occur over the focus dimension. This is expected as smartphone cameras have a small apertures, demonstrating that a 5D blur field may be a sufficient blur characterization for certain applications. The 6D capture setup comprises a monitor, a phone stand, and a translating stage that controls positioning along the target distance dimension.}
\label{fig:6D-model}
\end{figure}

\edit{A 6D blur field extends a 5D blur field by including an additional dimension, \(d\), for target distance.} We demonstrate recovery of the 6D lens blur field for an iPhone 12 Pro's wide lens in \cref{fig:6D-model}, illustrating our method's adaptability to increased spatial dimensions. \edit{We see less variation in the PSF over target distance dimension than in other dimensions, such as focus position. This behavior is consistent with that of a thin lens, where variations in target distance can be largely predicted from a 5D blur field.}

\subsection{Partial-Field-of-View Blur Field Acquisition}
\begin{figure}[!t]
  \centering
  \includegraphics[width=0.48\textwidth]{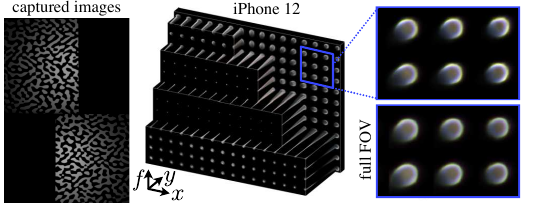}
  \caption{Demonstration of partial-field-of-view capture using an iPhone 12 Pro wide lens. Two sets of focal stacks are captured, spanning roughly 55\% of the image plane's width. We follow the same post-capture processing scheme for each set separately and train the network on both datasets. The reconstructed blur field yields similar results to the full-field-of-view (FOV) reconstruction (bottom right).}
\label{fig:tiling}
\end{figure}

Our method successfully acquires a full-sensor blur field, even in scenarios where the display spans a partial field of view of the sensor. ~\cref{fig:tiling} shows an experiment where one blur field was trained on two sets of captures, with the patterns partially observed in each set. 

\section{Rendering with Blur Fields}
Lens blur fields can be used as lightweight computational proxies for underlying device optics.
We demonstrate this principle with a few examples where we use it to render synthetic scenes with device-specific blur.

\subsection{Lens Blur Field Dataset}
To facilitate applications of lens blur fields in computational photography and rendering, we make available a dataset of lens blur fields calibrated across a range of smartphone cameras and SLR lenses. The dataset includes all lens blur fields described thus far (a complete listing of cameras and lenses included in the dataset is provided in Supplement Tab. S-III).

\subsection{Resolution Chart Experiment}
\begin{figure*}[!ht]
    \includegraphics[width=0.99\textwidth]{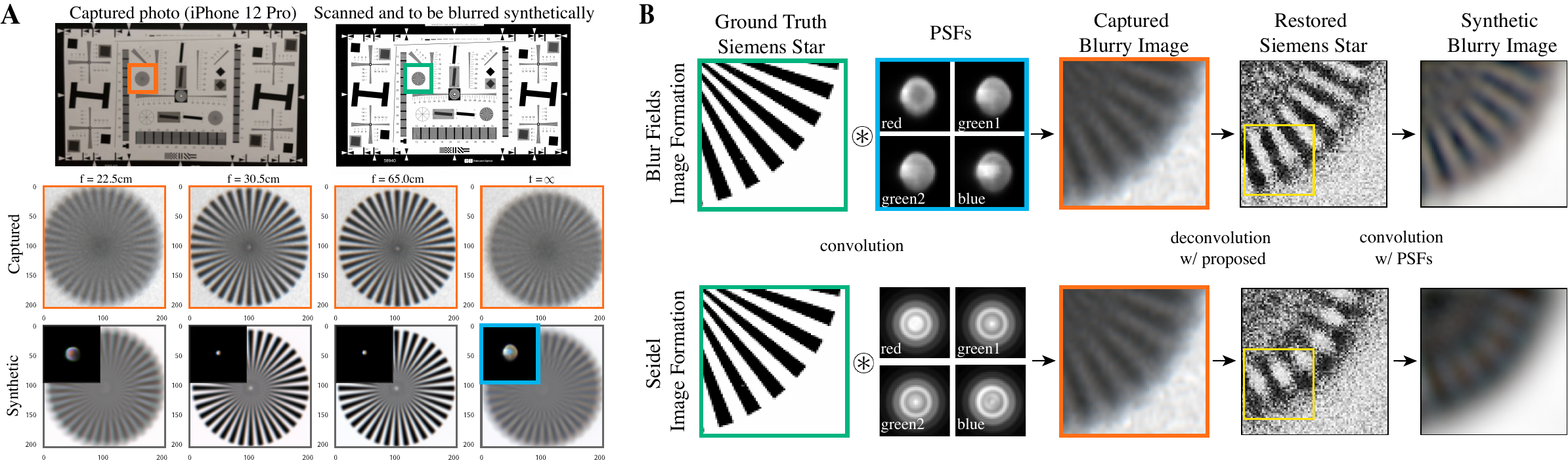}
    \caption{\edit{(A) Comparison of real captures and synthetically generated renders using a learned blur field from the iPhone 12 Pro wide lens, and (B) restoration results from these captures via non-blind deconvolution using our blur field PSFs versus a Seidel model fitted to our PSFs.} A resolution chart (Edmund Optics I3A/ISO 12233 Digital Camera Resolution Test Chart, \#58-940) is positioned at a distance of 36 cm from an iPhone 12 Pro. The chart is scanned at a resolution of 600 DPI and subsequently blurred at the different focus settings indicated, using the previously-captured blur field (visualized in ~\cref{fig:6D-model}). The 2D PSF at the center of each region is used for blurring and shown as insets. \edit{There is overall agreement between the captured and estimated images across the entire focus range.} \edit{In the restoration results, we perform deconvolution using the proposed method. Our forward model convolved the ground truth estimate with PSFs to create a synthetic image, which is then compared to the actual capture to compute the loss. The blurry capture is boxed in orange, and the PSFs from blur fields (boxed in blue) and Seidel are shown for each color channel. While the Seidel PSFs are comparable in size to the blur field PSFs, they do not capture real-life effects as effectively. The lens blur field preserves high-frequency details in the Siemens Star arms better than the Seidel PSF, as seen in the yellow-boxed region. The synthetic blurry image generated with blur field PSFs preserves more detail and more closely resembles the captured blurry image than that generated with the Seidel PSFs.}
 }
    \label{fig:siemens-star}
\end{figure*}

A comparison of captures and estimates rendered with the calibrated lens blur field of the iPhone 12 Pro visualized in ~\cref{fig:tiling} is presented in the Supplement. We capture a focal stack of a resolution chart and the captured images are compared with image restorations, showing overall agreement across the focus range (see \cref{fig:siemens-star}).

\edit{\subsection{Improved Image Restoration with Blur Fields Compared to Seidel Model}}
\edit{We compare the proposed lens blur fields with the parametric Seidel model by evaluating their performance for image restoration on a real-world capture of a Siemens Star test pattern using an iPhone 12 Pro wide camera (\cref{tab:image_valid}). After constructing the lens blur field, we follow Kohli et al.~\cite{kohli2024ring} to fit a Seidel model—with spherical, coma, astigmatism, field curvature, and defocus coefficients—to the PSFs recovered by the blur field by randomly sampling 32 image plane locations per lens position. The Seidel coefficients are optimized via gradient descent to match the lens blur field PSFs at the sampled locations using 300 iterations with a learning rate of $0.01$, the default parameters in Kohli et al.~\cite{kohli2024ring}. For image restoration, we use the same optimization approach as for PSF estimation (\cref{ssec:blurfield}). We solve for a restored image that, when convolved with the respective blurs, produces a synthetic blurry image. We then compute a loss between this synthetic image and the real-life blurry image.}

\edit{We compare the resulting restorations and find that blur field PSFs yield superior results. Qualitative results in \cref{fig:siemens-star} demonstrate that the lens blur field better preserves high-frequency details in the Siemens Star pattern, reconstructing longer portions of the star's arms than do restorations using PSFs from the Seidel model. A direct comparison with the ground truth Siemens Star in \cref{tab:restorations} shows that restorations using blur field PSFs are superior. Furthermore, the synthetic blurry image generated by convolving the restored Siemens Star with blur field PSFs more closely matches the real-life image than the synthetic image generated using Seidel PSFs.}

\begin{table}[h]
  \centering
  \resizebox{0.48\textwidth}{!}{
  \renewcommand{\arraystretch}{1.25}
  
  \begin{tabular}{l@{\quad}c@{\qquad}c@{\qquad}cccc}
    \toprule
    \multirow{2}{*}{\raisebox{-\heavyrulewidth}{\makecell[l]{\edit{Restoration} \\ \edit{Results}}}} & & \multirow{2}{*}{\edit{\makecell[c]{Siemens Star \\Restoration}}} & \multicolumn{4}{c}{\edit{Synthetic Blurry Image}} \\
    \cmidrule{4-7}
    &&  & \edit{Red} & \edit{Green 1} & \edit{Green 2} & \edit{Blue}\\
    \midrule
    \multirow{3}{*}{\raisebox{-\heavyrulewidth}{\edit{Proposed}}}
                & \edit{PSNR} $\uparrow$   & \edit{8.900} & \edit{20.335} & \edit{20.392} & \edit{21.479} & \edit{17.700}\\
                & \edit{SSIM} $\uparrow$   & \edit{0.293} & \edit{0.747}  & \edit{0.780}  & \edit{0.786}  & \edit{0.600}\\
                & \edit{RMSE} $\downarrow$ & \edit{0.353} & \edit{0.050}  & \edit{0.054}  & \edit{0.047}  & \edit{0.059}\\

    \midrule
    \multirow{3}{*}{\raisebox{-\heavyrulewidth}{\makecell[l]{\edit{Seidel} \\ \edit{Coefficients}}}}
                & \edit{PSNR} $\uparrow$   & \edit{7.415} & \edit{20.184} & \edit{19.660} & \edit{20.528} & \edit{17.754}\\
                & \edit{SSIM} $\uparrow$   & \edit{0.127} & \edit{0.534} & \edit{0.588} & \edit{0.595} & \edit{0.515}\\
                & \edit{RMSE} $\downarrow$ & \edit{0.421} & \edit{0.051} & \edit{0.059} & \edit{0.053} & \edit{0.058}\\
    \bottomrule
  \end{tabular}
  }
  \caption{\edit{Restoration results for both the estimated Siemens Star and the synthetic blurry image obtained by convolving the estimated Star with either the blur fields PSF or the Seidel PSF. Both the restored Siemens Star and the synthetic blurry image show that the blur field PSFs recover real-life effects more effectively than the Seidel model.}
}
  \label{tab:restorations}
\end{table}

\subsection{Rendering Scenes with Blur}
To facilitate synthetic rendering, we create a Blender~\cite{blender} postprocessing add-on that allows one to take advantage of our pre-trained lens blur field models as part of their rendering pipeline. We demonstrate this technique on synthetic scenes shown in Supplement Sec. S-V, rendered with the occlusion-aware model of Ikoma et al.~\cite{hayota2021depth} (which itself is agnostic to whether Blender or another engine is used). Different lens blur fields result in a diverse set of bokeh patterns and we show an additional result on captured data in Supplement Sec. S-V.

\section{Concluding Remarks}

Although billions of people carry around smartphone cameras in their pockets, the precise optical behavior of individual devices is inaccessible, or entirely unknown, to the owner. 
Proprietary databases exist which characterize optical behavior generally for a make and model of a lens,\footnote{For example, DxO has one such database described at \url{https://www.dxo.com/technology/how-dxo-corrects-lens-flaws/}.} but calibrating device-specific high-dimensional models of optical blur with commodity hardware has largely been an intractable problem. 
We take steps towards democratizing accurate optical characterization via the lens blur field, and provide a robust, lightweight method by which it can be acquired.

We are excited about several avenues of future work. For example, given that we find significant differences between lens blur fields from devices of the same make and model, how distinct are these differences across a wider cross-section of device types and instances? What can we learn about the devices from these aberrations? Additionally, we wish to explore how the lens blur field can be extended to even higher dimensions (\emph{e.g.,} light fields,  hyperspectral data, polarization imaging, \emph{etc.}).

It may also be possible to improve the optimization procedure by incorporating pre-processing steps (such as geometric alignment and auto-exposure) into the main optimization loop.
Tackling the blind deconvolution problem would also enable optical characterization in entirely unstructured environments.

Finally, neural blur fields may open up new applications as proxy functions for emulating optical systems.
Here, we envision uses for photorealistic rendering and synthetic depth of field for mobile photography.

\ifCLASSOPTIONcompsoc
  \section*{Acknowledgments}
\else
  \section*{Acknowledgment}
\fi

DBL and KNK acknowledge support of NSERC under the RGPIN, RTI, and USRA programs. DBL also acknowledges support from the Canada Foundation for Innovation and the Ontario Research Fund.

\bibliographystyle{IEEEtran}
\bibliography{IEEEabrv,references}

\begin{thebibliography}{10}
\providecommand{\url}[1]{#1}
\csname url@samestyle\endcsname
\providecommand{\newblock}{\relax}
\providecommand{\bibinfo}[2]{#2}
\providecommand{\BIBentrySTDinterwordspacing}{\spaceskip=0pt\relax}
\providecommand{\BIBentryALTinterwordstretchfactor}{4}
\providecommand{\BIBentryALTinterwordspacing}{\spaceskip=\fontdimen2\font plus
\BIBentryALTinterwordstretchfactor\fontdimen3\font minus \fontdimen4\font\relax}
\providecommand{\BIBforeignlanguage}[2]{{%
\expandafter\ifx\csname l@#1\endcsname\relax
\typeout{** WARNING: IEEEtran.bst: No hyphenation pattern has been}%
\typeout{** loaded for the language `#1'. Using the pattern for}%
\typeout{** the default language instead.}%
\else
\language=\csname l@#1\endcsname
\fi
#2}}
\providecommand{\BIBdecl}{\relax}
\BIBdecl

\bibitem{cossairt2010spectral}
O.~Cossairt and S.~Nayar, ``Spectral focal sweep: Extended depth of field from chromatic aberrations,'' in \emph{IEEE Int. Conf. Comput. Photography}.\hskip 1em plus 0.5em minus 0.4em\relax IEEE, 2010, pp. 1--8.

\bibitem{tang2012utilizing}
H.~Tang and K.~N. Kutulakos, ``Utilizing optical aberrations for extended-depth-of-field panoramas,'' in \emph{Asian Conf. Comput. Vision}.\hskip 1em plus 0.5em minus 0.4em\relax Springer, 2012, pp. 365--378.

\bibitem{heide2013modern}
F.~Heide, M.~Rouf, M.~B. Hullin, B.~Labitzke, W.~Heidrich, and A.~Kolb, ``High-quality computational imaging through simple lenses,'' \emph{ACM Trans. Graph.}, vol.~32, no.~5, oct 2013.

\bibitem{levoy2006light}
M.~Levoy, R.~Ng, A.~Adams, M.~Footer, and M.~Horowitz, ``Light field microscopy,'' \emph{ACM Trans. Graph.}, vol.~25, no.~3, p. 924–934, 2006.

\bibitem{tao2013depth}
M.~W. Tao, S.~Hadap, J.~Malik, and R.~Ramamoorthi, ``Depth from combining defocus and correspondence using light-field cameras,'' in \emph{Proc. Int. Conf. Comput. Vis.}, 2013, pp. 673--680.

\bibitem{tao2015depth}
M.~W. Tao, P.~P. Srinivasan, J.~Malik, S.~Rusinkiewicz, and R.~Ramamoorthi, ``Depth from shading, defocus, and correspondence using light-field angular coherence,'' in \emph{Proc. IEEE Conf. Comput. Vis. Pattern Recog.}, 2015, pp. 1940--1948.

\bibitem{pasqual1998use}
A.~A. Pasqual, K.~Aizawa, and M.~Hatori, ``Use of multiple visual features for object tracking,'' in \emph{Vis. Commun. Image Process.}, vol. 3653.\hskip 1em plus 0.5em minus 0.4em\relax SPIE, 1998, pp. 946--955.

\bibitem{suwajanakorn2015depthff}
S.~Suwajanakorn, C.~Hern{\'a}ndez, and S.~M. Seitz, ``Depth from focus with your mobile phone,'' \emph{Proc. IEEE Conf. Comput. Vis. Pattern Recog.}, pp. 3497--3506, 2015.

\bibitem{tang2017depth}
H.~Tang, S.~Cohen, B.~Price, S.~Schiller, and K.~N. Kutulakos, ``Depth from defocus in the wild,'' in \emph{Proc. IEEE Conf. Comput. Vis. Pattern Recog.}, 2017, pp. 2740--2748.

\bibitem{punnappurath2020modeling}
A.~Punnappurath, A.~Abuolaim, M.~Afifi, and M.~S. Brown, ``Modeling defocus-disparity in dual-pixel sensors,'' in \emph{IEEE Int. Conf. Comput. Photography}.\hskip 1em plus 0.5em minus 0.4em\relax IEEE, 2020, pp. 1--12.

\bibitem{abuolaim2022multi}
A.~Abuolaim, M.~Afifi, and M.~S. Brown, ``Multi-view motion synthesis via applying rotated dual-pixel blur kernels,'' in \emph{Proc. IEEE Winter Conf. Appl. Comput. Vis.}, 2022, pp. 701--708.

\bibitem{xin2021defocus}
S.~Xin, N.~Wadhwa, T.~Xue, J.~T. Barron, P.~P. Srinivasan, J.~Chen, I.~Gkioulekas, and R.~Garg, ``Defocus map estimation and deblurring from a single dual-pixel image,'' in \emph{Proc. Int. Conf. Comput. Vis.}, 2021, pp. 2228--2238.

\bibitem{levin2007image}
A.~Levin, R.~Fergus, F.~Durand, and W.~T. Freeman, ``Image and depth from a conventional camera with a coded aperture,'' \emph{ACM Trans. Graph.}, vol.~26, no.~3, p. 70–es, jul 2007.

\bibitem{asif2016flatcam}
M.~S. Asif, A.~Ayremlou, A.~Sankaranarayanan, A.~Veeraraghavan, and R.~G. Baraniuk, ``Flatcam: Thin, lensless cameras using coded aperture and computation,'' \emph{IEEE Trans. Comput. Imaging}, vol.~3, no.~3, pp. 384--397, 2016.

\bibitem{antipa2018diffusercam}
N.~Antipa, G.~Kuo, R.~Heckel, B.~Mildenhall, E.~Bostan, R.~Ng, and L.~Waller, ``Diffusercam: lensless single-exposure 3d imaging,'' \emph{Optica}, vol.~5, no.~1, pp. 1--9, 2018.

\bibitem{levin20094d}
A.~Levin, S.~W. Hasinoff, P.~Green, F.~Durand, and W.~T. Freeman, ``4d frequency analysis of computational cameras for depth of field extension,'' \emph{ACM Trans. Graph.}, vol.~28, no.~3, pp. 1--14, 2009.

\bibitem{pavani2008high}
S.~R.~P. Pavani and R.~Piestun, ``High-efficiency rotating point spread functions,'' \emph{Opt. Express}, vol.~16, no.~5, pp. 3484--3489, 2008.

\bibitem{born1999principles}
M.~Born and E.~Wolf, \emph{Principles of Optics: Electromagnetic Theory of Propagation, Interference and Diffraction of Light (7th Edition)}, 7th~ed.\hskip 1em plus 0.5em minus 0.4em\relax Cambridge University Press, 1999.

\bibitem{tang2018modeling}
H.~Tang, ``Modeling and analysis of optical blur for everyday photography,'' Ph.D. dissertation, University of Toronto, 2018.

\bibitem{mahajan1994aberration}
V.~N. Mahajan, ``Zernike circle polynomials and optical aberrations of systems with circular pupils,'' \emph{Appl. Optics}, vol.~33, no.~34, pp. 8121--8124, Dec 1994.

\bibitem{smith2008modern}
W.~J. Smith, \emph{Modern optical engineering: the design of optical systems}.\hskip 1em plus 0.5em minus 0.4em\relax McGraw-Hill Education, 2008.

\bibitem{kee2011modeling}
E.~Kee, S.~Paris, S.~Chen, and J.~Wang, ``Modeling and removing spatially-varying optical blur,'' in \emph{IEEE Int. Conf. Comput. Photography}.\hskip 1em plus 0.5em minus 0.4em\relax IEEE, 2011, pp. 1--8.

\bibitem{jang2016modeling}
J.~Jang, J.~D. Yun, and S.~Yang, ``Modeling non-stationary asymmetric lens blur by normal sinh-arcsinh model,'' \emph{IEEE Trans. Image Process.}, vol.~25, no.~5, pp. 2184--2195, 2016.

\bibitem{schuler2012blind}
C.~J. Schuler, M.~Hirsch, S.~Harmeling, and B.~Sch{\"o}lkopf, ``Blind correction of optical aberrations,'' in \emph{Proc. Eur. Conf. Comput. Vis.}\hskip 1em plus 0.5em minus 0.4em\relax Springer, 2012, pp. 187--200.

\bibitem{simpkins2014parameterized}
J.~Simpkins and R.~L. Stevenson, ``Parameterized modeling of spatially varying optical blur,'' \emph{J. Electron. Imaging}, vol.~23, no.~1, pp. 013\,005--013\,005, 2014.

\bibitem{zernicke1934physica}
von F.~Zernike, ``Beugungstheorie des schneidenver-fahrens und seiner verbesserten form, der phasenkontrastmethode,'' \emph{Physica}, vol.~1, no.~7, pp. 689--704, 1934.

\bibitem{song2019full}
P.~Song, S.~Jiang, H.~Zhang, X.~Huang, Y.~Zhang, and G.~Zheng, ``{Full-field Fourier ptychography (FFP): Spatially varying pupil modeling and its application for rapid field-dependent aberration metrology},'' \emph{APL Photonics}, vol.~4, no.~5, 05 2019, 050802.

\bibitem{bezdidko1974use}
S.~Bezdid'ko, ``Use of zernicke polynomials in optics,'' \emph{Soviet J. Opt. Technol.}, vol.~41, no.~9, p. 425 – 429, 1974.

\bibitem{tango1977thecp}
W.~J. Tango, ``The circle polynomials of zernike and their application in optics,'' \emph{Appl. Physics}, vol.~13, pp. 327--332, 1977.

\bibitem{debarnot2021learning}
V.~Debarnot, P.~Escande, T.~Mangeat, and P.~Weiss, ``Learning low-dimensional models of microscopes,'' \emph{IEEE Trans. Comput. Imaging}, vol.~7, pp. 178--190, 2021.

\bibitem{joshi2008psf}
N.~Joshi, R.~Szeliski, and D.~J. Kriegman, ``Psf estimation using sharp edge prediction,'' in \emph{Proc. IEEE Conf. Comput. Vis. Pattern Recog.}\hskip 1em plus 0.5em minus 0.4em\relax IEEE, 2008, pp. 1--8.

\bibitem{mannan}
F.~Mannan and M.~S. Langer, ``Blur calibration for depth from defocus,'' in \emph{Conf. Comput. Robot Vision}, 2016, pp. 281--288.

\bibitem{Gwak:20}
M.~Gwak and S.~Yang, ``Modeling nonstationary lens blur using eigen blur kernels for restoration,'' \emph{Opt. Express}, vol.~28, no.~26, pp. 39\,501--39\,523, Dec 2020.

\bibitem{hirsch2015self}
M.~Hirsch and B.~Scholkopf, ``Self-calibration of optical lenses,'' in \emph{Proc. Int. Conf. Comput. Vis.}, 2015, pp. 612--620.

\bibitem{gwak2020modeling}
M.~Gwak and S.~Yang, ``Modeling nonstationary lens blur using eigen blur kernels for restoration,'' \emph{Opt. Express}, vol.~28, no.~26, pp. 39\,501--39\,523, 2020.

\bibitem{hu2011psf}
W.~Hu, J.~Xue, and N.~Zheng, ``Psf estimation via gradient domain correlation,'' \emph{IEEE Trans. Image Process.}, vol.~21, no.~1, pp. 386--392, 2011.

\bibitem{cho2011blur}
T.~S. Cho, S.~Paris, B.~K. Horn, and W.~T. Freeman, ``Blur kernel estimation using the radon transform,'' in \emph{Proc. IEEE Conf. Comput. Vis. Pattern Recog.}\hskip 1em plus 0.5em minus 0.4em\relax IEEE, 2011, pp. 241--248.

\bibitem{rav2005two}
A.~Rav-Acha and S.~Peleg, ``Two motion-blurred images are better than one,'' \emph{Pattern Recognit. Lett.}, vol.~26, no.~3, pp. 311--317, 2005.

\bibitem{yuan2007image}
L.~Yuan, J.~Sun, L.~Quan, and H.-Y. Shum, ``Image deblurring with blurred/noisy image pairs,'' \emph{ACM Trans. Graph.}, vol.~26, no.~3, p. 1–es, 2007.

\bibitem{brauers2010direct}
J.~Brauers, C.~Seiler, and T.~Aach, ``Direct psf estimation using a random noise target,'' in \emph{Digital Photography VI}, vol. 7537.\hskip 1em plus 0.5em minus 0.4em\relax SPIE, 2010, pp. 96--105.

\bibitem{schuler2011non}
C.~J. Schuler, M.~Hirsch, S.~Harmeling, and B.~Sch{\"o}lkopf, ``Non-stationary correction of optical aberrations,'' in \emph{Proc. Int. Conf. Comput. Vis.}\hskip 1em plus 0.5em minus 0.4em\relax IEEE, 2011, pp. 659--666.

\bibitem{heide2013high}
F.~Heide, M.~Rouf, M.~B. Hullin, B.~Labitzke, W.~Heidrich, and A.~Kolb, ``High-quality computational imaging through simple lenses,'' \emph{ACM Trans. Graph.}, vol.~32, no.~5, pp. 1--14, 2013.

\bibitem{delbracio2012non}
M.~Delbracio, P.~Mus{\'e}, A.~Almansa, and J.-M. Morel, ``The non-parametric sub-pixel local point spread function estimation is a well posed problem,'' \emph{Int. J. Comput. Vis.}, vol.~96, no.~2, pp. 175--194, 2012.

\bibitem{hansen2006deblurrinq}
P.~C. Hansen, \emph{Deblurring images: matrices, spectra, and filtering}.\hskip 1em plus 0.5em minus 0.4em\relax Society for Industrial and Applied Mathematics, 2006.

\bibitem{gkioulekas2015micron}
I.~Gkioulekas, A.~Levin, F.~Durand, and T.~Zickler, ``Micron-scale light transport decomposition using interferometry,'' \emph{ACM Trans. Graph.}, vol.~34, no.~4, Jul 2015.

\bibitem{yang2021designing}
A.~Yang and A.~C. Sankaranarayanan, ``Designing display pixel layouts for under-panel cameras,'' \emph{IEEE Trans. Pattern Anal. Mach. Intell.}, vol.~43, no.~7, pp. 2245--2256, 2021.

\bibitem{shih2012image}
Y.~Shih, B.~Guenter, and N.~Joshi, ``Image enhancement using calibrated lens simulations,'' in \emph{Proc. Eur. Conf. Comput. Vis.}, 2012.

\bibitem{tseng2021differentiable}
E.~Tseng, A.~Mosleh, F.~Mannan, K.~St-Arnaud, A.~Sharma, Y.~Peng, A.~Braun, D.~Nowrouzezahrai, J.-F. Lalonde, and F.~Heide, ``Differentiable compound optics and processing pipeline optimization for end-to-end camera design,'' \emph{ACM Trans. Graph.}, vol.~40, no.~2, pp. 1--19, 2021.

\bibitem{zemax}
\BIBentryALTinterwordspacing
Zemax, ``Opticstudio,'' 2023. [Online]. Available: \url{https://www.zemax.com/}
\BIBentrySTDinterwordspacing

\bibitem{sitzmann2018end2end}
V.~Sitzmann, S.~Diamond, Y.~Peng, X.~Dun, S.~Boyd, W.~Heidrich, F.~Heide, and G.~Wetzstein, ``End-to-end optimization of optics and image processing for achromatic extended depth of field and super-resolution imaging,'' \emph{ACM Trans. Graph.}, vol.~37, no.~4, jul 2018.

\bibitem{metzler2020deep}
C.~A. Metzler, H.~Ikoma, Y.~Peng, and G.~Wetzstein, ``Deep optics for single-shot high-dynamic-range imaging,'' in \emph{Proc. IEEE Conf. Comput. Vis. Pattern Recog.}, 2020, pp. 1375--1385.

\bibitem{baek2021single}
S.-H. Baek, H.~Ikoma, D.~S. Jeon, Y.~Li, W.~Heidrich, G.~Wetzstein, and M.~H. Kim, ``Single-shot hyperspectral-depth imaging with learned diffractive optics,'' in \emph{Proc. Int. Conf. Comput. Vis.}, 2021, pp. 2651--2660.

\bibitem{yang2022sub}
H.~Yang, E.~Y. Lin, K.~N. Kutulakos, and G.~V. Eleftheriades, ``Sub-wavelength passive single-shot computational super-oscillatory imaging,'' \emph{Optica}, vol.~9, no.~12, pp. 1444--1447, 2022.

\bibitem{xu2017motion}
X.~Xu, J.~Pan, Y.-J. Zhang, and M.-H. Yang, ``Motion blur kernel estimation via deep learning,'' \emph{IEEE Trans. Image Process.}, vol.~27, no.~1, pp. 194--205, 2017.

\bibitem{li2018learning}
L.~Li, J.~Pan, W.-S. Lai, C.~Gao, N.~Sang, and M.-H. Yang, ``Learning a discriminative prior for blind image deblurring,'' in \emph{Proc. IEEE Conf. Comput. Vis. Pattern Recog.}, 2018, pp. 6616--6625.

\bibitem{ren2020neural}
D.~Ren, K.~Zhang, Q.~Wang, Q.~Hu, and W.~Zuo, ``Neural blind deconvolution using deep priors,'' in \emph{Proc. IEEE Conf. Comput. Vis. Pattern Recog.}, 2020, pp. 3341--3350.

\bibitem{shajkofci2020spatially}
A.~Shajkofci and M.~Liebling, ``Spatially-variant cnn-based point spread function estimation for blind deconvolution and depth estimation in optical microscopy,'' \emph{IEEE Trans. Image Process.}, vol.~29, pp. 5848--5861, 2020.

\bibitem{sun2015learning}
J.~Sun, W.~Cao, Z.~Xu, and J.~Ponce, ``Learning a convolutional neural network for non-uniform motion blur removal,'' in \emph{Proc. IEEE Conf. Comput. Vis. Pattern Recog.}, 2015, pp. 769--777.

\bibitem{ren_2020_cvpr}
D.~Ren, K.~Zhang, Q.~Wang, Q.~Hu, and W.~Zuo, ``Neural blind deconvolution using deep priors,'' in \emph{Proc. IEEE/CVF Conf. Comput. Vis. Pattern Recognit. (CVPR)}, June 2020.

\bibitem{sureau2020deep}
F.~Sureau, A.~Lechat, and J.-L. Starck, ``Deep learning for a space-variant deconvolution in galaxy surveys,'' \emph{Astron. \& Astrophys.}, vol. 641, p. A67, 2020.

\bibitem{rego2021robust}
J.~D. Rego, K.~Kulkarni, and S.~Jayasuriya, ``Robust lensless image reconstruction via psf estimation,'' in \emph{Proc. IEEE Winter Conf. Appl. Comput. Vis.}, 2021, pp. 403--412.

\bibitem{yanny2022deep}
K.~Yanny, K.~Monakhova, R.~W. Shuai, and L.~Waller, ``Deep learning for fast spatially varying deconvolution,'' \emph{Optica}, vol.~9, no.~1, pp. 96--99, 2022.

\bibitem{zhang2017learning}
J.~Zhang, J.~Pan, W.-S. Lai, R.~W. Lau, and M.-H. Yang, ``Learning fully convolutional networks for iterative non-blind deconvolution,'' in \emph{Proc. IEEE Conf. Comput. Vis. Pattern Recog.}, 2017, pp. 3817--3825.

\bibitem{dong2020deep}
J.~Dong, S.~Roth, and B.~Schiele, ``Deep wiener deconvolution: Wiener meets deep learning for image deblurring,'' \emph{Adv. Neural Inf. Process. Syst.}, vol.~33, pp. 1048--1059, 2020.

\bibitem{xie2022neuralfields}
Y.~Xie, T.~Takikawa, S.~Saito, O.~Litany, S.~Yan, N.~Khan, F.~Tombari, J.~Tompkin, V.~Sitzmann, and S.~Sridhar, ``Neural fields in visual computing and beyond,'' \emph{Comput. Graph. Forum}, 2022.

\bibitem{takikawa2023compact}
T.~Takikawa, T.~M{\"u}ller, M.~Nimier-David, A.~Evans, S.~Fidler, A.~Jacobson, and A.~Keller, ``Compact neural graphics primitives with learned hash probing,'' in \emph{Proc. SIGGRAPH Asia 2023}, 2023.

\bibitem{ramasinghe2022regularizingcoordinatemlps}
S.~Ramasinghe, L.~MacDonald, and S.~Lucey, ``On regularizing coordinate-mlps,'' 2022.

\bibitem{ramasinghi2022frequency}
S.~Ramasinghe, L.~E. MacDonald, and S.~Lucey, ``On the frequency-bias of coordinate-mlps,'' in \emph{Adv. Neural Inf. Process. Syst.}, vol.~35, 2022, pp. 796--809.

\bibitem{couture2011unstructured}
V.~Couture, N.~Martin, and S.~Roy, ``Unstructured light scanning to overcome interreflections,'' in \emph{Proc. Int. Conf. Comput. Vis.}, 2011, pp. 1895--1902.

\bibitem{herrmann2020learning}
C.~Herrmann, R.~S. Bowen, N.~Wadhwa, R.~Garg, Q.~He, J.~T. Barron, and R.~Zabih, ``Learning to autofocus,'' in \emph{Proc. IEEE Conf. Comput. Vis. Pattern Recog.}, 2020.

\bibitem{aggarwal2001cosine}
M.~Aggarwal, H.~Hua, and N.~Ahuja, ``On cosine-fourth and vignetting effects in real lenses,'' in \emph{Proc. Int. Conf. Comput. Vis.}, vol.~1.\hskip 1em plus 0.5em minus 0.4em\relax IEEE, 2001, pp. 472--479.

\bibitem{kolb1995realistic}
C.~Kolb, D.~Mitchell, and P.~Hanrahan, ``A realistic camera model for computer graphics,'' in \emph{Proc. 22nd Annu. Conf. SIGGRAPH}, ser. SIGGRAPH '95.\hskip 1em plus 0.5em minus 0.4em\relax New York, NY, USA: Association for Computing Machinery, 1995, p. 317–324.

\bibitem{tancik2020fourier}
M.~Tancik, P.~Srinivasan, B.~Mildenhall, S.~Fridovich-Keil, N.~Raghavan, U.~Singhal, R.~Ramamoorthi, J.~Barron, and R.~Ng, ``Fourier features let networks learn high frequency functions in low dimensional domains,'' \emph{Adv. Neural Inform. Process. Syst.}, vol.~33, pp. 7537--7547, 2020.

\bibitem{tiny-cuda-nn}
\BIBentryALTinterwordspacing
T.~M\"uller, ``{tiny-cuda-nn},'' 4 2021. [Online]. Available: \url{https://github.com/NVlabs/tiny-cuda-nn}
\BIBentrySTDinterwordspacing

\bibitem{hirsch2010efficient}
M.~Hirsch, S.~Sra, B.~Sch{\"o}lkopf, and S.~Harmeling, ``Efficient filter flow for space-variant multiframe blind deconvolution,'' in \emph{Proc. IEEE Conf. Comput. Vis. Pattern Recog.}\hskip 1em plus 0.5em minus 0.4em\relax IEEE, 2010, pp. 607--614.

\bibitem{scharstein2003high}
D.~Scharstein and R.~Szeliski, ``High-accuracy stereo depth maps using structured light,'' in \emph{Proc. IEEE Conf. Comput. Vis. Pattern Recog.}, vol.~1.\hskip 1em plus 0.5em minus 0.4em\relax IEEE, 2003, pp. I--I.

\bibitem{fischler1981random}
M.~A. Fischler and R.~C. Bolles, ``Random sample consensus: a paradigm for model fitting with applications to image analysis and automated cartography,'' \emph{Commun. ACM}, vol.~24, no.~6, pp. 381--395, 1981.

\bibitem{opencv_library}
G.~Bradski, ``{The OpenCV Library},'' \emph{Dr. Dobb's Journal of Software Tools}, 2000.

\bibitem{malacara2003handbook}
D.~Malacara-Hern{\'a}ndez, Z.~Malacara-Hern{\'a}ndez, and Z.~Malacara, \emph{Handbook of optical design}.\hskip 1em plus 0.5em minus 0.4em\relax CRC Press, 2003.

\bibitem{kohli2024ring}
A.~Kohli, A.~N. Angelopoulos, D.~McAllister, E.~Whang, S.~You, K.~Yanny, F.~M. Gasparoli, and L.~Waller, ``Ring deconvolution microscopy: An exact solution for spatially-varying aberration correction,'' 2024.

\bibitem{blender}
\BIBentryALTinterwordspacing
{Blender Online Community}, \emph{Blender—a 3D modeling and rendering package}, Blender Foundation, 2023. [Online]. Available: \url{http://www.blender.org}
\BIBentrySTDinterwordspacing

\bibitem{hayota2021depth}
H.~Ikoma, C.~M. Nguyen, C.~A. Metzler, Y.~Peng, and G.~Wetzstein, ``Depth from defocus with learned optics for imaging and occlusion-aware depth estimation,'' in \emph{IEEE Int. Conf. Comput. Photography}, 2021, pp. 1--12.

\end{thebibliography}


\begin{thebibliography}{10}
\providecommand{\url}[1]{#1}
\csname url@samestyle\endcsname
\providecommand{\newblock}{\relax}
\providecommand{\bibinfo}[2]{#2}
\providecommand{\BIBentrySTDinterwordspacing}{\spaceskip=0pt\relax}
\providecommand{\BIBentryALTinterwordstretchfactor}{4}
\providecommand{\BIBentryALTinterwordspacing}{\spaceskip=\fontdimen2\font plus
\BIBentryALTinterwordstretchfactor\fontdimen3\font minus \fontdimen4\font\relax}
\providecommand{\BIBforeignlanguage}[2]{{%
\expandafter\ifx\csname l@#1\endcsname\relax
\typeout{** WARNING: IEEEtran.bst: No hyphenation pattern has been}%
\typeout{** loaded for the language `#1'. Using the pattern for}%
\typeout{** the default language instead.}%
\else
\language=\csname l@#1\endcsname
\fi
#2}}
\providecommand{\BIBdecl}{\relax}
\BIBdecl

\bibitem{scharstein2003high}
D.~Scharstein and R.~Szeliski, ``High-accuracy stereo depth maps using structured light,'' in \emph{Proc. IEEE Conf. Comput. Vis. Pattern Recog.}, vol.~1.\hskip 1em plus 0.5em minus 0.4em\relax IEEE, 2003, pp. I--I.

\bibitem{couture2011unstructured}
V.~Couture, N.~Martin, and S.~Roy, ``Unstructured light scanning to overcome interreflections,'' in \emph{Proc. Int. Conf. Comput. Vis.}, 2011, pp. 1895--1902.

\bibitem{tiny-cuda-nn}
\BIBentryALTinterwordspacing
T.~M\"uller, ``{tiny-cuda-nn},'' 4 2021. [Online]. Available: \url{https://github.com/NVlabs/tiny-cuda-nn}
\BIBentrySTDinterwordspacing

\bibitem{mannan}
F.~Mannan and M.~S. Langer, ``Blur calibration for depth from defocus,'' in \emph{Conf. Comput. Robot Vision}, 2016, pp. 281--288.

\bibitem{joshi2008psf}
N.~Joshi, R.~Szeliski, and D.~J. Kriegman, ``Psf estimation using sharp edge prediction,'' in \emph{Proc. IEEE Conf. Comput. Vis. Pattern Recog.}\hskip 1em plus 0.5em minus 0.4em\relax IEEE, 2008, pp. 1--8.

\bibitem{punnappurath2020modeling}
A.~Punnappurath, A.~Abuolaim, M.~Afifi, and M.~S. Brown, ``Modeling defocus-disparity in dual-pixel sensors,'' in \emph{IEEE Int. Conf. Comput. Photography}.\hskip 1em plus 0.5em minus 0.4em\relax IEEE, 2020, pp. 1--12.

\bibitem{xin2021defocus}
S.~Xin, N.~Wadhwa, T.~Xue, J.~T. Barron, P.~P. Srinivasan, J.~Chen, I.~Gkioulekas, and R.~Garg, ``Defocus map estimation and deblurring from a single dual-pixel image,'' in \emph{Proc. Int. Conf. Comput. Vis.}, 2021, pp. 2228--2238.

\bibitem{hayota2021depth}
H.~Ikoma, C.~M. Nguyen, C.~A. Metzler, Y.~Peng, and G.~Wetzstein, ``Depth from defocus with learned optics for imaging and occlusion-aware depth estimation,'' in \emph{IEEE Int. Conf. Comput. Photography}, 2021, pp. 1--12.

\bibitem{wadhwa2018synthetic}
N.~Wadhwa, R.~Garg, D.~E. Jacobs, B.~E. Feldman, N.~Kanazawa, R.~Carroll, Y.~Movshovitz-Attias, J.~T. Barron, Y.~Pritch, and M.~Levoy, ``Synthetic depth-of-field with a single-camera mobile phone,'' \emph{ACM Trans. Graph.}, vol.~37, no.~4, pp. 1--13, 2018.

\bibitem{hasinoff2007layer}
S.~W. Hasinoff and K.~N. Kutulakos, ``A layer-based restoration framework for variable-aperture photography,'' in \emph{Proc. Int. Conf. Comput. Vis.}, 2007, pp. 1--8.

\bibitem{zhang2017learning}
J.~Zhang, J.~Pan, W.-S. Lai, R.~W. Lau, and M.-H. Yang, ``Learning fully convolutional networks for iterative non-blind deconvolution,'' in \emph{Proc. IEEE Conf. Comput. Vis. Pattern Recog.}, 2017, pp. 3817--3825.

\bibitem{ren_2020_cvpr}
D.~Ren, K.~Zhang, Q.~Wang, Q.~Hu, and W.~Zuo, ``Neural blind deconvolution using deep priors,'' in \emph{Proc. IEEE/CVF Conf. Comput. Vis. Pattern Recognit. (CVPR)}, June 2020.

\bibitem{dong2020deep}
J.~Dong, S.~Roth, and B.~Schiele, ``Deep wiener deconvolution: Wiener meets deep learning for image deblurring,'' \emph{Adv. Neural Inf. Process. Syst.}, vol.~33, pp. 1048--1059, 2020.

\end{thebibliography}

\begin{IEEEbiography}[{\includegraphics[width=1in,height=1.25in,clip,keepaspectratio]{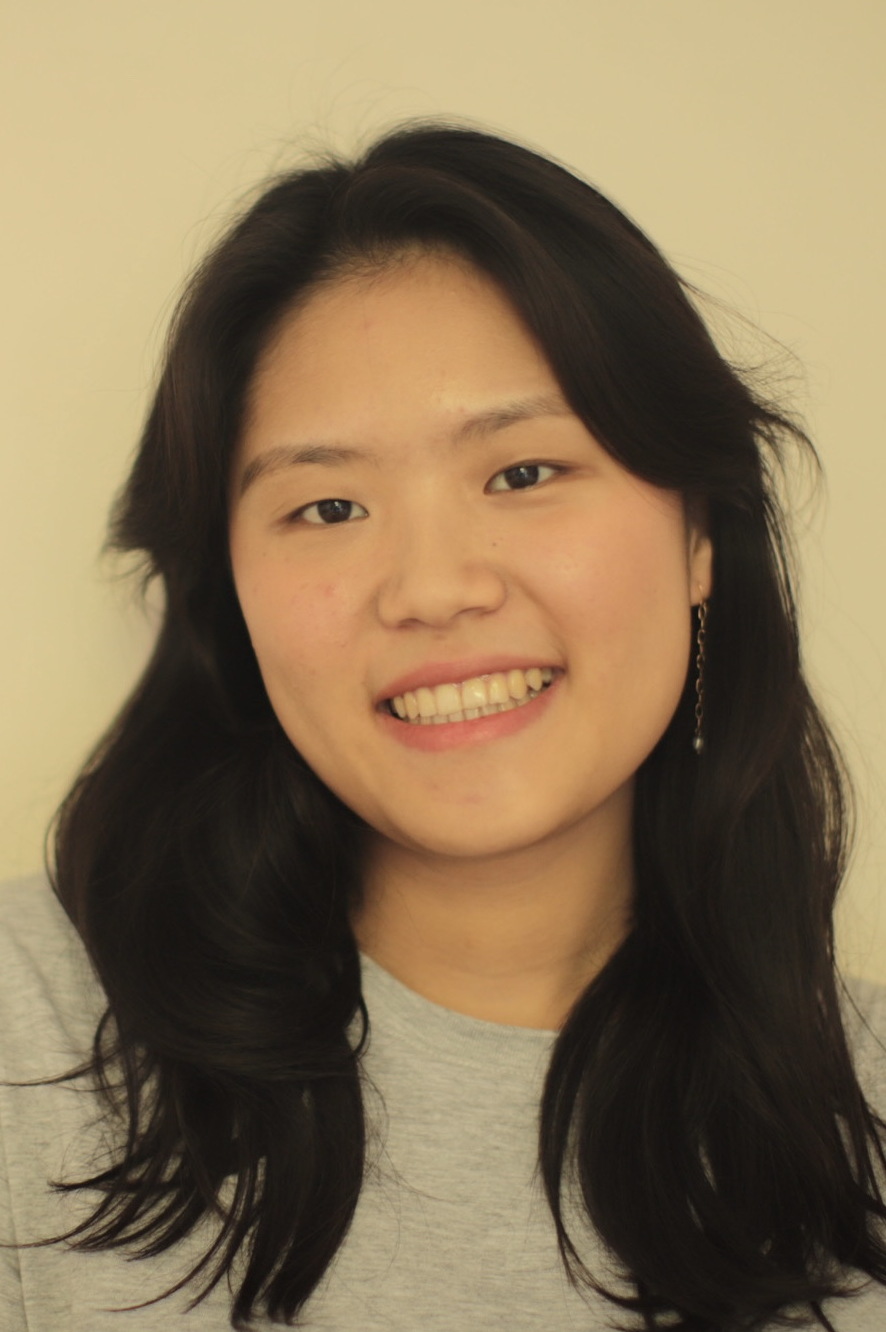}}]{Esther Y. H. Lin}
is a Computer Science Ph.D. student at the University of Toronto, co-advised by David Lindell and Kyros Kutulakos. Her research focuses on optical modelling, sub-diffraction imaging, and signal processing. She is supported by the SPIE Optical Design and Engineering Scholarship and the NSERC Canada Graduate Scholarship. 
\end{IEEEbiography}

\begin{IEEEbiography}[{\includegraphics[width=1in,height=1.25in,clip,keepaspectratio]{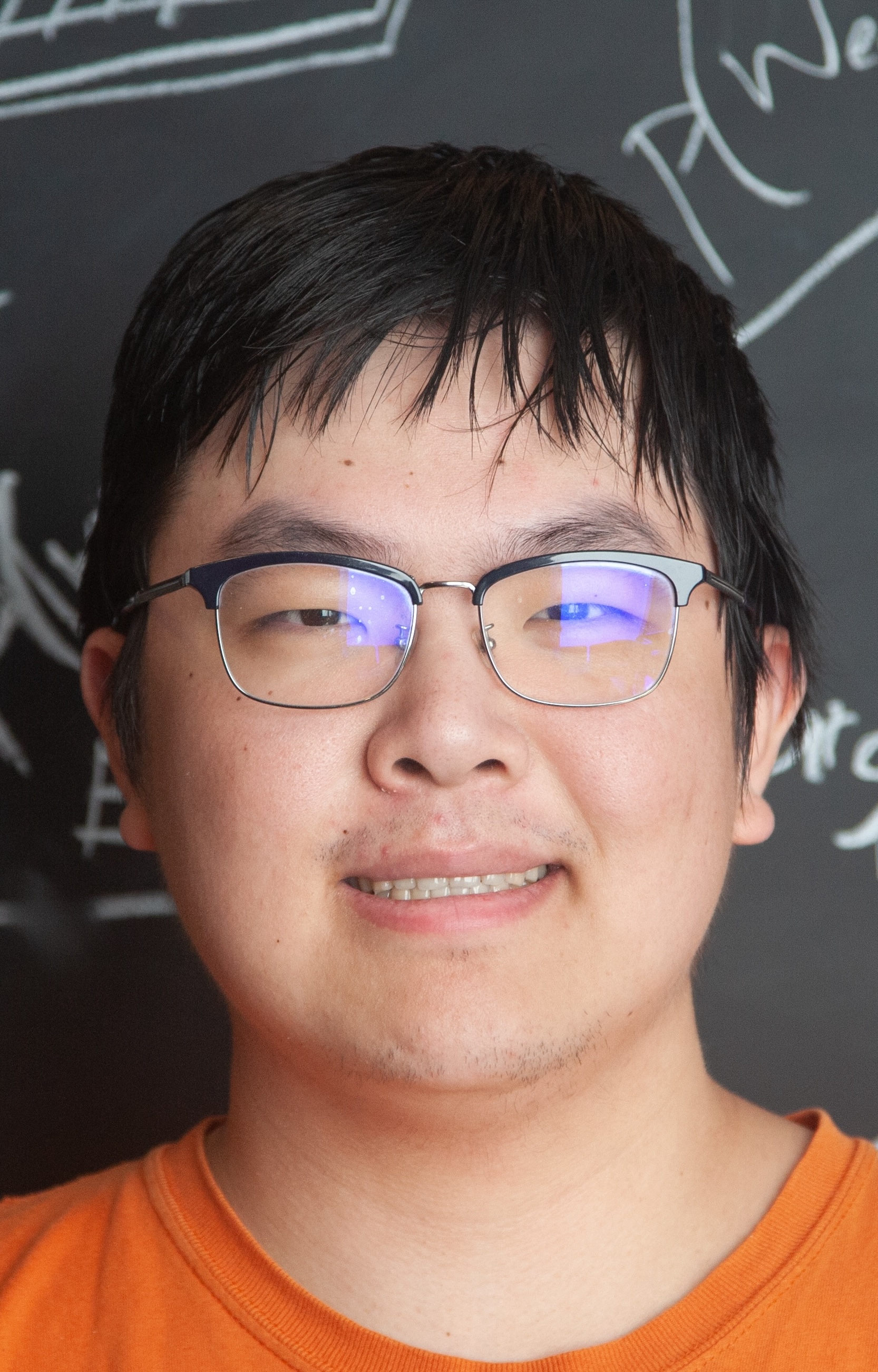}}]{Zhecheng Wang}
is a Ph.D. student in Computer Science at the University of Toronto. He received B.S. degrees in Electrical Engineering, Mathematics, and Radio-Television-Film, and an M.S. in Data Science from UT Austin. His research focuses on physics-based animation, geometry processing, and neural representations for simulation and fabrication.
\end{IEEEbiography}

\begin{IEEEbiography}[{\includegraphics[width=1in,height=1.25in,clip,keepaspectratio]{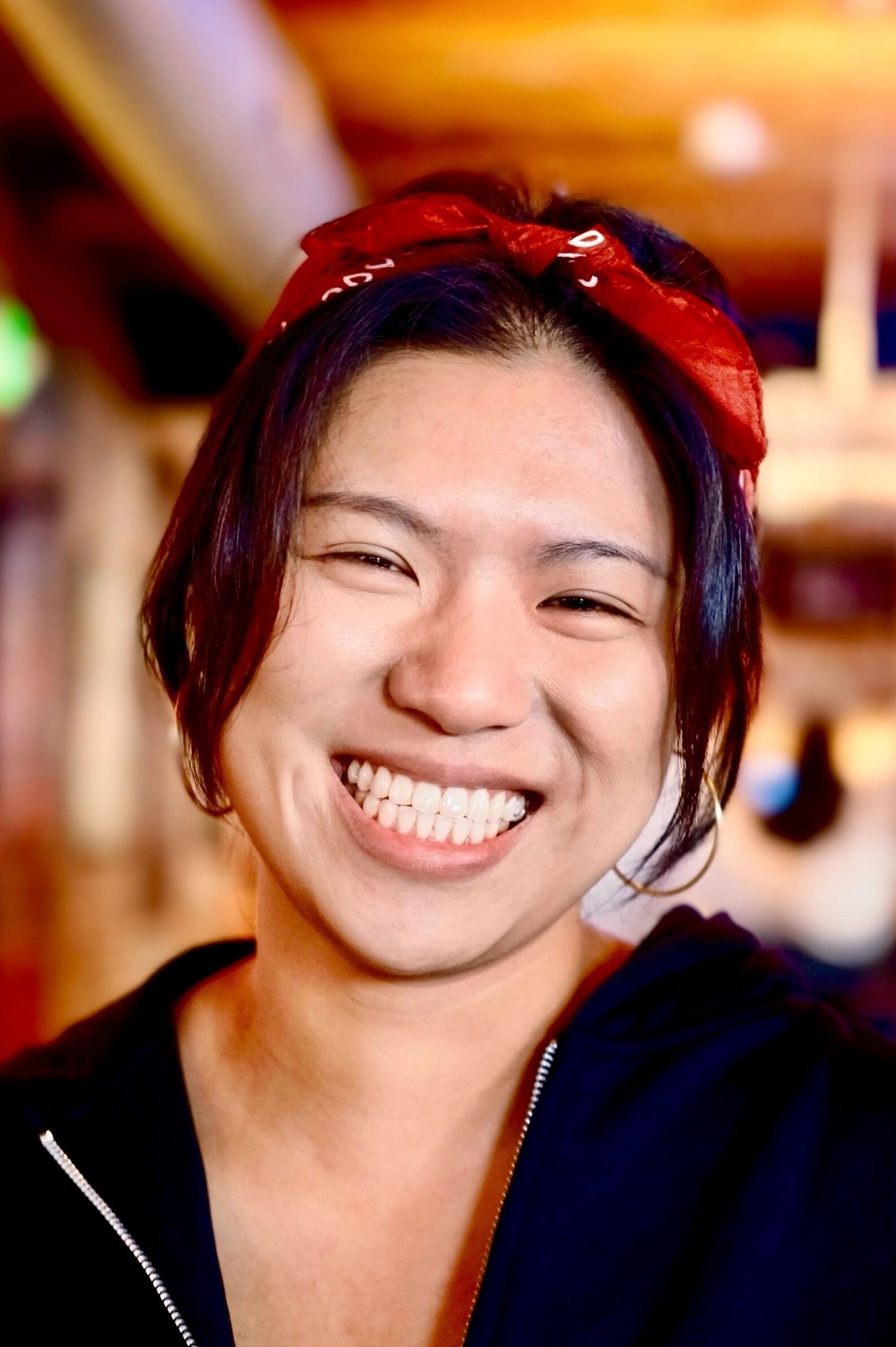}}]{Rebecca Lin}
is an EECS Ph.D. student at MIT, co-advised by Erik Demaine in CSAIL and Zach Lieberman at the Media Lab. Her research focuses on mathematical abstractions and computational tools that support expressivity in art, design, and fabrication. She is supported by the MIT MAD Design Fellowship, an NSERC Postgraduate Scholarship, and the MIT Stata Family Presidential Fellowship.
\end{IEEEbiography}

\begin{IEEEbiography}[{\includegraphics[width=1in,height=1.25in,clip,keepaspectratio]{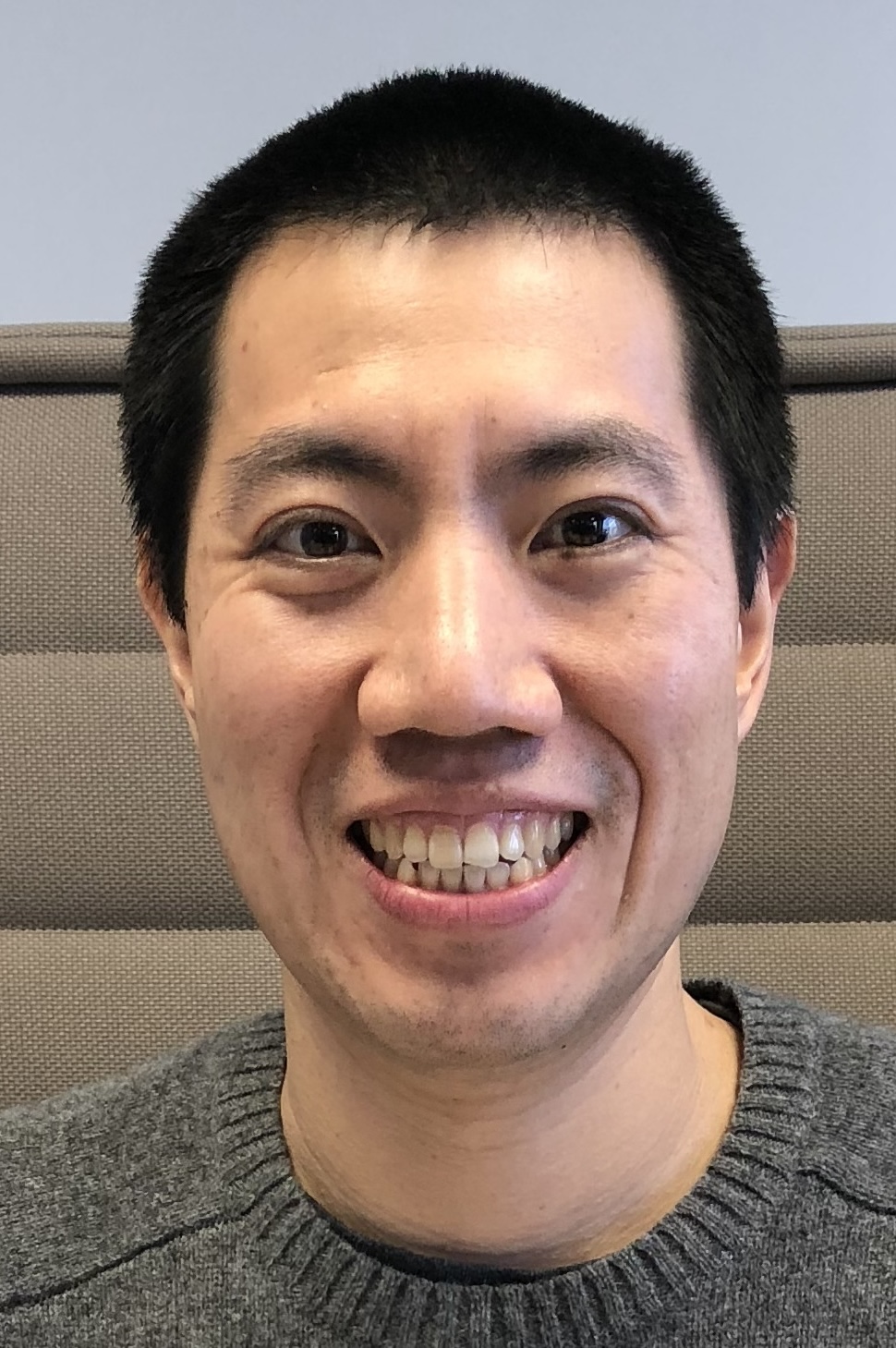}}]{Daniel Miau}
received his Ph.D. degree in computer science from Columbia University in 2018. He is currently a computer scientist at Adobe. His research interests include computational photography, machine learning, and human-computer interaction.
\end{IEEEbiography}

\begin{IEEEbiography}[{\includegraphics[width=1in,height=1.25in,clip,keepaspectratio]{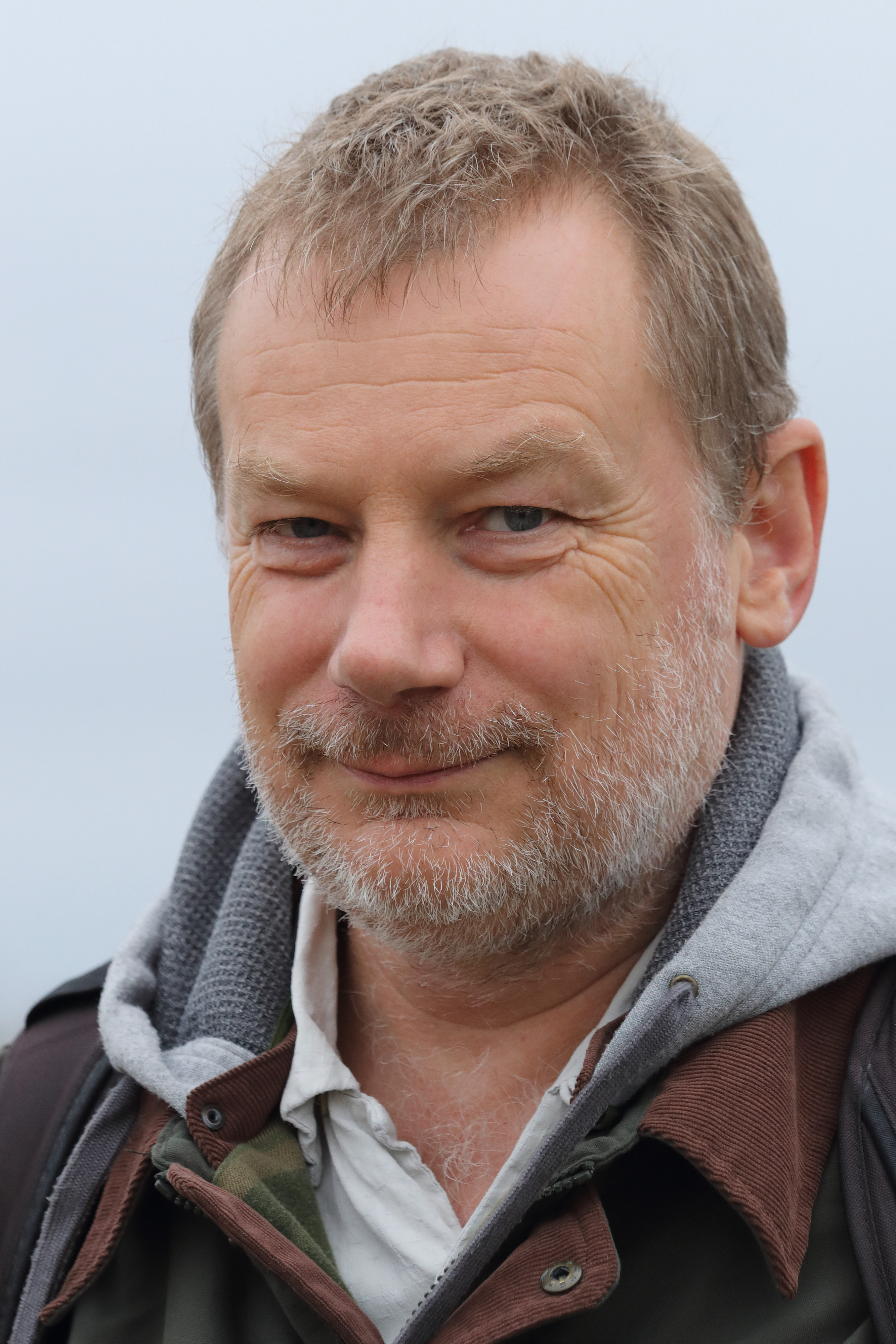}}]{Florian Kainz}
is a Senior Computer Scientist at Adobe, where he is a member of a computational photography team. Prior to joining Adobe he was at Google, working on image capture for virtual reality, and on the Astrophotography Mode for Pixel phones. Before that he developed software for visual effect production at Industrial Light \& Magic.
\end{IEEEbiography}

\begin{IEEEbiography}[{\includegraphics[width=1in,height=1.25in,clip,keepaspectratio]{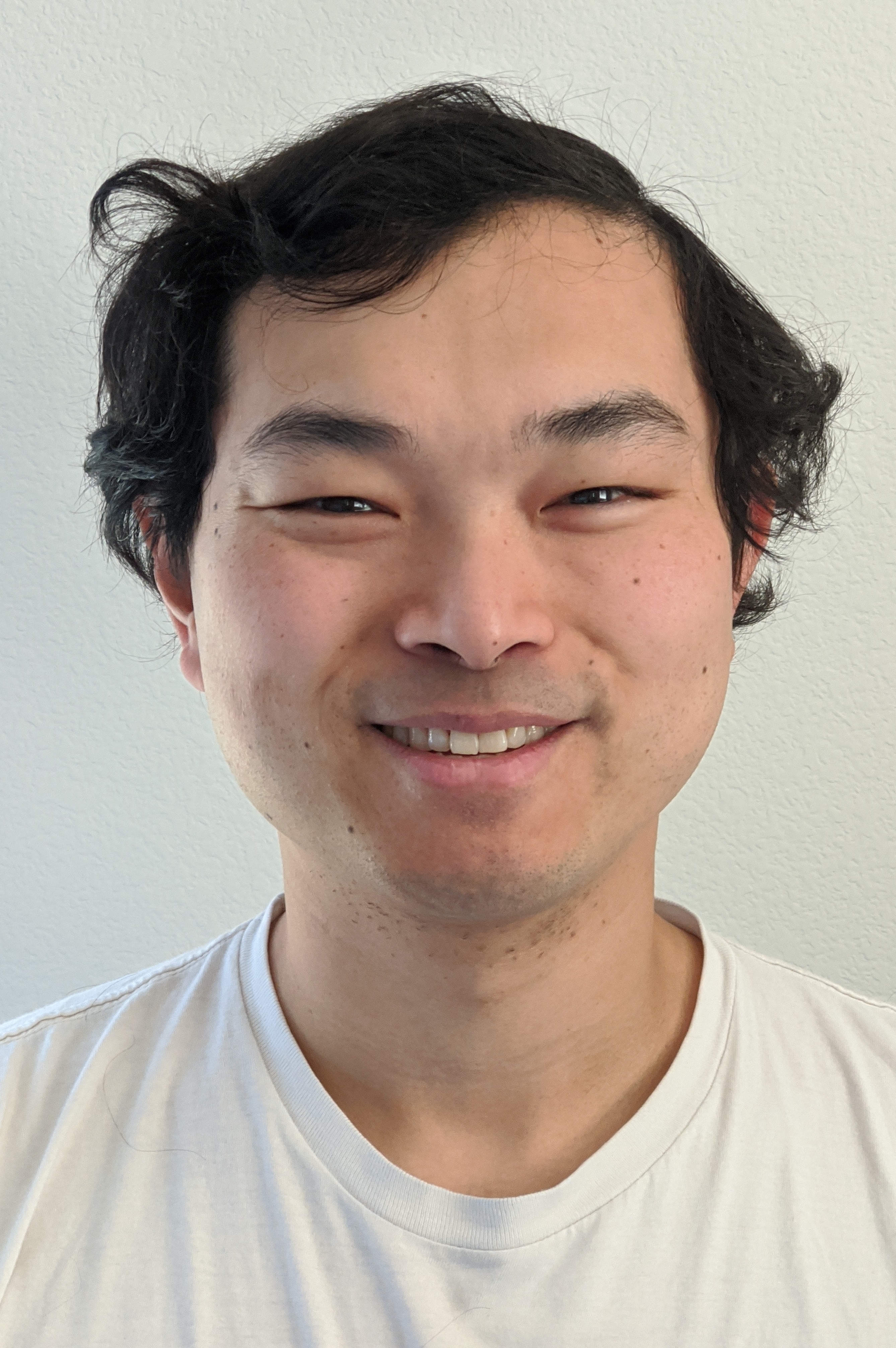}}]{Jiawen Chen}
is a Principal Research Scientist / Engineer at Adobe. He holds a PhD in Electrical Engineering and Computer Science from MIT where he studied under Frédo Durand. His research interests are in computational photography, signal processing, and high performance computation.
\end{IEEEbiography}

\begin{IEEEbiography}[{\includegraphics[width=1in,height=1.25in,clip,keepaspectratio]{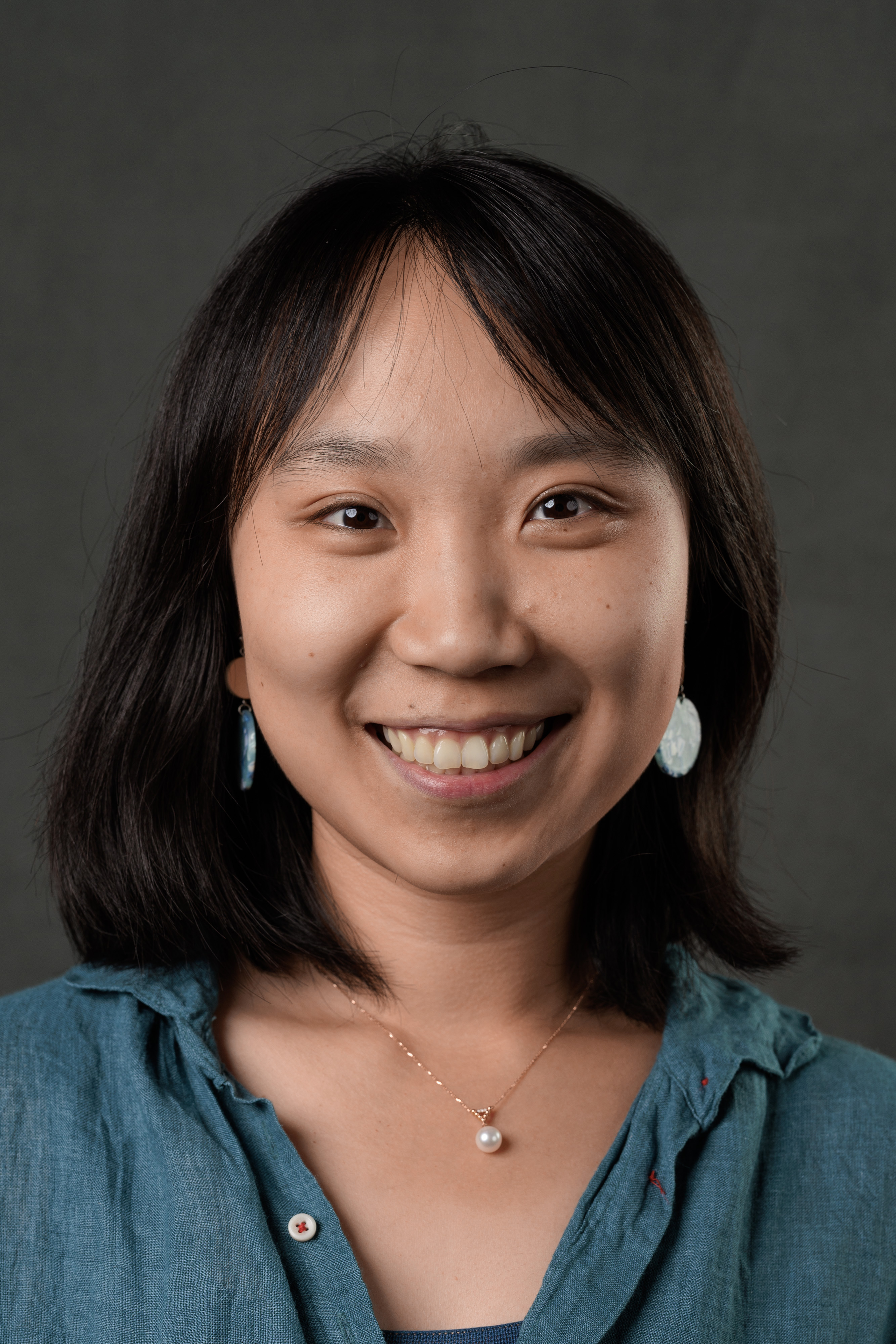}}]{Xuaner Zhang}
received the Ph.D. degree in Computer Science from the University of California, Berkeley, in 2020, advised by Ren Ng. She received the B.S. degree from Rice University, where she worked on computer vision research under the supervision of Ashok Veeraraghavan. She was a researcher at Adobe, where she was part of the Computational Photography team led by Marc Levoy. Her work focused on developing AI-first features for photographers and building a software-defined camera app using computational photography.
\end{IEEEbiography}

\begin{IEEEbiography}[{\includegraphics[width=1in,height=1.25in,clip,keepaspectratio]{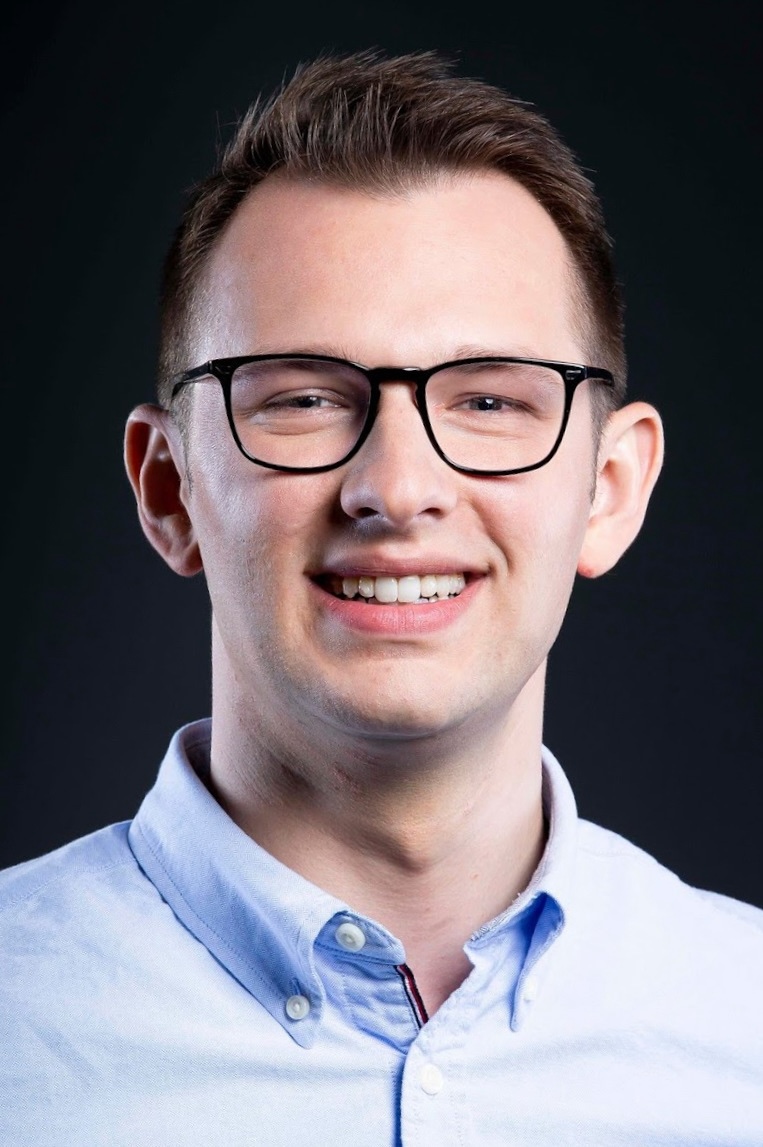}}]{David B. Lindell}
is an Assistant Professor in the Department of Computer Science at the University of Toronto. Prior to joining the University of Toronto, he received his Ph.D. from Stanford University. His work combines emerging sensors, machine learning, and physics-based models to enable new capabilities in visual computing. He is a recipient of the ACM SIGGRAPH Outstanding Dissertation Award Honorable Mention, a Google Research Scholar award, a Sony Faculty Innovation Award, and the 2023 Marr Prize.
\end{IEEEbiography}

\begin{IEEEbiography}[{\includegraphics[width=1in,height=1.25in,clip,keepaspectratio]{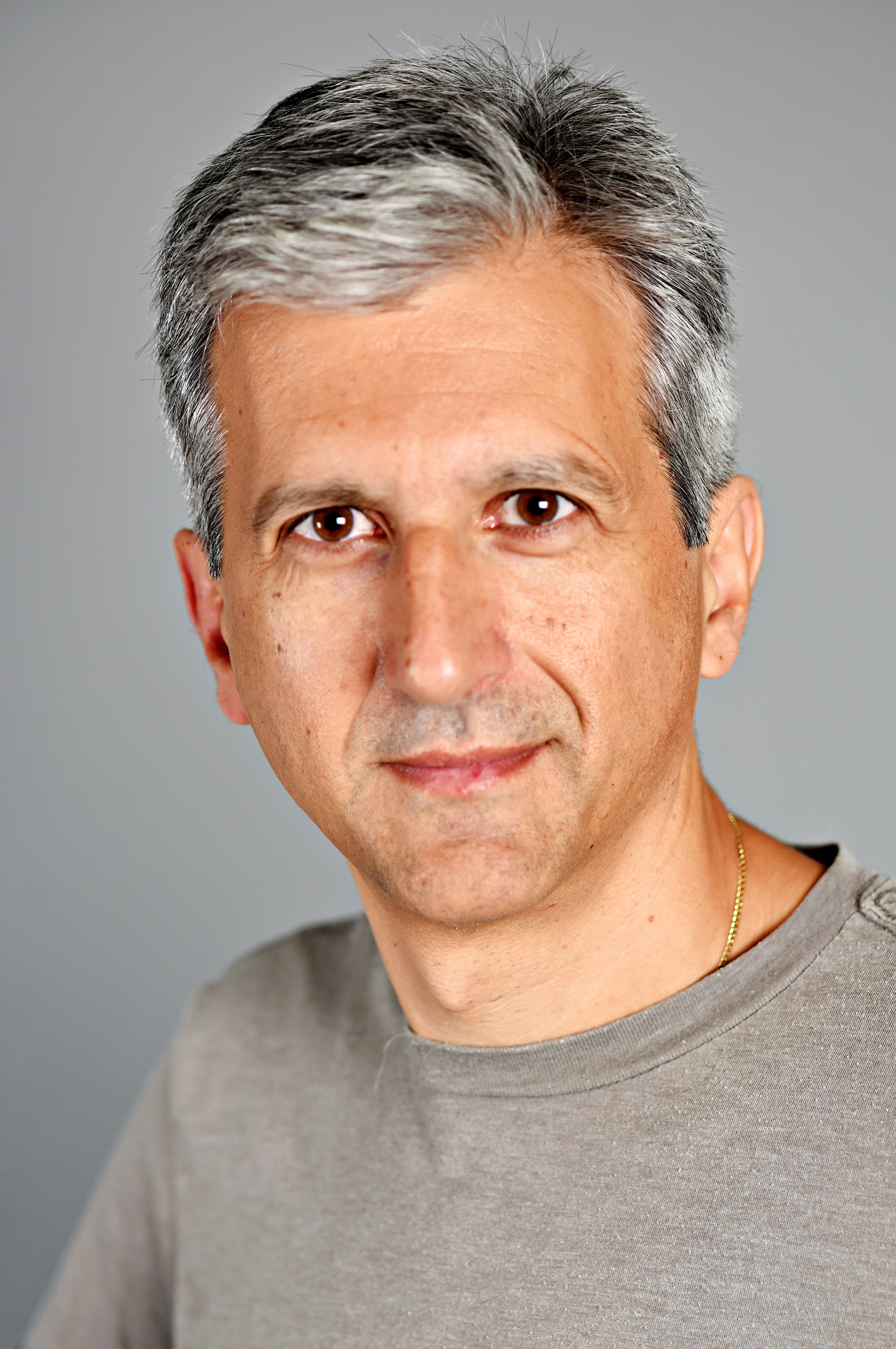}}]{Kiriakos N. Kutulakos}
 is a Professor of Computer Science at the University of
Toronto. He received his PhD degree from the University of Wisconsin-Madison in
1994 and his BS degree from the University of Crete in 1988, both in Computer
Science. His interests include computer vision and computational imaging. Kyros
is the recipient of an Alfred P. Sloan Fellowship, Marr Prizes in 1999 and
2023, a Marr Prize Honorable Mention in 2005 and five more paper prizes (CVPR
2019, CVPR 2017, CVPR 2014, ECCV 2006 and CVPR 1994). He was an Associate
Editor of IEEE Transactions on Pattern Analysis and Machine Intellegence and
was Program Co-Chair of CVPR 2003, ICCP 2010, and ICCV 2013. Kyros is also
serving as Program Co-Chair of ICCP 2025.
\end{IEEEbiography}

\end{document}

\title{Learning Lens Blur Fields \\ Supplementary Material}
\author{Esther Y. H. Lin,~\IEEEmembership{Student Member,~IEEE,}
Zhecheng Wang,~\IEEEmembership{Student Member,~IEEE,} \\
Rebecca Lin,~\IEEEmembership{Student Member,~IEEE,}
Daniel Miau,~\IEEEmembership{Member,~IEEE,}
Florian Kainz,~\IEEEmembership{Member,~IEEE,}\\
Jiawen Chen,~\IEEEmembership{Member,~IEEE,}
Xuaner Zhang,~\IEEEmembership{Member,~IEEE,}\\
David B. Lindell,~\IEEEmembership{Member,~IEEE,}
Kiriakos N. Kutulakos,~\IEEEmembership{Member,~IEEE,}
}

\maketitle

\definecolor{olivegreen}{HTML}{3C8031}

\newcommand{\PSF}{\text{PSF}}
\renewcommand{\S}{\text{S}}
\newcommand{\T}{\text{T}}

\newcommand{\fignum}[1]{\ref{#1}}
\newcommand{\sectnum}[1]{\ref{#1}}

\newcommand{\edit}[1]{\textcolor{black}{#1}} %

\newcommand{\finaledit}[1]{\textcolor{green}{#1}} %

\section{Supplemental Capture and Processing Details}
In this section, we provide additional details about the capture setup and preprocessing pipelined to prepared training imaged for learning lens blur fields.

\subsection{Image Capture}
Our captures can be split into three main categories: captures done with iPhones running iOS, Pixel 4 with Android, and DLSRs. For iPhones, we capture focal stacks for patterns. The sequence of operations is for each pattern, we capture a focal stack. For Android and DSLR, we flip the sequence to capture the full set of patterns for each focus setting. These sequences are based on our app implementations and control of cameras. Since we capture focal stacks of each pattern for our iPhone captures, we may be susceptible to variances in the focus distance each time we capture a focal stack. Our SLR lenses are captured with a Canon EOS 6D Mark II. 

\paragraph{Overview of Capture Dataset} Our dataset contains a collection of smartphone and SLR lens models. We also provide the code for an Android application we developed to capture focal stacks of raw dual-pixel left and right photodiode images. Our Android application can only read out the left and right photodiodes of the green pixels. Hence, we do not have data for the red and blue channels, and our models only output estimates for the green left and right dual-pixels.

\paragraph{Remote Control for Capturing} To minimize motion blur, the smartphones are remotely controlled through the use of a Bluetooth mouse. Similarly, SLR cameras are remotely controlled via manufacturer-provided software or a remote control switch. Smartphones are mounted on phone holders, and SLR cameras are mounted on tripods. For SLR lenses, we find software-controlled positioning to be less repeatable than simply adjusting the lens focus ring, so we configured the focus position manually.

\paragraph{\edit{Monitor Choice}}

\edit{This method requires a high-DPI monitor (e.g., a 4K display) that is sufficiently large (at least 32 inches) so that the camera can be positioned at a distance that minimizes moiré artifacts resulting from the RGB subpixel structure. In addition, most modern LCD monitors include a built-in diffuser (often implemented as a light guide plate) between the LED backlight and the LCD panel, which further mitigates the risk of moiré patterns. We would also like to note that if the monitor is not large enough, our partial field of view experiment in Sec. IV-I in the main paper demonstrates that we can leverage the interpolating nature of MLPs to relax the monitor size constraint by piecing together the image plane from multiple monitor positions.
}

\paragraph{Monitor Refresh} To avoid capturing screen refresh, it is important to use an appropriate exposure time. We set ISO to the lowest possible value. Our iPhone calibrations use an exposure time of 100 ms (the max allowable exposure time in our app), the Pixel 4 uses an exposure time of 250 ms, and SLR captures uses 250 ms. We capture bursts of three images of the same exposure time for smartphone captures. Our radiometry compensation procedure of capturing black and white images at the same position across the full focal stack allows us to effectively compensate for this characteristic. It corrects for any spatial non-uniformity due to the monitor's illumination.

\paragraph{Focal Stack Sampling} We sample 7 to 24 focus positions depending on the control we have over focus control. For example, the Android platform allows for more granular control, enabling our app to sample 24 lens positions, while our iPhone app is constrained to 20. For SLR lenses, we were constrained to the number of markings on each lens such that we could identify focus distance. Overall, the acquisition process is efficient, with the limiting factor being exposure time. For example, our iPhone app automates the process of capturing a focal stack. Efficiency can be further improved by synchronizing the patterns displayed on the monitor with the camera, allowing for the automatic capture of a focal stack upon a change in pattern on the monitor. For our smartphone captures (iPhone and Pixel), we sample our focal stack uniformly in diopter space. Our 6D experiment also samples the sensor-object distance uniformly in diopter space. 

\paragraph{Camera Placement} To capture training images, cameras are positioned in a fronto-parallel orientation facing the screen, with distance chosen such that the screen just fills the sensor's field of view when the monitor is in focus. For experiments where sensor-screen distance is varied, we mount the camera on a linear translation stage for systematic control of distance. 

\subsection{Overview of Registering Blur-Free Images to Captured Data}
\label{ssec:registering-sharp}

\begin{figure*}[!t]
\centering
\includegraphics[width=1\textwidth]{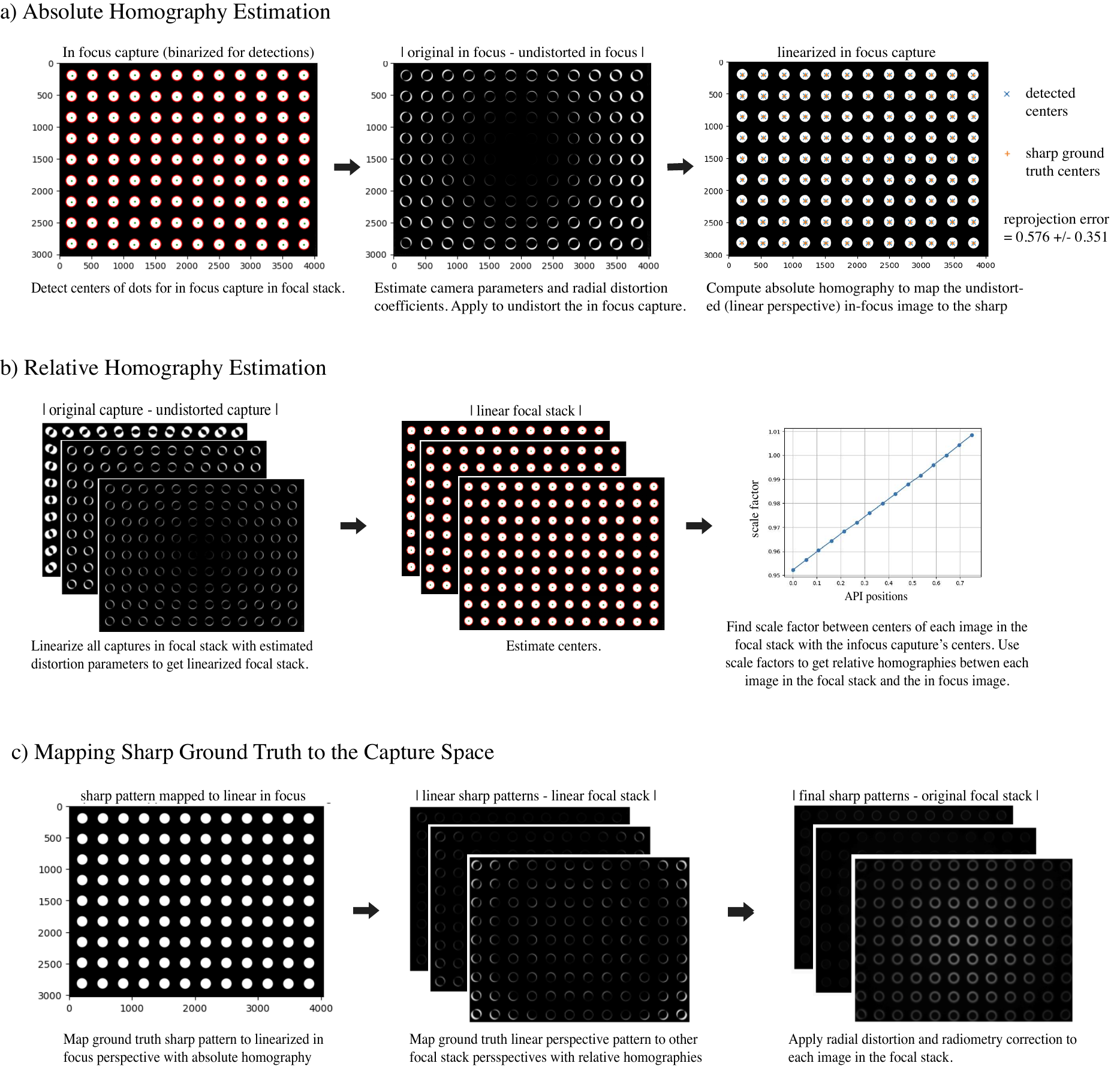}
\caption{\edit{Overview of our processing pipeline for registering sharp ground truth images to captured data, as outlined in~\cref{ssec:registering-sharp}. Example images are from the preprocessing of an iPhone 12 Pro wide camera.}}
\label{fig:registering-sharp}
\end{figure*}

\begin{figure}[!t]
\centering
\includegraphics[width=\linewidth]{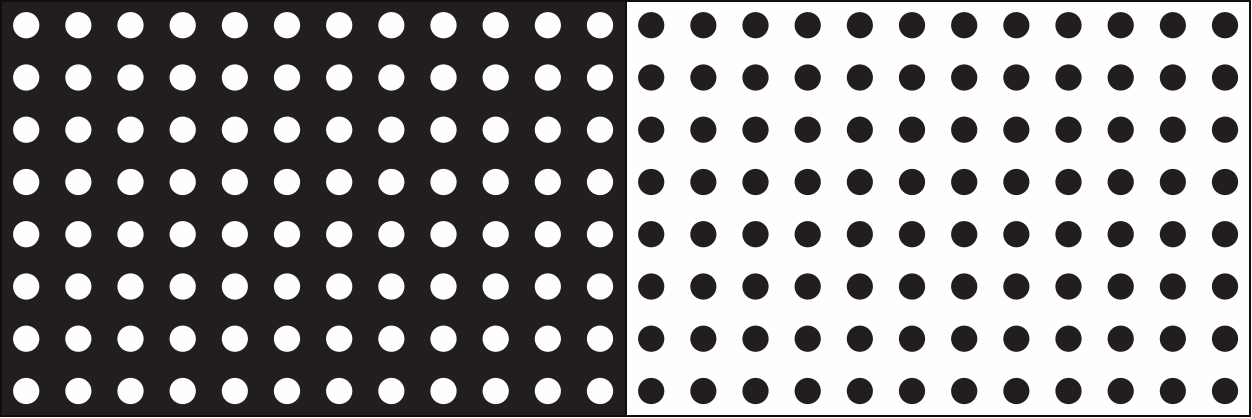}
\caption{Grid of dots and its inverse for homography correction. The spacing between the centers of each dot is four times the radius of each dot.}
\label{fig:homography-patterns}
\end{figure}

\begin{figure}[!t]
\centering
\includegraphics[width=\linewidth]{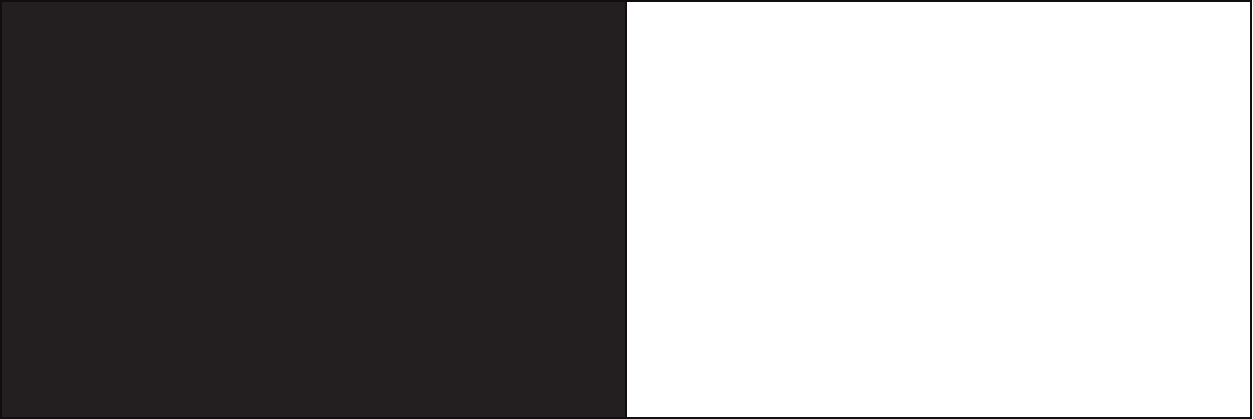}
\caption{Black and white binary images for radiometry and vignetting correction. }
\label{fig:radiometry-patterns}
\end{figure}

In this section, we detail the preprocessing steps taken to prepare raw focal stack captures as input to our training. An overview of this process with examples on an iPhone 12 Pro wide camera are shown in ~\cref{fig:registering-sharp}. All operations are performed on the full sensor resolution. For input, we capture a focal stack of a grid of dots and its conjugate image. 

\paragraph{Step 1: Absolute Homography Estimation}
The purpose of this step is to 1) map raw captured images to linear perspective by accounting for radial distortion, and 2) registering our sharp digital pattern to the captured focal stack. The input to this step is a focal stack of a grid of dots. First, the most in-focus image from the input raw focal stack is identified. Then, camera parameters and radial distortion coefficients are estimated for this in-focus image. Using these parameters, the most in-focus image is mapped to a linear perspective. A homography is then computed, mapping the linear perspective in-focus image to the sharp ground truth pattern. The outputs of this step are camera parameters, radial distortion coefficients, and the homography matrix that aligns our sharp pattern to the in-focus image in the focal stack.

\paragraph{Step 2: Relative Homography Estimation}
The radial distortion coefficients are applied to each image in the focal stack to create a linear perspective focal stack. For each linear perspective image in the stack, an image scaling-only homography is found to map it to the in-focus image. The scaling is performed around the principal point to ensure accurate alignment between images in the focal stack. This step enables the creation of a linear perspective focal stack that can be compared with the in-focus image.

\paragraph{Step 3: Mapping Sharp Ground Truth to the Capture Space of Each Image in Focal Stack}
First, the sharp ground truth is mapped to the in-focus image's linear perspective using the previously computed homography. Next, a sharp linear perspective focal stack is created by applying the scaling-only homography to the transformed ground truth. Radial distortion is then applied to to generate sharp images in the raw capture space. Finally, radiometry compensation is performed using the white and black images captured in the lens positions' raw space. This process ensures that the sharp ground truth is correctly mapped to the raw capture space of each image in the focal stack.

\paragraph{Output of Processing Procedure}
The final outputs to the training process include the raw blurry focal stack without any modifications and the sharp focal stack mapped to the raw capture space. These datasets can be used for training algorithms that aim to improve the sharpness of captured images.

\subsection{Additional Processing Details}

\paragraph{Choice of grid of dots pattern}
We display binary preprocessing grid-of-dots pattern on the screen or monitor for the preprocessing operations mentioned above. We use the dot centers as keypoints for our homogrpahy and radial distortion estimations. We chose the grid-of-dots to be our homography pattern because it is more robust to defocus. At extreme levels of defocus, corner-detection-based patterns like the checkerboard fail because of large defocus sizes. We use dots grids of 9x12, where the center-to-center spacing between each dot is four times the radius of each dot.

\paragraph{Identifying the centers of the dots}
Our circle detection is composed of OpenCV's Hough Transform, refined by computing centroid coordinates within the neirborhood of each centre. The neighborhood is chosen to be the radius of the dots. For more accurate estimations for both radial and homography estimations, we perform circle detection on the binarization of our captured images. We perform binarization following the method outlined by ~\cite{scharstein2003high}. We display the grid-of-dots and its inverse (\cref{fig:homography-patterns}) on the screen. We then label each pixel based on whether the pattern or its inverse appears brighter. This methods helps us avoid manual tunings of a binarization threshold value.

\paragraph{Radial distortion correction} 
The dot centers are then used as input to OpenCV's camera calibration function \edit{cv2.calibrateCamera()} to find radial distortion coefficients of the most in focus image.

\subsection{iOS Camera Lens Position API Calibration}
Unlike Android systems where the focal point can be set with metric units, iOS returns the focus position of the lens as a scalar value between $[0.0,1.0]$ with $0.0$ being the shortest distance at which the lens can focus and $1.0$ the furthest. We experimentally calibrate in $[0.1,0.8]$ for our experiments as we find the behaviour outside this range to have little to no change.

To calibrate these API values, we mount the iPhone on a high precision linear rail and tabulate the API outputs as we focus on the screen while moving the camera away from the screen. A dense sampling is captured and used as a lookup table and interpolation is used to get metric distances for API values that weren't included in the calibration. We perform this calibration for each camera on each iPhone. After we are able to map iOS len position API returns to a physical metric distance, we convert to diopter space for training.

\subsection{Radiometry Compensation and Vignetting Correction}
We display and capture completely black and white binary focal stacks for radiometry compensations. It is important for the images to be binary and not grayscale to sidestep calibrating for the monitor's transfer function. Since we capture raw images, they are linear with a known per-image black level. Overall, we calibrate in the camera's device-dependent color space with no non-linear contributions from the display. The focus positions in the focal stack correspond to the focus positions used for capturing training images. The binary black and white images we display are shown in \cref{fig:radiometry-patterns}. These images are directly used without any alterations, this allows us to also account for monitor and lens vignetting. 

\begin{figure*}[!h]
\centering
\includegraphics[width=0.99\textwidth]{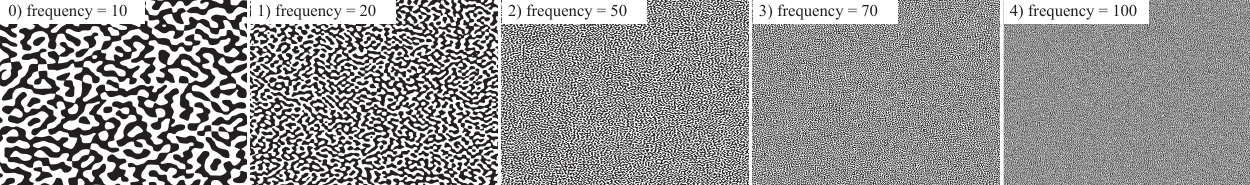}
\caption{\edit{Synthetic noise patterns used in trainings generated using the method of Couture et al. \cite{couture2011unstructured} with controlled spatial frequency content. The five patterns correspond to frequency parameters of 10, 20, 50, 70, and 100, where lower frequency values yield coarser textures and higher values finer details. We provide code for generating these patterns.} }
\label{fig:simulation-patterns}
\end{figure*}

\edit{\subsection{Generating Synthetic Noise Patterns}}
\edit{We follow Couture et al. \cite{couture2011unstructured} to generate synthetic noise patterns with controlled spatial frequency content, as shown in \cref{fig:simulation-patterns}. First, we generate random noise and then apply a Fourier transform to filter the frequency components, thereby defining a target frequency band for the texture. This band is controlled by a parameter \(f\) that sets the lower bound, while the upper bound is automatically set to \(2f\) by zeroing out components below \(f\) and above \(2f\). For example, when \(f=10\), the pattern retains lower frequencies and produces coarser textures; when \(f=20\), the band shifts to higher frequencies, yielding finer details. Finally, the image is binarized based on its mean value.}

\section{Supplemental Experimental Details}

\begin{table}[h]
  \centering
  \label{tab:experimental-hardware}
  \resizebox{0.98\linewidth}{!}{
  \begin{tabular}{lcccc}
    \toprule
    Description & Setup(s) & Quantity & Model Name & Company\\
    \midrule
    4K 32 inch monitor & SLR, smartphone & 1 & UltraSharp & Dell\\
    Linear actuator & smartphone & 1 & X-BLQ1045-E01 & Zaber\\
    Phone holder & smartphone & 1 & VTM7-ALX-1 & Vastar \\
    Tripod legs & SLR & 1 & 3258 & Manfrotto \\
    Tripod head & SLR & 1 & 3047 & Manfrotto \\
    \bottomrule
  \end{tabular}
  }
  \vspace{0.8em}
  \caption{List of hardware used in experimental setups.}
  \label{tab:tab-harware}
\end{table}

\subsection{6D Varying Sensor-Object Distance Experiment}
The 6D experiment (shown in main paper) aims to illustrate our method's adaptability to increased spatial dimensions. Our 6D experiments were captured with an iPhone 12 Pro Wide camera. We capture a sensor-object stack with six slices roughly sampled uniformly in diopter space, where the closest sensor-object distance corresponds to the nearest distance our iPhone 12 Pro Wide camera can consistently focus on, 20.5 cm, and the farthest to 41.5 cm. Focal stacks of 15 images are captured at each sensor placement. The camera is mounted on a Zaber linear actuator for systematic control of the positions. 

\subsection{Device Repeatability Experiments}
We identify three categories of repeatability experiments:  
\begin{enumerate}
    \item Without remount (stationary) repeatability: Comparing how estimated blur fields vary when capturing and recapturing data for a single camera, where the only difference between each capture is the camera/phone is on sleep mode. The placement of the camera is not altered between captures.  
    \item With remount repeatability: Comparing how estimated blur fields vary when capturing and recapturing data for a single camera, where the difference between each capture is the phone being put on sleep mode and a dismount and remount. 
    \item Lens remount repeatability: Comparing how estimated blur fields vary when capturing and recapturing data for a single lens when capturing and recapturing data for a single SLR camera, where the only difference between each capture is the lens being unmounted and remounted.
\end{enumerate}

\begin{figure*}[!t]
\centering
\includegraphics[width=0.99\textwidth]{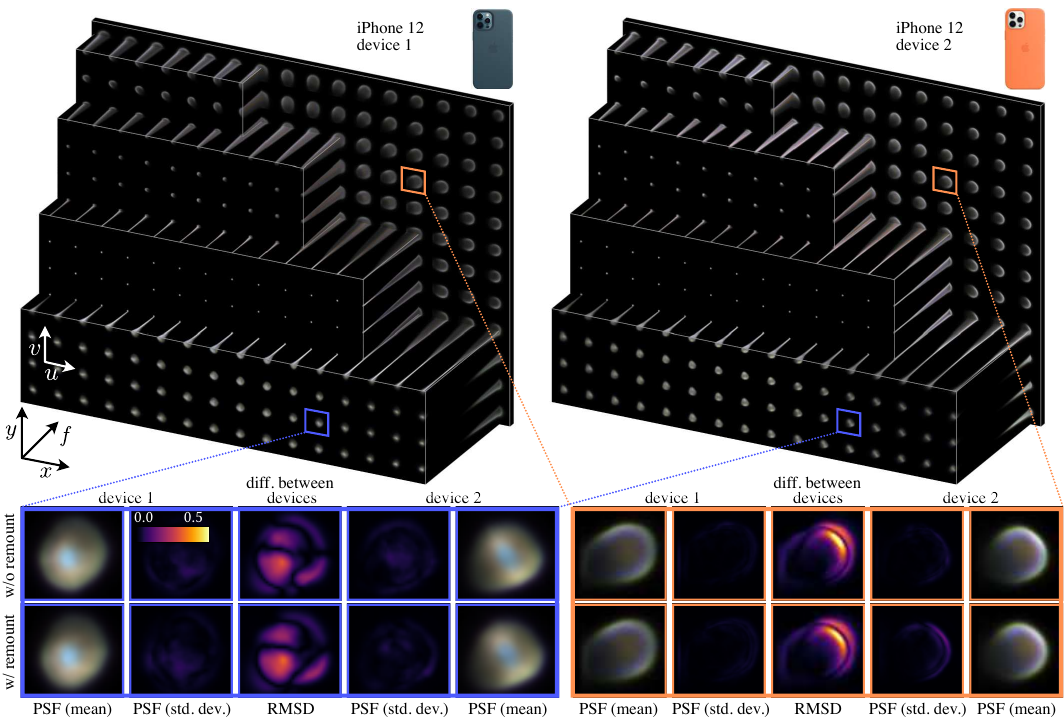}
\caption{Comparison betw een calibrated lens blur fields from two iPhone 12 Pro devices visualized as different slices of the 5D blur field. We estimate blur fields using three separate sets of calibration measurements captured independently for each device. Visualizing slices of the mean estimated 5D PSF function for each device (A, B, top) reveals broad agreement in the PSF structure; however, significant variations can be seen upon close inspection (insets, bottom). While imperfections in the calibration and capture process result in differences between PSF estimates for the same device (see standard deviation of PSFs; A, B, bottom left and right), this effect is generally weaker than PSF differences between devices (center insets, shown as root-mean-square deviation).}
\label{fig:device-id}
\end{figure*}

Together, we show that our method can resolve the differences in optical behavior between the wide-angle cameras on two separate iPhone 12 Pro devices (\cref{fig:device-id}) and two separate iPhone 14 Pro devices (see main paper).

\paragraph{Without remount repeatability experiments} 
We use a pair of iPhone 12 Pro  wide cameras and a pair of iPhone 14 Pro wide cameras for this experiment. For each iPhone, we mount it on a phone clamp, placed 37 cm away from a 4K 32 inch monitor. The distance is chosen such that the patterns on the monitor just fill the entire field of view. For each capture repetition, we take focal stacks of the black and white radiometry compensation patterns, the two grid of dots for homography correction, and several frequency patterns. For each lens position, we take a burst of three images. In processing, we operate on the average over the three images for each lens position.

We provide supplemental visualizations illustrating variations in the lens blur field between two calibrated iPhone 12 Pro devices.
In addition to the 5D visualization shown in the main paper, \cref{fig:device-id} (bottom) contains a $(u, v)$ cross section of the blur field, where each sub-image of the visualization indicates regions on the image plane where the PSF is active (for a given $(u, v)$ coordinate).

\paragraph{With remount repeatability experiments} 

We use a pair of iPhone 12 Pro wide cameras and a pair of iPhone 14 Pro wide cameras for this experiment. Capture settings are similar to that of the without remount experiments, except after each capture repetition, the phone is dismounted from the holder and placed on a nearby tabletop for at least 1 minute with the display turned off. Then, the phone is remounted onto the holder in preparation for the next capture repetition. Therefore, the phone placements between each repetition is different as all alignments are done by hand.

\paragraph{Lens repeatability experiments} 
Our methodology displays robustness against dismounting and remounting of SLR lenses, indicating that optical axis misalignment is negligible between lens removals. This is evidenced through repeatability experiments (both without remount and with remount) conducted with a Canon EF 50mm f/1.4 lens. For the without remount experiment, we power off the camera between trials, whereas for the with remount experiment, we dismount, remount, and restart the camera between trials. Each experiment involves capturing 5 distinct focal stack datasets and training an individual model for each, resulting in a total of ten datasets. The comparison of blur fields between without remount and with remount experiments reveals insignificant differences, as shown in \cref{fig:device-id}.

We use a Canon EF 50mm f/1.2L USM lens for this experiment. Between each capture repetition, we unmount the lens and remount it onto the SLR camera. The position of the camera is 1 meter away from a 4K 32 inch monitor and is fixed for each capture repetition. The distance is chosen such that the patterns on the monitor just fill the entire field of view. For each capture repetition, we take 8-lens position focal stacks of the black and white radiometry compensation patterns, the two grid of dots for homography correction, and several frequency patterns. For each lens position, we take a single image of exposure time 250ms. The camera is turned off between each capture repetition. The camera is also mounted on a tripod. The focus position is controlled manually.

\paragraph{Registration between trials for iOS}
There can be uncertainty in iOS API readings, particularly in the starting and ending lens positions. To ensure comparability between trials, we register them to a reference trial. This registration is carried out by reusing the scale factors during post-capture processing. We designate the first trial as the reference trial and translate and scale the API versus scale factor plot, ensuring proper alignment with the reference trial.

\subsection{Partial Field-of-View Experiment} 

While we have primarily focused on lenses with constrained fields of view for which the display projects across the entire sensor. For wide field of view lenses, the display may illuminate only part of the sensor. 

In this experiment, we illustrate the scalability of our method to lenses with expansive field of views. Rather than capturing focal stacks that cover the entire image plane, we apply a sparse scanning regime. This enables us to assemble focal stacks taken from various image plane locations, effectively covering the full image plane, demonstrating the versatility of our approach. We capture multiple focal stacks by repositioning the phone (while maintaining the same distance $d$ to the scene plane) so that the display projects to different regions of the sensor. We then recover the lens blur field by training on the sensor regions that image the display across the multiple captured focal stacks. It is important to note that the processing procedure for registering blur-free images to the captured data can be applied separatey to each partial view.

\subsection{Pixel 4 Experiment}
In our Pixel experiments, the raw dual pixel values read out were binned, using a factor of 2 for the width and 4 for the height of the raw image resolution. This binning process is inherent to the sensor and cannot be adjusted through app development. To guarantee training on the raw full pixel grid resolution, we upsampled with linear interpolation. Similar to previous experiments, the MLP was trained on pixels where we had data, in this case, solely the green channel, and interpolated for other areas.
 
\subsection{Summary of Experimental Setups}
We detail the hardware used in our captures (\cref{tab:experimental-hardware}, \cref{tab:tab-harware}), physical distances (\cref{tab:experimental-distances}, \cref{tab:tab-dataset-params}) , and frequency patterns used. Besides the 6D experiment, the camera and screen distances remains constant for each experiment throughout the capture of all patterns.

\edit{\subsection{Focal Stack Acquisition Times}}

\edit{The total acquisition time to capture the input focal stack on smartphones is approximately 15 minutes, assuming the use of a streamlined smartphone capture app that supports burst captures and a secure mounting solution (e.g., a tripod or selfie stick). Specifically, the actual data capture—estimated at about 200 ms per image for 3 images per burst, over roughly 8 patterns and 15 focal positions—takes around 1.5 minutes, while the remaining 5 to 10 minutes are spent on alignments and setup. Alignment and setup involve mounting the camera and positioning the monitor within its field of view. This may include centering the monitor in the camera's view or arranging the tiling placements for the partial field-of-view experiments in Sec. IV-I of the main paper. Note that this estimate applies exclusively to smartphones, which allow programmatic acquisition of the entire dataset; we built apps in iOS and Android to automate the capture process for our smartphone experiments. We were able to set the number of focus positions, images in each burst per focus position, ISO, and exposure time. Acquisition times for our SLR cameras were longer (average 30 mins per lens, including setup) because we did not have such an interface and would need to set the focus positions manually.  }

\begin{table*}
  \centering
  \label{tab:experimental-distances}
  \resizebox{0.90\textwidth}{!}{
  \begin{tabular}{lcccc}
    \toprule
    Experiment/ & Capture & Sensor-Screen & Aperture & Quantity of \\
    Dataset Item & Device & Distance(s) [cm] & Setting & Training Patterns $N_i$\\
    \midrule
    with remount repeatability & iPhone 12 Pro Wide & 37 & n/a & 4\\
    without remount repeatability & iPhone 12 Pro Wide & 37 & n/a & 4\\
    lens remount repeatability & Canon EOS 6D Mark II & 100 & f/2.5 & 4\\
    6D & iPhone 12 Pro Wide & $[20.5 - 41.5]$ & n/a & 4\\
    Pixel4 & Pixel 4 Wide & 37 & n/a & 3\\
    Canon EF 50mm f1.2 & Canon EOS 6D Mark II & 100 & f/1.2 & 4\\
    Canon EF 50mm f1.4 & Canon EOS 6D Mark II & 100 & f/2.5 & 4\\
    Canon EF 24-70mm f/2.8L & Canon EOS 6D Mark II & 100 & f/2.8 & 4\\
    Canon EF 14mm f/2.8L & Canon EOS 6D Mark II & 35 & f/2.8 & 4\\
    \bottomrule
  \end{tabular}
  }
  \vspace{0.8em}
  \caption{List of experiments and database items, the device they were captured with, their capture distances from screen.} 
  \label{tab:tab-dataset-params}
\end{table*}

\section{Supplemental Implementation Details}

In this section, we provide additional details about the training process for learning lens blur fields.

\edit{\subsection{Training Details}}

Lens blur fields are trained for $4 - 10 \times 10^6$ iterations with the Adam optimizer ($\beta_1 = 0.5$,  $\beta_2 = 0.999$, $\eta = 10^{-5}$ 
$\epsilon = 10^{-15}$, $||\mathbf{w}||^2 = 10^{-6}$, $\gamma = 0.98$, update step $= 10000$) and an exponential decay learning rate scheduler on a single NVIDIA RTX A6000 using the \texttt{tiny-cuda-nn} framework~\cite{tiny-cuda-nn}. For the SLR lenses, they were trained on half-resolution, downsampled data. We use 7 layers, 512 units each for all models. 

\edit{We find that the inference time—i.e., the time to recover a PSF from the MLP—is $277 \pm 19$ microseconds for a blur kernel tensor of size 80×80×4, and the rendering time—i.e., the time to apply the blur field to an image via convolution—for a patch size 1.5× the blur size is $730 \pm 27$ microseconds for a blur kernel tensor of size 80×80×4 on a NVIDIA RTX A6000. Values and mean and standard deviation over 1000 measurements. Overall, our measurements indicate that the convolution operation is the most time-consuming component, not the model’s inference time to recover a PSF. Also, not that recovering a PSF from the MLP requires a fixed amount of time for a given image resolution, while the convolution time scales quadratically with the dimensions of the patch size. 
}

\edit{\subsection{Selecting Synthetic Noise Training Pattern Frequencies}}
There are several considerations when selecting pattern frequencies. \edit{While \cref{ssec:frequency-ablation} discusses the relationship between estimation results and input pattern frequencies, application of the finding to real-life experimental settings requires prior knowledge of the maximum size of the blurs.} 

\edit{In practice,} we follow a rough guideline for determining the set of frequencies that should be used, \edit{inspired by the analysis in \cref{ssec:frequency-ablation}}. First, we begin by selecting a low-frequency pattern that ensures the visibility of at least one edge in local patches, given the size of the largest blur. To identify the maximum frequency, capture in-focus images of a sequence of frequency patterns with increasing \edit{frequencies bandlimits}, increasing approximately on a logarithmic scale \edit{(e.g., 10, 20, 50, 70, 100, \ldots)}. \edit{Next,} we choose the maximum frequency as the highest frequency at which the most in focus image can no longer resolve the frequency pattern on the screen. \edit{Finally}, we construct the final sequence of pattern frequencies using the original sequence up to and including the determined maximum frequency.

\subsection{Training Patch Size} 
We set the training patch size to be 1.5 times the extent of the 2D PSF along each dimension, a factor that is consistent across all lenses and camera systems. Convolution is performed within the valid region of the patch to avoid boundary artifacts. \edit{The training patch is chosen so that, for the largest PSFs in the focal stack, the non-zero (active) region of the PSF is centered with a surrounding padding equal to one-quarter of its width. This configuration ensures that the valid convolution region—excluding the padded zero areas—exactly corresponds to the non-zero portion of the PSF on which the loss is evaluated. }

\section{Dual-Pixel Simulations and Comparisons}

In this section, we show all of the PSFs in our simulation evaluation, the ground truth parametric models, estimations from our method, and that from Mannan \& Langer~\cite{mannan} and Joshi et al.~\cite{joshi2008psf}.

\subsection{Baseline Comparisons}

\begin{figure*}[!p]
\centering
\includegraphics[width=\linewidth]{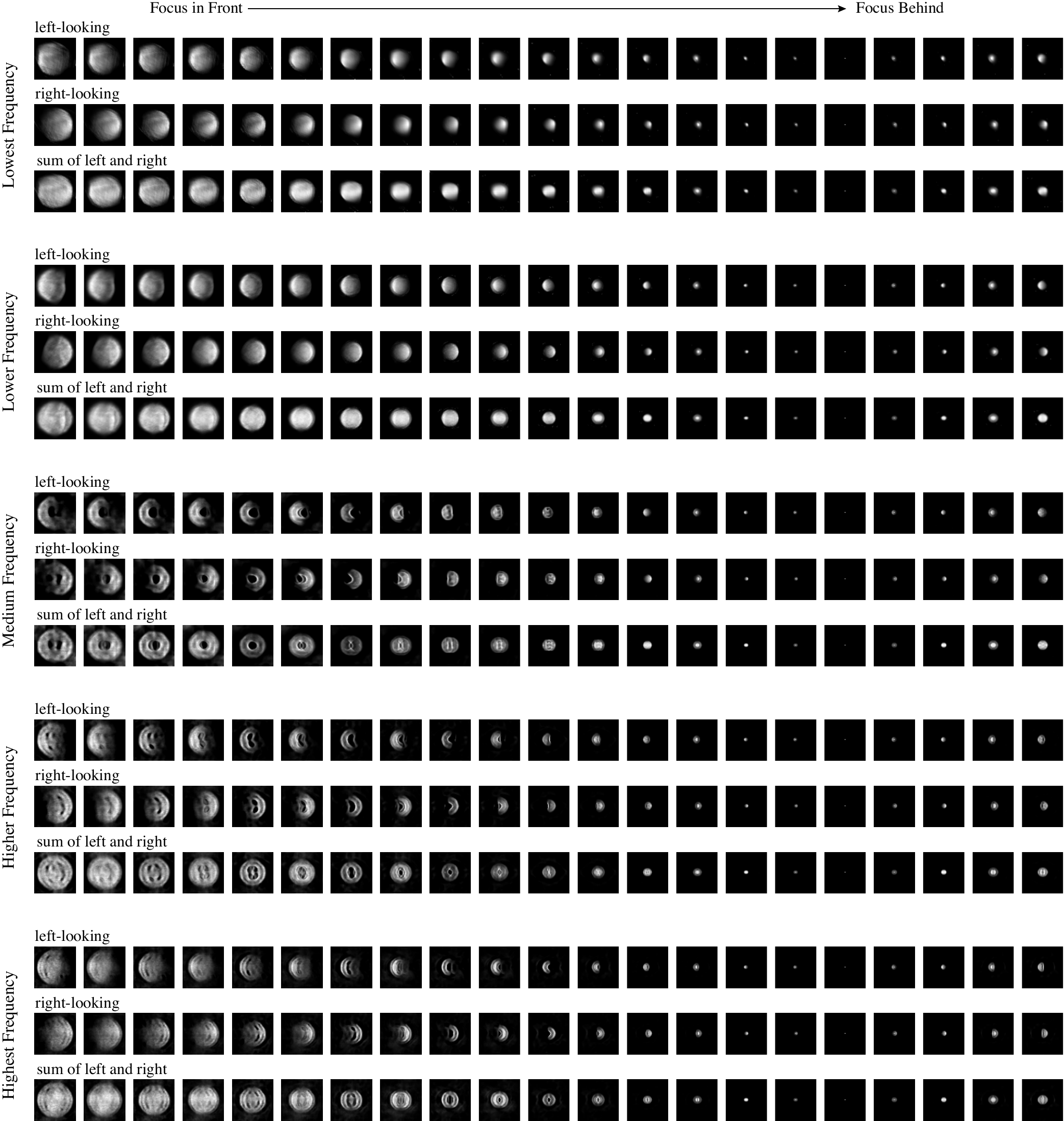}
\caption{Joshi et al.~\cite{joshi2008psf} grid of blurs estimates solved from patches extracted from the patterns in \cref{fig:simulation-patterns}. Estimates at each frequency (low, lower, medium, higher, high) are presented. This grid shows both estimated and interpolated results used in baselines in the main text. Every evenly-indexed (starting at 0) image is estimated from the algorithm, and every odd-index image is the result of interpolation between the two adjacent estimations. }
\label{fig:joshi-solves}
\end{figure*}

\begin{figure*}[!p]
\centering
\includegraphics[width=\linewidth]{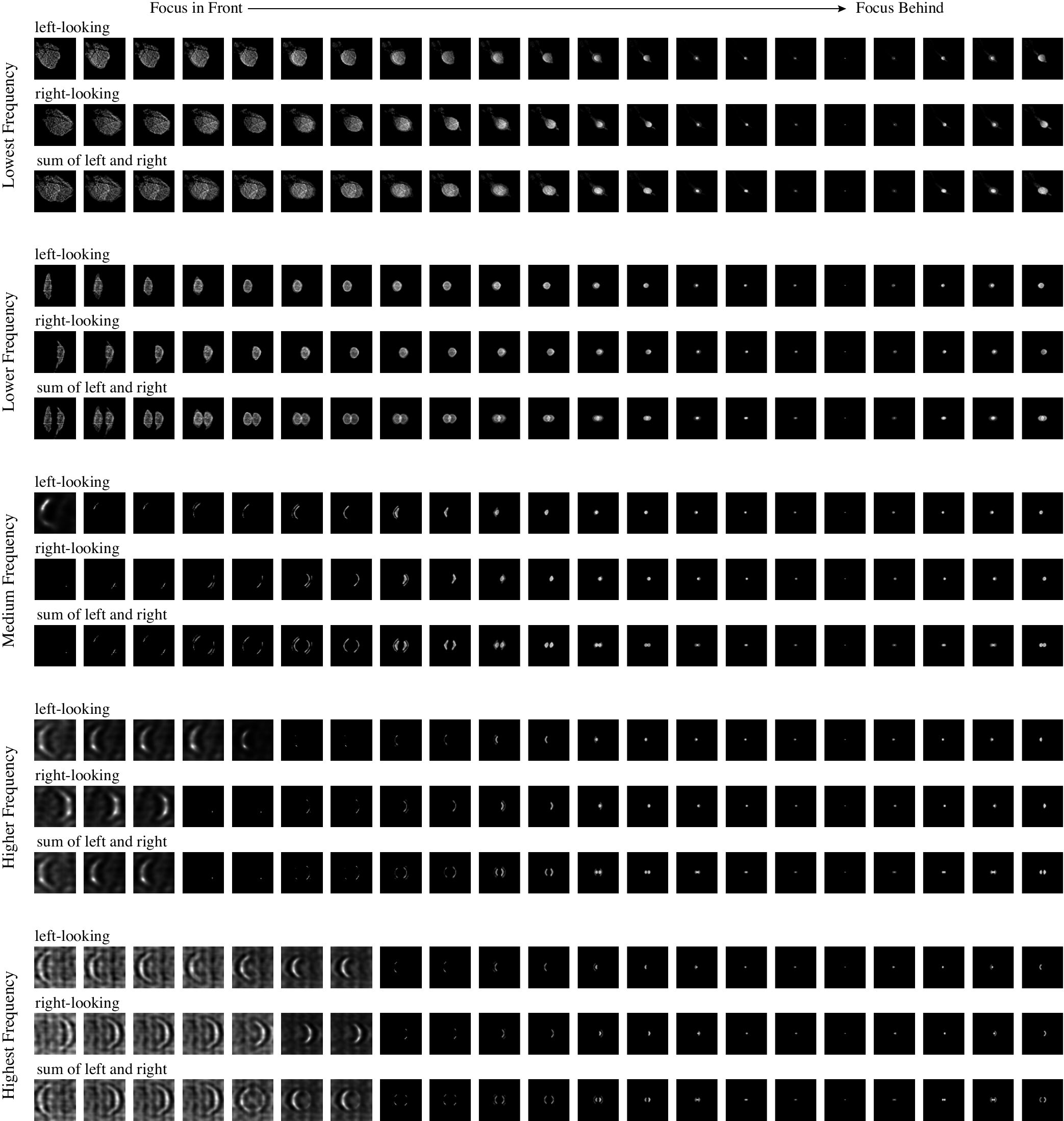}
\caption{Mannan and Langer~\cite{mannan} grid of blur estimates solved from patches extracted from the patterns in \cref{fig:simulation-patterns}. Estimates at each frequency (low, medium, high) are presented. This grid shows both estimated and interpolated results used as baselines in the main text. Every evenly-indexed (starting at 0) image is estimated from the algorithm, and every odd-index image is the result of interpolation between the two adjacent estimations. }
\label{fig:mannan-solves}
\end{figure*}

For simulations, we synthetically generate captures with three patterns shown in \cref{fig:simulation-patterns} with a single spatially-invariant dual pixel blur ~\cite{punnappurath2020modeling}. The procedure for creating these synthetic captures is described in the main text. We train our method jointly on all five patterns. For our baselines, since they are not joint optimization methods, we solve for blurs from each pattern, with results presented in \cref{fig:joshi-solves} and \cref{fig:mannan-solves}. We choose the best result for comparisons in the main text. 

We solve for $99 \times 99$ pixel kernels given input patches of $201 \times 201$ for both Mannan \& Langer~\cite{mannan} and Joshi et al.~\cite{joshi2008psf}. Patch sizes were chosen to be two times the size of the kernels. We let each method run till convergence. 

\subsection{Aliasing Behavior of Mannan \& Langer~\cite{mannan} and Joshi et al.~\cite{joshi2008psf}}
In general, we found that Mannan \& Langer~\cite{mannan} performs poorly on large kernels, with our estimations shown in \cref{fig:mannan-solves}. Joshi et al.~\cite{joshi2008psf} also suffers in large kernel situations. We show our estimations in \cref{fig:joshi-solves}. Both methods suffer more from noise than our method, where we are qualitatively able to see a stronger similarity between our estimations and the ground truth than Joshi et al.~\cite{joshi2008psf} and Mannan \& Langer~\cite{mannan}. We also observed both methods yielding high frequency artifacts in estimated blurs when the input patches contain higher frequencies.

\subsection{\edit{Comparison with Mannan \& Langer~\cite{mannan} on Dot Patterns}}

\edit{In this section, we run Mannan \& Langer's~\cite{mannan} method on dot patterns, specifically a centered dot pattern and a partial dot pattern where the inputs are composed of two half circles. We present two experiments to compare Mannan \& Langer's~\cite{mannan} and the proposed methods on dot patterns.}

\begin{figure*}[!t]
    \centering
    \includegraphics[width=\textwidth]{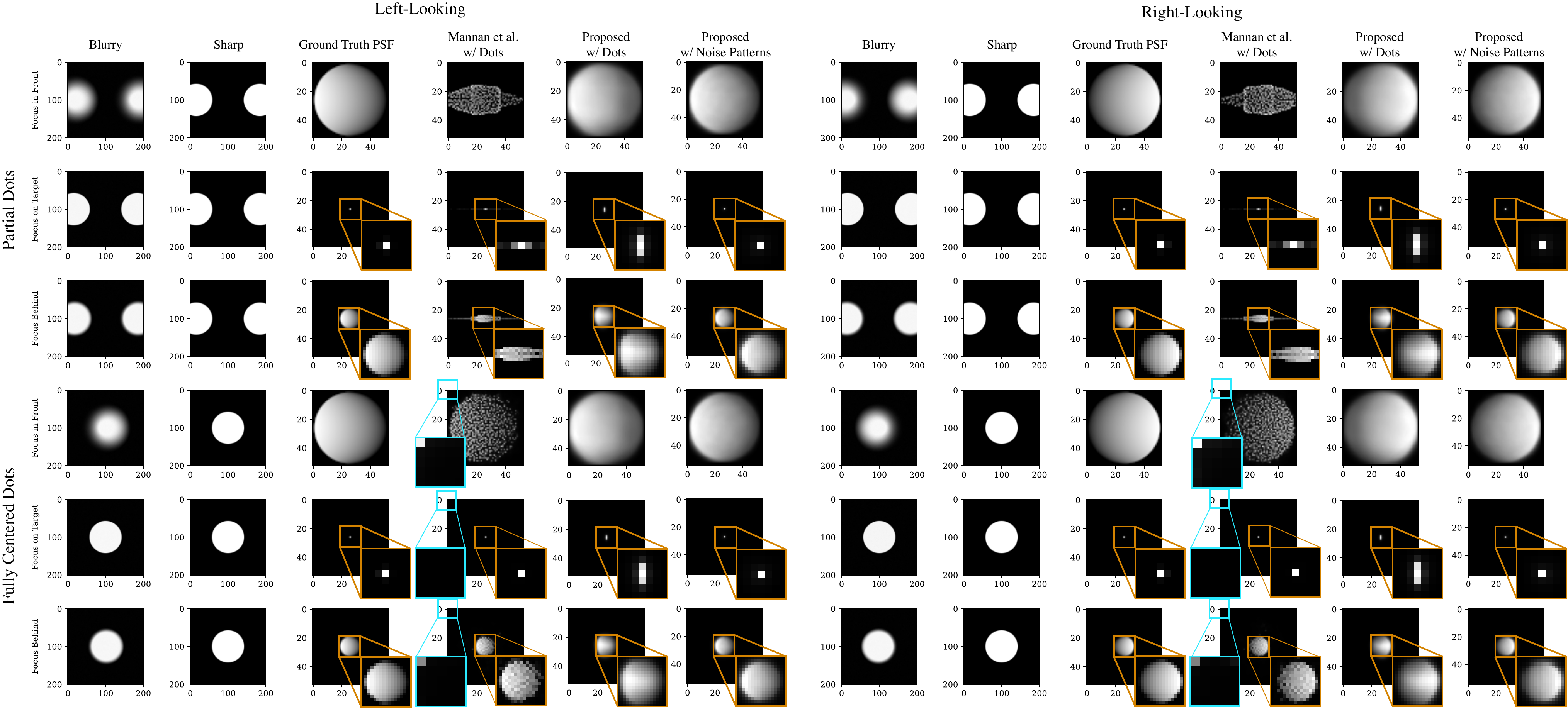}
    \caption{\edit{Simulated recovery of spatially invariant dual-pixel blur kernels generated using the parametric models proposed by Punnappurath et al.~\cite{punnappurath2020modeling} with Mannan \& Langer's~\cite{mannan} method. The method takes sharp and blurry images as inputs and outputs an estimated PSF, which we compare to the ground truth. Inset regions highlight high-energy artifacts present in Mannan \& Langer's~\cite{mannan} estimation method. Mannan \& Langer's~\cite{mannan} performs well when the PSF is small and the input patterns are centered dots, even better than the proposed method with the dot pattern. However, partial dots are more challenging to estimate with. Additionally, large PSFs are harder to estimate with Mannan \& Langer's~\cite{mannan}, with artifacts appearing (shown in the top left corners of the blue insets). Overall, Mannan \& Langer's~\cite{mannan} performs well for small PSFs and centered dot patterns but struggles with partial dots pattern inputs and larger PSFs.} }
    \label{fig:mannan-issues}
\end{figure*}

\begin{figure*}[!p]
  \centering
  \includegraphics[width=1\linewidth]{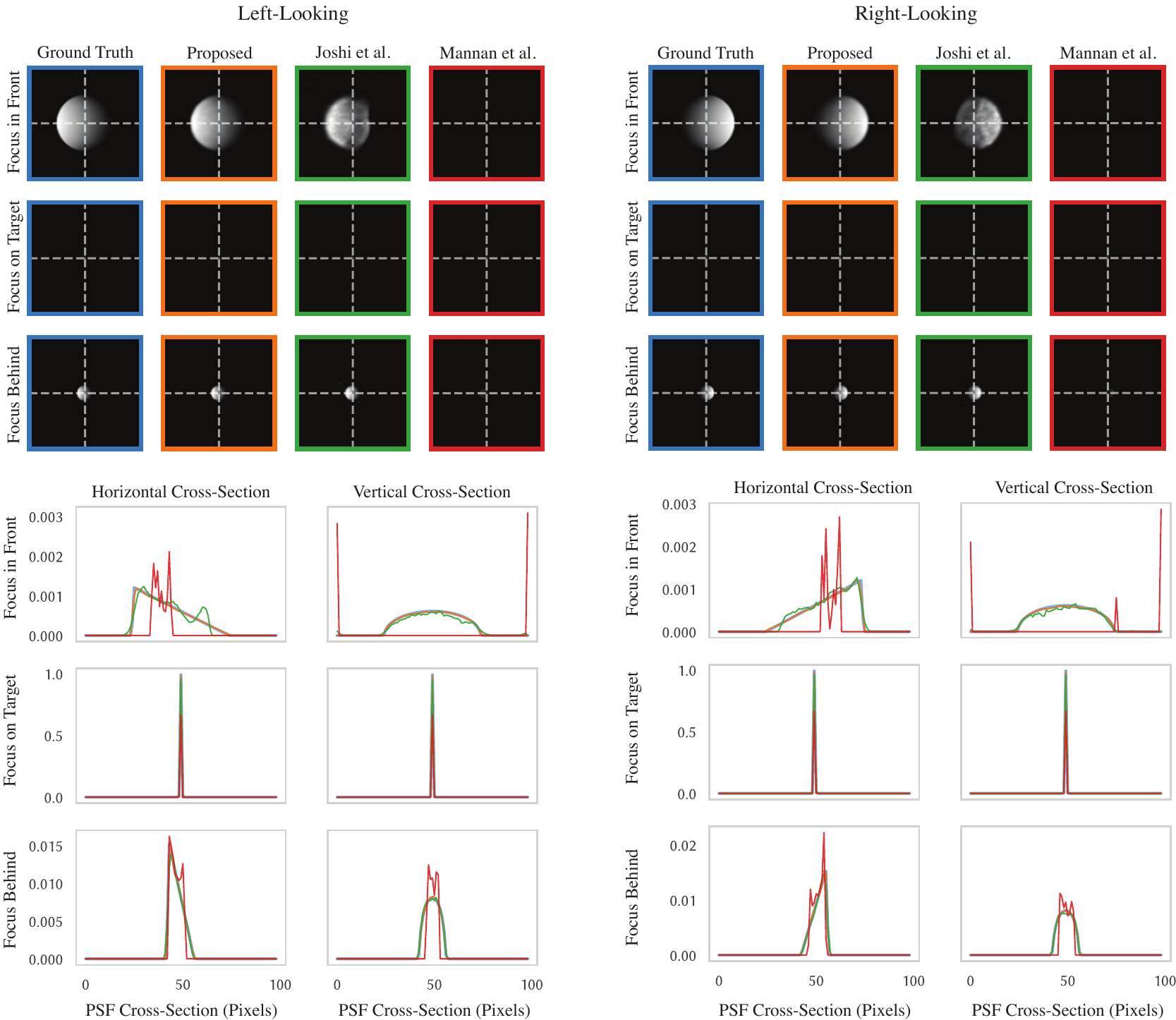}
  \caption{Cross sections of the PSFs shown in \edit{Fig.} 3 of the main text. We analyzed cross-sections of the in-focus PSFs of \edit{Fig.} 3 in the main document. Our method fits the peak and low-amplitude regions of this PSF better than Mannan and Langer~\cite{mannan} and Joshi et al.~\cite{joshi2008psf}.}
  \label{fig:simulations-cross-sections}
\end{figure*}

\edit{First, we solve for the ground truth PSFs created in Sec. IV-A of the main paper by using the dot patterns as our patterns. We compare the PSF estimations qualitatively in ~\cref{fig:mannan-issues}, where we find Mannan \& Langer's~\cite{mannan} method to work well on the centered dot pattern versus the partial dot pattern. We also find that PSFs estimated with Mannan \& Langer's~\cite{mannan} method tend to display high frequency artifacts either in the body of the PSF, or in the corners of the of the PSF. This issue is especially apparent as the PSF gets larger. Note that these corner artifacts can also be seen in \edit{Fig.} 2 of the Supplement of Xin et al.~\cite{xin2021defocus}. These corner artifacts are also found when running Mannan \& Langer's~\cite{mannan} on the noise patterns, as seen in ~\cref{fig:simulations-cross-sections}.}

\edit{Quantitatively, we compare the estimated PSFs to the ground truth by calculating image metrics over the entire PSF kernel, as presented in~\cref{tab:mannan_psf_metrics}. Overall, in terms of average performance, we observe that the proposed method using noise patterns yields the highest performance across all metrics, followed by the proposed method using the centered dot pattern, and finally the PSFs estimated by Mannan \& Langer~\cite{mannan} on the dot pattern. We note that the standard deviation for Mannan \& Langer~\cite{mannan}'s method is quite high. This is because their method performs well at estimating small PSFs (less than $10\times10$ pixels)—even better than the proposed method using dot patterns—but suffers from high-frequency artifacts on larger PSFs.
}

\edit{Next, we present a simulation experiment that corresponds to the ``Training Patterns'', and ``Training Lens Positions'' portion of Table 1 and with methodology from Sec. IV-A of the main paper. We create synthetic blurry images by convolving the dot pattern blurred with a known PSF, a spatially-invariant parametric model of lens blur for dual-pixel cameras by Punnappurath et al.~\cite{punnappurath2020modeling}. Then, we solve for the PSF using Mannan \& Langer's~\cite{mannan} method and evaluate, with results presented in~\cref{tab:mannan_psfs}. We focus on two cases: (1) when the dot pattern is centered versus having a cropped dot, and (2) when the PSF to be estimated varies in size. Qualitative results are presented in \cref{fig:mannan-issues}, where we can see artifacts in the corners for large PSF estimations. We also compute quantitative results to compare with the ``Training Patterns'' and  ``Training Lens Positions'' portion of Table 1 from the main paper. We find that Mannan \& Langer's~\cite{mannan} performs better than the proposed method when both solving on a single dot pattern for in focus PSFs, the proposed method performs significantly better when given the noise patterns as inputs. }

\begin{table}[h]
\centering
  \resizebox{1\linewidth}{!}{
  \renewcommand{\arraystretch}{1.25}
        \begin{tabular}{lccc}
        \toprule
        \edit{Metric} & \edit{Proposed w/ Noise} & \edit{Proposed w/ Dots} & \edit{Mannan \& Langer's~\cite{mannan}} \\
        \midrule
        \edit{PSNR $\uparrow$} & \edit{$46.786 \pm 12.626$} & \edit{$36.994 \pm 4.728$} & \edit{$33.306 \pm 17.224$} \\
        \edit{SSIM $\uparrow$} & \edit{$0.994 \pm 0.008$}  & \edit{$0.982 \pm 0.020$}  & \edit{$0.940 \pm 0.075$} \\
        \edit{RMSE $\downarrow$} & \edit{$(6.368 \pm 3.901)\times10^{-5}$} & \edit{$(8.395 \pm 21.365)\times10^{-4}$} & \edit{$(5.610 \pm 2.670)\times10^{-2}$} \\
        \bottomrule
        \end{tabular}
    }
\caption{\edit{We evaluate the PSF reconstruction accuracy of Mannan \& Langer~\cite{mannan} on the dot pattern and compare it to our proposed method by directly comparing the recovered PSFs to the ground truth PSFs. The ground truth PSFs are created following the process described in Sec.~IV-A of the main paper. Examples of the estimated PSFs, as well as the inputs used for the estimation (blurry and sharp images), are shown in~\cref{fig:mannan-issues}. We compute these metrics over the entire kernel to capture the effects of any artifacts. Overall, we observe that the proposed method using noise patterns yields the highest performance across all metrics, followed by the proposed method using the centered dot pattern, and finally the PSFs estimated by Mannan \& Langer~\cite{mannan} on the dot pattern.
} }
\label{tab:mannan_psf_metrics}
\end{table}

\begin{table*}
  \centering
  \resizebox{1\textwidth}{!}{
  \renewcommand{\arraystretch}{1.25}
  \begin{tabular}{lc@{\qquad}cccc@{\qquad}cccc}
    \toprule
    \multirow{2}{*}{\raisebox{-\heavyrulewidth}{}} && \multicolumn{4}{c}{Training Lens Positions} & \multicolumn{4}{c}{Validation Lens Positions} \\
    \cmidrule{3-10}
    && \edit{\makecell{Proposed Estimated w/\\ Centered Dots}} & \edit{\makecell{Proposed Estimated w/\\ Noise Patterns}} & \edit{\makecell{Mannan \& Langer~\cite{mannan}\\ Estimated w/ Centered Dots}} & \edit{\makecell{Mannan \& Langer~\cite{mannan}\\Estimated w/ Noise Patterns}} 
    & \edit{\makecell{Proposed Estimated w/\\ Centered Dots}} & \edit{\makecell{Proposed Estimated w/\\ Noise Patterns}} & \edit{\makecell{Mannan \& Langer~\cite{mannan}\\ Estimated w/ Centered Dots}} & \edit{\makecell{Mannan \& Langer~\cite{mannan}\\Estimated w/ Noise Patterns}} \\
    \midrule
    \multirow{3}{*}{\raisebox{-\heavyrulewidth}{\makecell{Training \\ Patterns}}}
                & PSNR $\uparrow$   & \edit{$ 31.8760 \pm 5.072 $} & $ 32.109 \pm 1.201$ & \edit{$ 33.252 \pm 1.856$} & $ 15.420 \pm 0.381$ 
                                    & \edit{$ 32.447 \pm 3.656 $} & $ 30.462 \pm 3.850$ & \edit{$ 28.792 \pm 4.752$} & $ 15.291 \pm 0.513$ \\
                & SSIM $\uparrow$   & \edit{$ 0.907 \pm 0.081$} & $ 0.929 \pm 0.072$ & \edit{$ 0.894 \pm 0.110$} & $ 0.601 \pm 0.021$ 
                                    & \edit{$ 0.920 \pm 0.074 $} & $ 0.924 \pm 0.066$ & \edit{$ 0.852 \pm 0.102$} & $ 0.600 \pm 0.024$ \\
                & RMSE $\downarrow$ & \edit{$ 0.032\pm 0.029$} & $ 0.022 \pm 0.004$ & \edit{$ 0.019 \pm 0.001$} & $ 0.166 \pm 0.010$ 
                                    & \edit{$ 0.026 \pm 0.012 $} & $ 0.031 \pm 0.023$ & \edit{$ 0.039 \pm 0.031$} & $ 0.169 \pm 0.013$ \\
    \midrule
    \multirow{3}{*}{\raisebox{-\heavyrulewidth}{\makecell{Training \\ Patterns}}}
                & PSNR $\uparrow$   & \edit{$ 32.047 \pm 5.102$} & $ 32.051 \pm 1.179$ & \edit{$ 33.285 \pm 1.938$} & $ 16.753 \pm 3.189$ 
                                    & \edit{$ 32.920 \pm 3.666$} & $ 30.396 \pm 3.840$ & \edit{$ 28.838 \pm 4.817$} & $ 16.400 \pm 3.086$ \\
                & SSIM $\uparrow$   & \edit{$ 0.899 \pm 0.091 $} & $ 0.924 \pm 0.078$ & \edit{$ 0.885 \pm 0.121$} & $ 0.461 \pm 0.290$ 
                                    & \edit{$ 0.912 \pm 0.084 $} & $ 0.919 \pm 0.072$ & \edit{$ 0.836 \pm 0.119$} & $ 0.452 \pm 0.278$ \\
                & RMSE $\downarrow$ & \edit{$ 0.031 \pm 0.030 $} & $ 0.022 \pm 0.004$ & \edit{$ 0.018 \pm 0.001$} & $ 0.139 \pm 0.079$ 
                                    & \edit{$ 0.025 \pm 0.012 $} & $ 0.031 \pm 0.023$ & \edit{$ 0.039 \pm 0.032$} & $ 0.145 \pm 0.077$ \\
    \bottomrule
  \end{tabular}}
  \caption{\edit{Evaluation of Mannan \& Langer~\cite{mannan} PSF reconstruction accuracy using dot patterns versus synthetic noise patterns on synthetic images generated by convolving the estimated PSFs with synthetic noise patterns, following the protocol of Table 1 in the main paper. Mannan \& Langer's method performs better when using the dot pattern for restoration, achieving higher PSNR and RMSE than our proposed method on the centered dot pattern. However, the proposed method with noise patterns performs better on validation lens positions. }}
\label{tab:mannan_psfs}
\end{table*}

\subsection{\edit{Ablation of Unstructured Noise Pattern Frequencies on Baselines}}
\label{ssec:frequency-ablation}
\begin{figure*}
    \centering
    \includegraphics[width=0.95\textwidth]{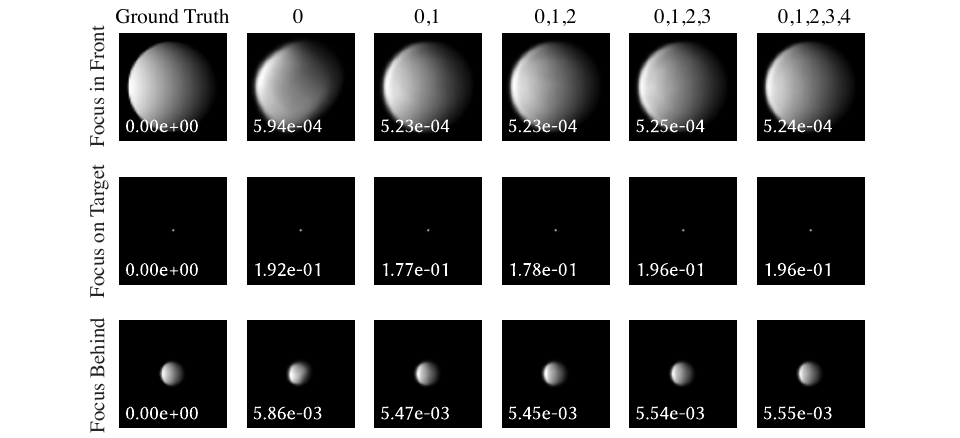}
    \includegraphics[width=0.95\textwidth]{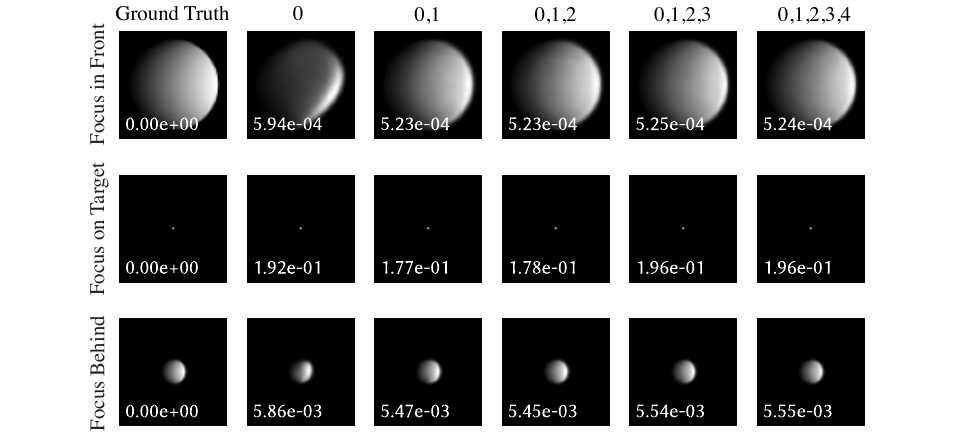}
    \caption{\edit{Ablation study of our optimization procedure on different calibration patterns. We simulate measurements using the same model as Sec. IV-A in the main document, following \edit{Punnappurath et al.~\cite{punnappurath2020modeling}}. The input patterns, indexed 0 through 4, correspond to increasing spatial frequency content—with frequency parameters 10, 20, 50, 70, and 100 respectively—where 0 denotes the lowest and 4 the highest frequency pattern as shown in Fig.~\ref{fig:simulation-patterns}. Each image represents the average of 10 runs. We see that to achieve an accurate PSF estimation, training requires unstructured noise patterns with features comparable to or smaller than the largest PSF radius. Higher-frequency patterns yield comparable results. Thus, we find pattern 2 (frequency = 50) to meet the minimum requirement for smartphones and pattern 3 (frequency = 70) for SLR lenses.}
    }
    \label{fig:freq-ablation}
\end{figure*}

\edit{We ablate different spatial frequency components of our joint training on multiple synthetic noise patterns in \cref{fig:freq-ablation}. We use calibration patterns blurred with a known PSF. Specifically, we use the same synthetic noise patterns outlined before and blur them with a spatially-invariant parametric model of lens blur for dual-pixel cameras, as described in Sec. IV-A of the main paper. The maximum and minimum defocus blur radii are set to 24 and 0.5 pixels, respectively. We run 10 trials for each set of frequencies. We observe that using only a single low-frequency pattern (\(\text{frequency} = 10\)) yields the poorest PSF estimation, and variations in estimation performance are more pronounced when the PSF is large and defocused.}

\edit{The accuracy improvement diminishes when using a noise pattern with \(\text{frequency} = 50\). For \(\text{frequency} = 50\), the generated noise pattern retains spatial frequencies roughly between 50 and 100, resulting in a dominant frequency of about \(1.5 \times 50 = 75\). Consequently, the dominant feature size is approximately the image height divided by 75. With an image height of 1512 pixels, the characteristic feature size is roughly \(1512/75 \approx 20\) pixels, which is comparable to our largest PSF radius of 24 pixels. This suggests that effective PSF estimation requires training with unstructured noise patterns whose features are comparable to or smaller than the largest PSF radius. Simulation results confirming this conclusion are shown in~\cref{fig:freq-ablation}.}

\edit{Additionally, from our simulation results in ~\cref{fig:freq-ablation} we see that including additional higher-frequency patterns yield similar results once the minimum frequency bar is met. With this, we find pattern 2 (frequency = 50) to meet the minimum requirement for smartphones and pattern 3 (frequency = 70) to meet the requirement for SLR lenses. In practice, we incorporate an additional higher-frequency pattern (pattern 3 for smartphones and pattern 4 for SLR lenses) as a precaution, especially as it does not compromise accuracy.}

\subsection{\edit{Ablation of MLP Architecture}}
\edit{In this section, we present ablation results for our MLP architecture and explain how the insights derived from these experiments informed our architectural choices for modeling the blur fields of real-life cameras.}

\begin{figure*}[!p]
    \centering
    \includegraphics[width=0.99\textwidth]{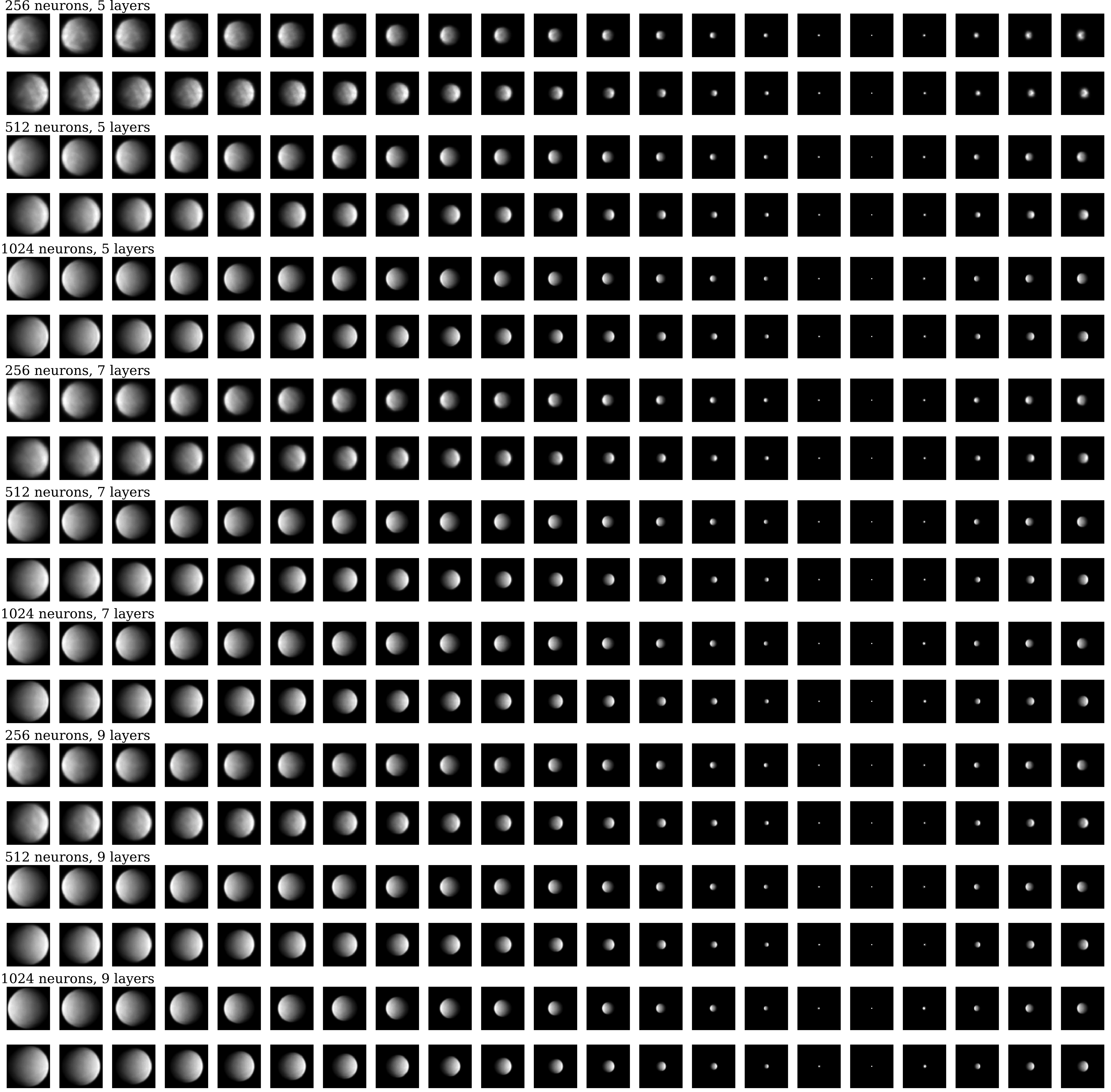}
    \caption{\edit{Left and right-looking PSF estimations from varying MLP architectures. We ablate number of layers (5, 7, 9) and number of neurons per layer (256, 512, 1024). As seen in these qualitative results, 256 neurons per layer yields the poorest results, regardless of number of layers. 512 versus 1024 neurons per layer yeild comparable results. We provide quantitative comparisons of these PSFs directly with the ground truth in ~\cref{fig:mlp-ablation-psfs}. }
    }
    \label{fig:mlp-ablation}
\end{figure*}

\begin{figure*}
    \centering
    \includegraphics[width=0.99\textwidth]{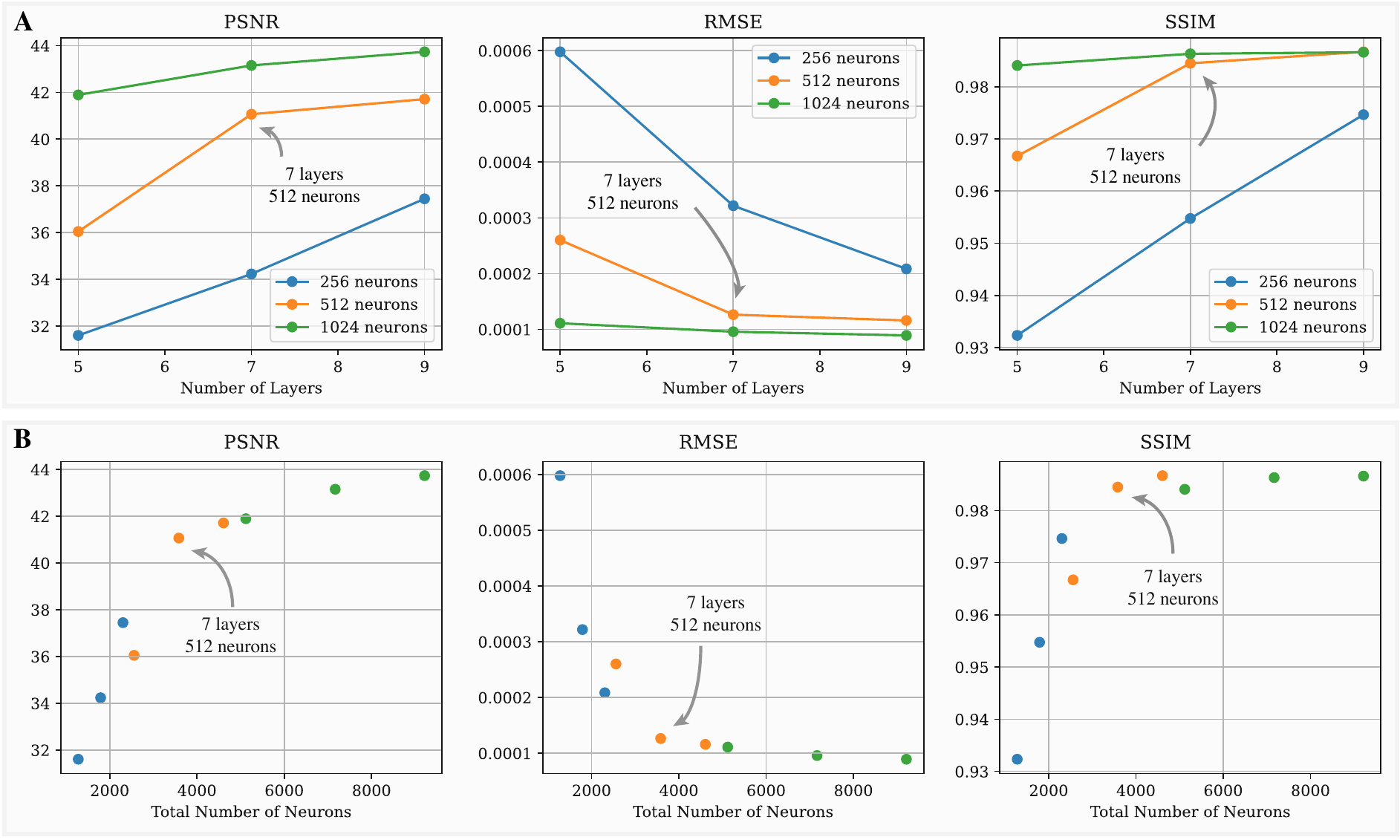}
    
    \caption{\edit{We ablate number of layers (5, 7, 9) and number of neurons per layer (256, 512, 1024) and compare the estimated PSFs with ground truth PSFs. We find that both training memory and training time grow exponentially as the number of layers and neurons increase. However, the quantitative metrics (e.g., PSNR, RMSE, SSIM) plateau once the network structure surpasses our default configuration of 7 layers and 512 neurons per layer, as highlighted by arrows in the plot. Based on these observations, we conclude that our chosen MLP architecture achieves high fidelity while maintaining a lower memory and computation budget.}
    }
    \label{fig:mlp-ablation-psfs}
\end{figure*}

\begin{figure*}[!t]
\centering
\includegraphics[width=\linewidth]{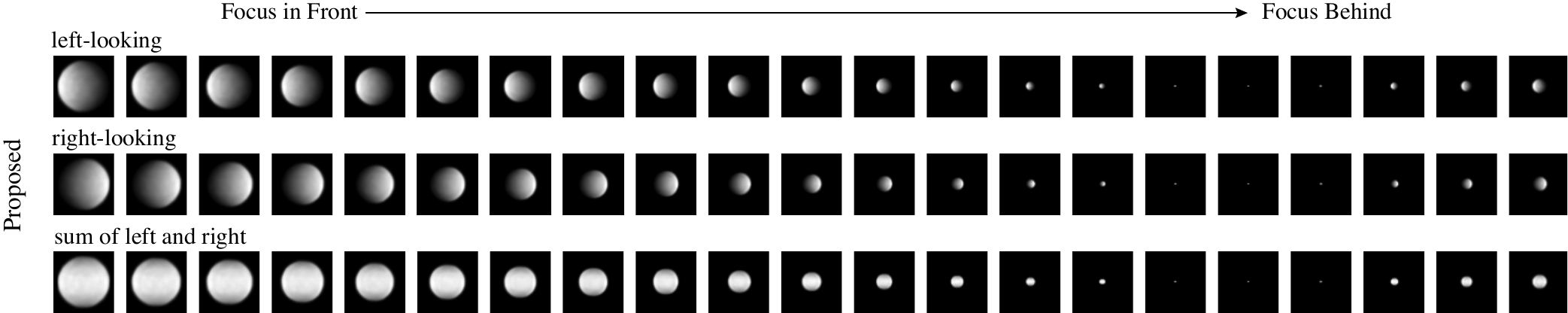}
\caption{Our method grid of PSFs, solved jointly on the patterns shown in \cref{fig:simulation-patterns}. This grid of estimations shows both kernels that were trained on the input focal positions and results that were interpolated by the MLP. Every evenly-indexed (starting at 0) image occurs at focus positions supervised in training, and every odd-index image is the result of interpolation after training. 
}
\label{fig:proposed-solves}
\end{figure*}

\begin{figure*}[!t]
\centering
\includegraphics[width=\linewidth]{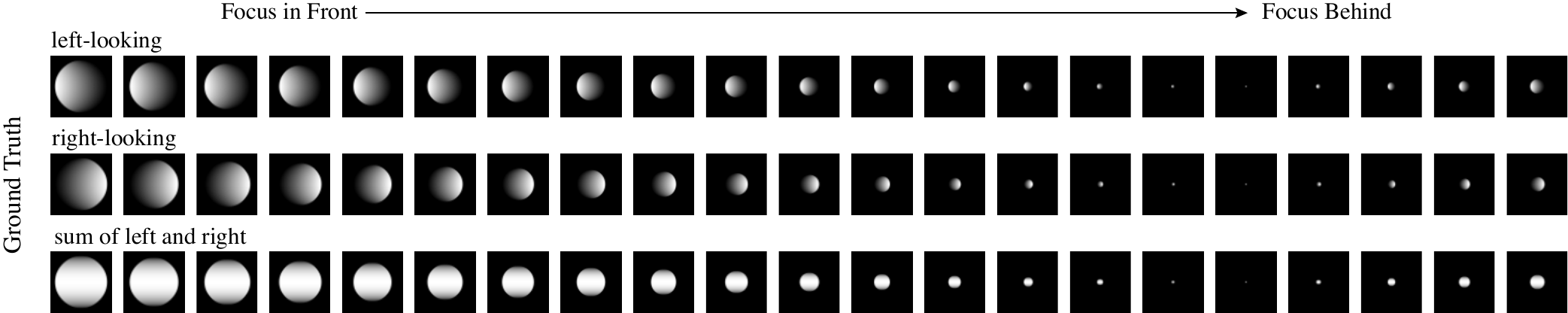}
\caption{Ground truth grid of blurs used to create simulation images.}
\label{fig:gnd-truth-blurs}
\end{figure*}

\edit{We present results on the restoration of the PSFs directly by comparing the estimated PSFs with ground truth in ~\cref{fig:mlp-ablation} (qualitative) and ~\cref{fig:mlp-ablation-psfs} (quantitative). As seen in both quantitative and qualitative results, 256 neurons per layer yields the poorest results, regardless of number of layers. 512 versus 1024 neurons per layer yield comparable results. We observe that while PSF restoration performance increases with the number of layers and neurons per layer, the improvements do not scale linearly (i.e., they diminish) as the model grows larger. As the network becomes larger (i.e., deeper and/or wider), its memory requirements and inference time increase. We ultimately chose 7 layers, 512 neurons per layer to be our MLP architecture (also shown in ~\cref{fig:proposed-solves}) as it yields comparable SSIM to the ground truth as the larger models, despite having a smaller size, as seen in panel B of ~\cref{fig:mlp-ablation-psfs}. } The ground truth PSFs can be found in ~\cref{fig:gnd-truth-blurs}.

\begin{figure*}
    \centering
    \includegraphics[width=0.99\textwidth]{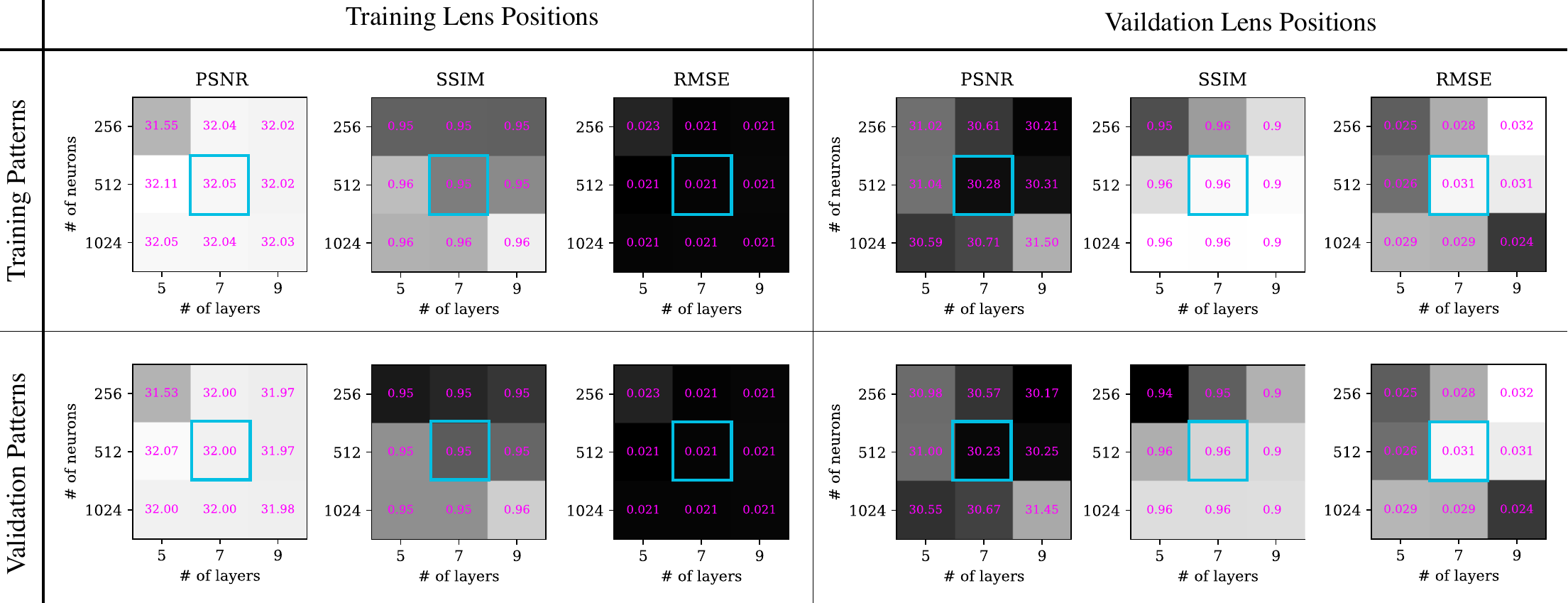}
    \caption{\edit{We ablate number of layers (5, 7, 9) and number of neurons per layer (256, 512, 1024) and render synthetic images with the estimated PSFs, then compare with ground truth renders, following the procedure described in Sec. IV-A of the main paper. Besides the cases when the number of neurons is 256, the presented image metrics are all comparable. The chosen configuration of 7 layers of 512 neurons each is highlighted in the cyan boxes. }
    }
    \label{fig:mlp-ablation-synthetic}
\end{figure*}

\edit{We also compare synthetic images rendered with the estimated PSFs with ground truth renders, following the procedure described in Sec. IV-A of the main paper. As shown in ~\cref{fig:mlp-ablation-synthetic}, besides the cases when the number of neurons is 256, the presented image metrics are all comparable. }

\section{Rendering with Blur Fields}
We outline the rendering procedure used as follows (which itself is agnostic to whether Blender or another engine is used).
Given a 3D scene, we first render an all-in-focus image with a $z$-buffer, which we quantize into a discrete number of fronto-parallel planes.
At this point, one can apply ~\cref{eq:image_formation} to compute a blurred image for each image plane, and composite back to front to create the final result.
However, since this does not account for occlusions, there will be artifacts around depth discontinuities.
In practice, we adopt the occlusion-aware model of Ikoma et al.~\cite{hayota2021depth}, which uses alpha compositing to merge the depth planes, resulting in a convincing if not entirely physically correct image. 
Similar techniques are used to add portrait mode effects for mobile photography, though typically with simple parametric blur kernels~\cite{wadhwa2018synthetic}.

\subsection{Rendering with Blur Fields}
\begin{figure*}
  \centering
  \includegraphics[width=\textwidth]{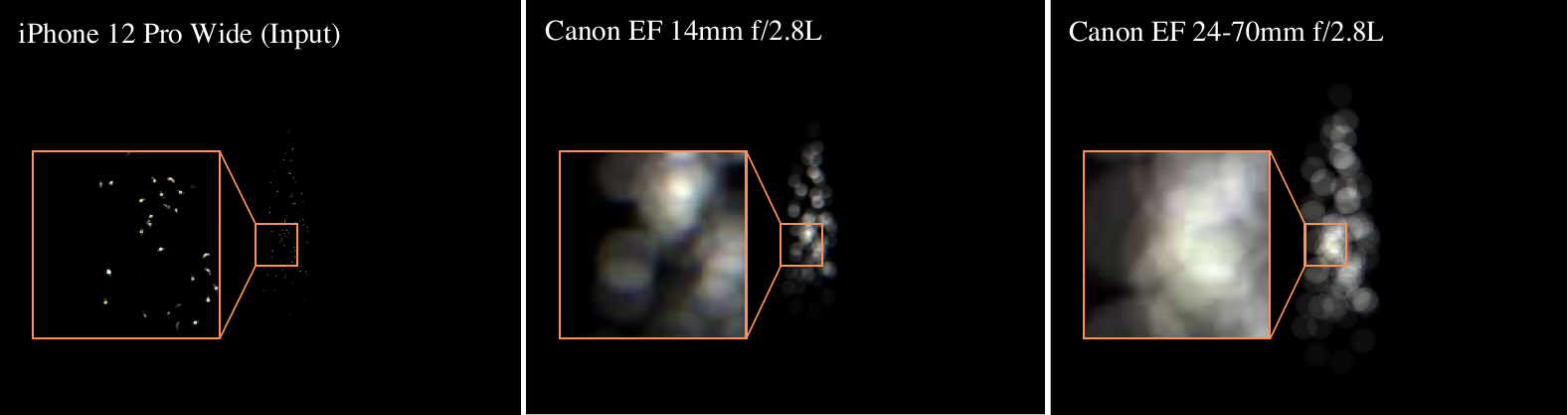}
  \caption{Synthetic defocus applied to a raw iPhone 12 Pro wide angle image to achieve the bokeh of SLR lenses. We capture a short exposure raw image of a string of fairy lights using an iPhone 12 Pro's main wide angle lens and render it using several blur fields from our database with the image formation model in ~\eqref{eq:image_formation}. }
  \label{fig:bokeh_render}
\end{figure*}

\begin{figure*}[!t]
    \centering
    \includegraphics[width=\textwidth]{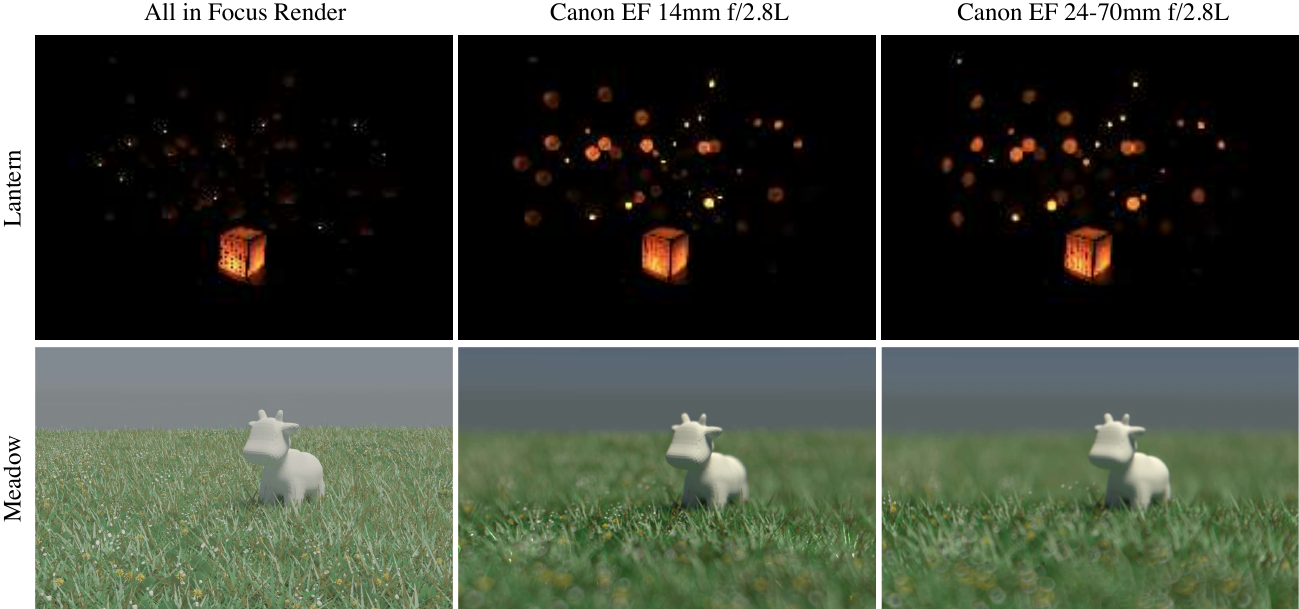}
    
    \caption{Synthetic renders in Blender using different blur field models from our database. We generate a high dynamic range all in focus render and mist pass (normalized depth also generated by Blender) and apply our blur fields. The final result is gamma corrected. Our models inject spatial and chromatic aberrations into the renders, allowing for greater photorealism. Scenes are taken from Blend Swap. We also leverage the dense depth maps in Blender to generate smooth portrait mode effects. The lantern scene is under a CC-BY license \texttt{https://blendswap.com/blend/17789}. The meadow scene is under a CC-0 license {\texttt{https://blendswap.com/blend/14433}}. The supplement contains more information on the rendering process.}
    \label{fig:blender}
\end{figure*}

Blur fields applied to captured imagery afford an opportunity for artistic control and emulation of realistic blur effects. For example, in ~\cref{fig:bokeh_render} we show a simple captured scene consisting of a string of fairy lights on a single depth plane. ~\cref{fig:blender} show examples of applying blur fields to scenes in Blender. Rendering these scenes using different lens blur fields results in a variety of convincing bokeh patterns.

\subsection{Comparing Image Formation Models on Scenes with Occlusion}

We compare three popular image formation models that do not use in-painting or neural networks: linear image formation~\cite{hayota2021depth}, approximate layered occlusion~\cite{hasinoff2007layer}, and nonlinear normalized image formation~\cite{hayota2021depth}. To begin with, we obtain a depth map and a corresponding image and then quantize them into $K$ depth layers ($l_k$) and corresponding binary masks ($a_k$). Lastly, we denote the refocused image as $I$. Note that we follow the convention of $k=0$ being the farthest depth layer, $\cdot$ being pixel-wise multiplication, and $*$ being convolution. Following this convention, we could interpret $\operatorname{PSF}_k$ as a slice of our 5D blur field at depth and lens position with index $k$.

To start our comparisons, we first consider the linear image formation described in ~\cite{hayota2021depth}, which simply masks the image depth layer $l_k$ at every layer $k$ with a binary mask $a_k$.
\begin{equation}
I = \sum_{k = 0}^{K-1} \operatorname{PSF}_{k} * (l_k \cdot a_k)
\label{eq:image_formation}
\end{equation}

Secondly, we consider the approximate layered occlusion model by adding a nonlinear weighting term for a smoother transition between two depth layers, according to~\cite{hasinoff2007layer}:
$$
   I = \sum_{k = 0}^{K-1} \operatorname{PSF}_{k} * (l_k \cdot a_k) \prod_{k'=k+1}^{K-1} (1-(\operatorname{PSF}_{k'} * a_{k'}))
$$

However, these two image formation models do not prevent the background from leaking into the foreground and causing halo artifacts. ~\cite{hayota2021depth} improves the model by adding a nonlinear term to normalize the amount of light exchanged between layers with the following convolution:
$$
    I = \sum_{k = 0}^{K-1} \tilde{l}_k \prod_{k'=k+1}^{K-1}(1-\tilde{a}_{k'}),
$$
where $\tilde{l}_k = (\operatorname{PSF}_k * (l_k \cdot a_k)) / E_k$, $\tilde{a}_k = (\operatorname{PSF}_k * a_k) / E_k$, and $E_k = \operatorname{PSF}_k * \sum_{k'=0}^{k}a_{k'}$.

\begin{figure*}[!t]
\centering
\includegraphics[width=\linewidth]{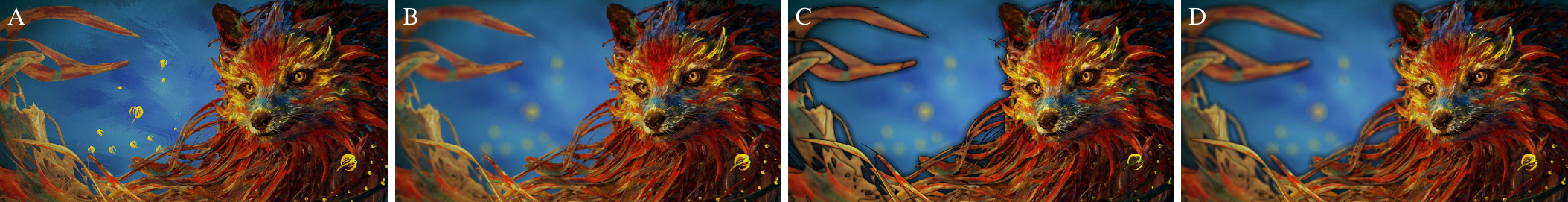}
\caption{Comparison of different image formation models. A is the all-in-focus image generated from Blender. B is rendered using the nonlinear normalized image formation model from ~\cite{hayota2021depth}. C is synthesized with the linear model mentioned in ~\cite{hayota2021depth}. D is composited using the approximate layered occlusion model from ~\cite{hasinoff2007layer}. Artwork by Daniel Bystedt under License CC-BY-SA 3.0 (https://cloud.blender.org/p/gallery/5f4cd4441b96699a87f70fcb).}
\label{fig:image_formation_models}
\end{figure*}

\begin{figure*}[!t]
\centering
\includegraphics[width=0.6\textwidth]{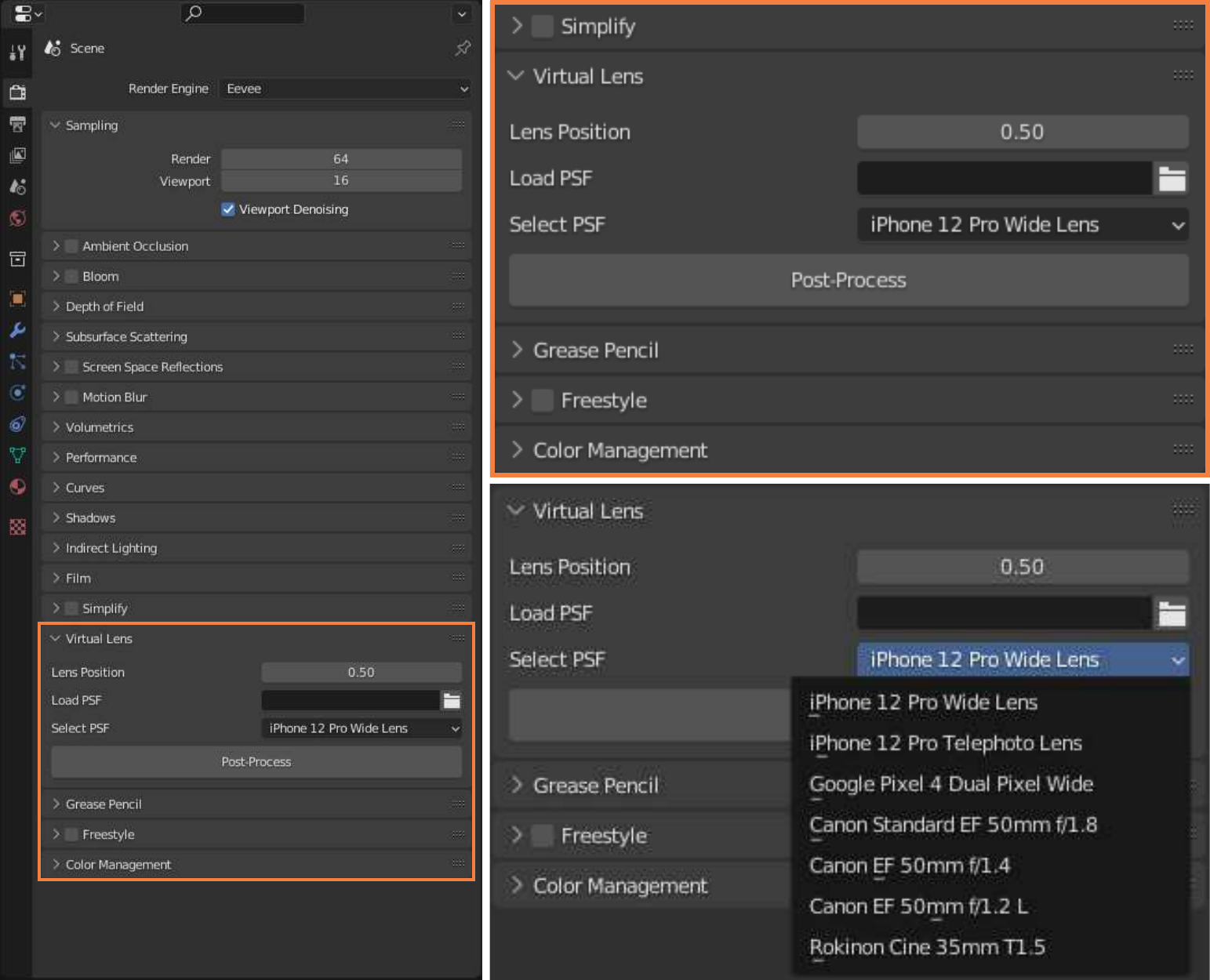}
\caption{Screenshot of proposed Blender postprocessing add-on. Users can choose from our database of learned 5D PSF fields or import one of their own. Pressing the apply button postprocess the scene using the PSF and the mist pass, focusing the virtual lens at a distance $f$ (normalized to $[-1, 1)$.}
\label{fig:blender-plugin}
\end{figure*}

In \cref{fig:image_formation_models}, we showcase synthetic defocus results rendered with the three image formation models. Observe that the nonlinear image formation model yields a plausible result without halo artifacts. A screenshot of our Blender postprocessing add-on is shown in \cref{fig:blender-plugin}.

\section{\edit{Comparison to Neural Deconvolution Approaches}}

\edit{While neural network non-blind deconvolution methods typically assume a spatially invariant PSF, our method addresses the challenge of estimating spatially varying PSFs. Moreover, our approach is not limited by the size of the PSF. 
}

\edit{For each neural deconvolution method (Zhang et al.~\cite{zhang2017learning}, Ren et al.~\cite{ren_2020_cvpr}, Dong et al.~\cite{dong2020deep}) in this section, we calculate image metrics following the same procedure as Table I in the main paper. We use the PSFs estimated from each neural deconvolution method, render a synthetic image with it and compare with ground truth. Quantitatively, we show in ~\cref{tab:neural-deconv} that the PSFs estimated by the neural deconvolution methods in this section do not perform as well as the proposed method. Specifically, these methods experience degraded performance in both PSF and sharp image estimation when blur size increases. Qualitatively, we present PSF and sharp image restoration results to highlight the challenges of neural deconvolution methods in generalizing to larger blur sizes and serving as a PSF estimator rather than solely a sharp image restorer.}

\begin{table*}
  \centering
  \resizebox{0.99\textwidth}{!}{
  \renewcommand{\arraystretch}{1.25}
  \begin{tabular}{lc@{\qquad}cccc@{\qquad}cccc}
    \toprule
    \multirow{2}{*}{\raisebox{-\heavyrulewidth}{}} && \multicolumn{4}{c}{Training Lens Positions} & \multicolumn{4}{c}{Validation Lens Positions} \\
    \cmidrule{3-10}
    && \edit{\makecell{Proposed }} & \edit{\makecell{Zheng et al.~\cite{zhang2017learning}\\ Noise=$0.03$}} & \edit{\makecell{Ren et al.~\cite{ren_2020_cvpr}}} & \edit{\makecell{Dong et al.~\cite{dong2020deep}}} 
    & \edit{\makecell{Proposed}} & \edit{\makecell{Zheng et al.~\cite{zhang2017learning}\\ Noise=$0.03$}} & \edit{\makecell{Ren et al.~\cite{ren_2020_cvpr}}} & \edit{\makecell{Dong et al.~\cite{dong2020deep}}} \\
    \midrule
    \multirow{3}{*}{\raisebox{-\heavyrulewidth}{\makecell{Training \\ Patterns}}}
                & PSNR $\uparrow$ & $ 32.109 \pm 1.201$ & \edit{$ 16.027 \pm 5.300$} & \edit{$ 19.221 \pm 6.297$} & \edit{$ 14.20 \pm 6.247$}
                                  & $ 30.462 \pm 3.850$ & \edit{$ 15.913 \pm 5.165$} & \edit{$ 20.681 \pm 5.713$} & \edit{$ 14.027 \pm 6.090$}\\
                & SSIM $\uparrow$   & $ 0.929 \pm 0.072$ & \edit{$ 0.304 \pm 0.187$} & \edit{$ 0.557 \pm 0.276$} & \edit{$ 0.233 \pm 0.135$}
                                    & $ 0.924 \pm 0.066$ & \edit{$ 0.323 \pm 0.209$} & \edit{$ 0.599 \pm 0.258$}  & \edit{$ 0.227 \pm 0.134$}\\
                & RMSE $\downarrow$ & $ 0.022 \pm 0.004$ & \edit{$ 0.175 \pm 0.112$} & \edit{$ 0.135 \pm 0.140$}  & \edit{$ 0.227 \pm 0.141 $}
                                    & $ 0.031 \pm 0.023$ & \edit{$ 0.176 \pm 0.107$} & \edit{$ 0.108 \pm 0.098$}  & \edit{$ 0.230 \pm 0.137$}\\
    \midrule
    \multirow{3}{*}{\raisebox{-\heavyrulewidth}{\makecell{Validation \\ Patterns}}}
                & PSNR $\uparrow$ & $ 32.051 \pm 1.179$ & \edit{$ 16.909 \pm 5.390$} & \edit{$ 19.274 \pm 6.386$} & \edit{$ 14.917 \pm 6.367$}
                                  & $ 30.396 \pm 3.840$ & \edit{$ 16.759 \pm 5.294$} & \edit{$ 20.823 \pm 5.937$} & \edit{$ 14.683 \pm 6.236$} \\
                & SSIM $\uparrow$   & $ 0.924 \pm 0.078$ & \edit{$ 0.323 \pm 0.209$} & \edit{$ 0.542 \pm 0.281$}  & \edit{$ 0.219 \pm 0.141$}
                                    & $ 0.919 \pm 0.072$ & \edit{$ 0.321 \pm 0.209$} & \edit{$ 0.584 \pm 0.270$}  & \edit{$ 0.213 \pm 0.137$}\\
                & RMSE $\downarrow$ & $ 0.022 \pm 0.004$ & \edit{$ 0.160 \pm 0.112$} & \edit{$ 0.135 \pm 0.142$}  & \edit{$ 0.213 \pm 0.140$}
                                    & $ 0.031 \pm 0.023$ & \edit{$ 0.162 \pm 0.108$} & \edit{$ 0.108 \pm 0.100$}  & \edit{$ 0.216 \pm 0.136$}\\
    \bottomrule
  \end{tabular}}
  \caption{\edit{Evaluation of reconstruction accuracy on synthetic noise patterns. The proposed approach quantitatively outperforms neural deconvolution methods in terms of PSNR, SSIM, and RMSE. Results were obtained following the same method in Sec. IV-A of the main paper. We present results for the Noise = 0.03 setting of Zhang et al.'s~\cite{zhang2017learning} work here in this table, as it is the highest-performing setting among those evaluated. For the results of the other settings in Zhang et al.~\cite{zhang2017learning}, see ~\cref{tab:zhang-deconv}. 
}}
\label{tab:neural-deconv}
\end{table*}

\subsection{\edit{Comparison to Zhang et al.~\cite{zhang2017learning}}}
\edit{Zhang et al.~\cite{zhang2017learning} proposes an iterative fully convolutional network for non-blind deconvolution, using learned image gradients as priors to guide deblurring.} 

\begin{figure*}[!t]
\centering
\includegraphics[width=0.8\linewidth]{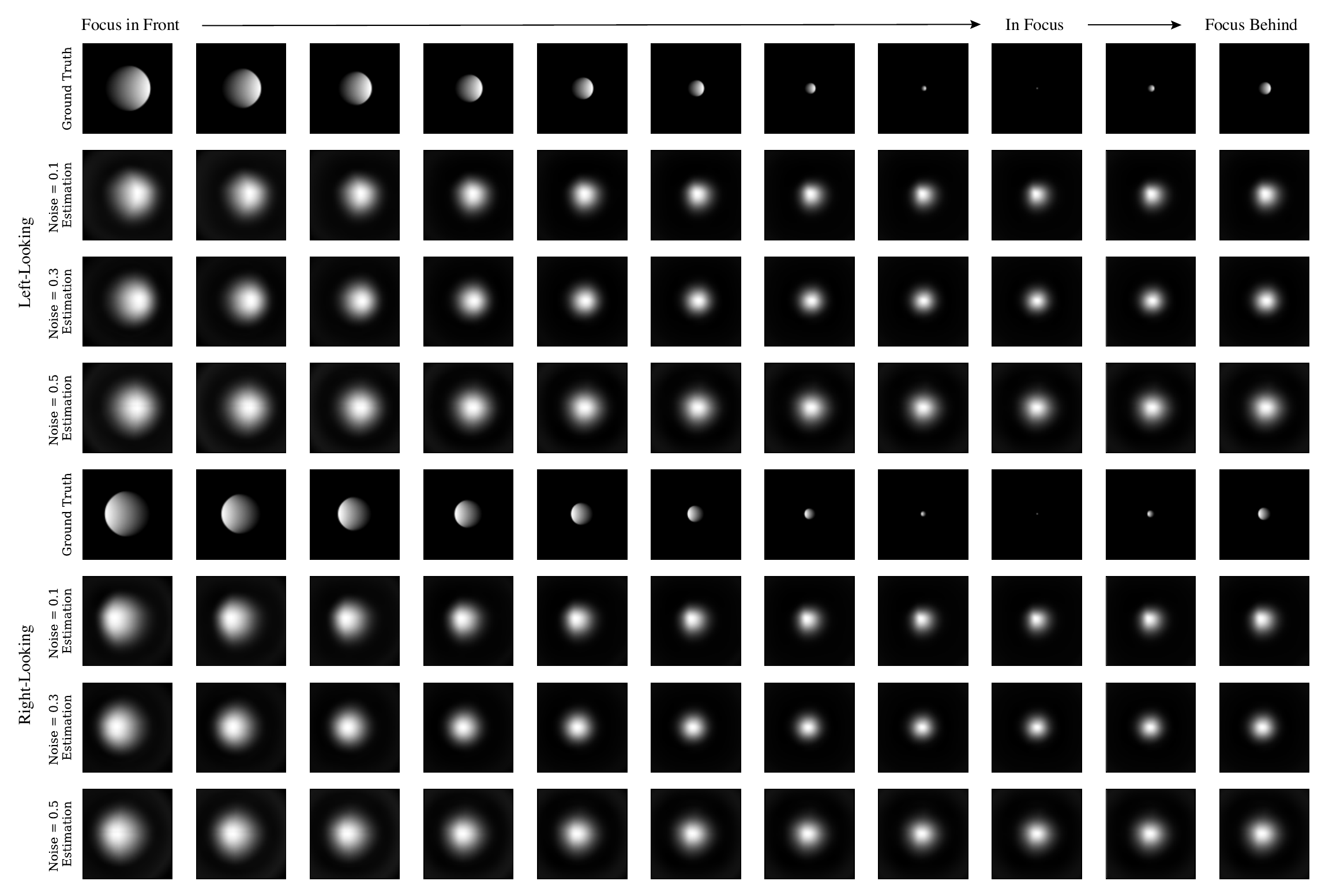}
\caption{\edit{Zhang et al.~\cite{zhang2017learning} PSF restorations from the full centered dot pattern shown in~\cref{fig:mannan-solves}. Zhang et al.~\cite{zhang2017learning} provides 3 pretrained models on different noise levels, $0.1, 0.3, 0.5$. We present results for all three, on the training lens positions corresponding to Table I of the main paper. Qualitatively, all three methods do not successfully capture the ground truth. When comparing synthetic images rendered with these PSFs with ground truth renders in~\cref{tab:zhang-deconv}, the proposed method outperforms all three settings across PSNR, SSIM and RMSE.}}
\label{fig:zhang-psf-dot}
\end{figure*}

\begin{figure*}[!t]
\centering
\includegraphics[width=0.8\linewidth]{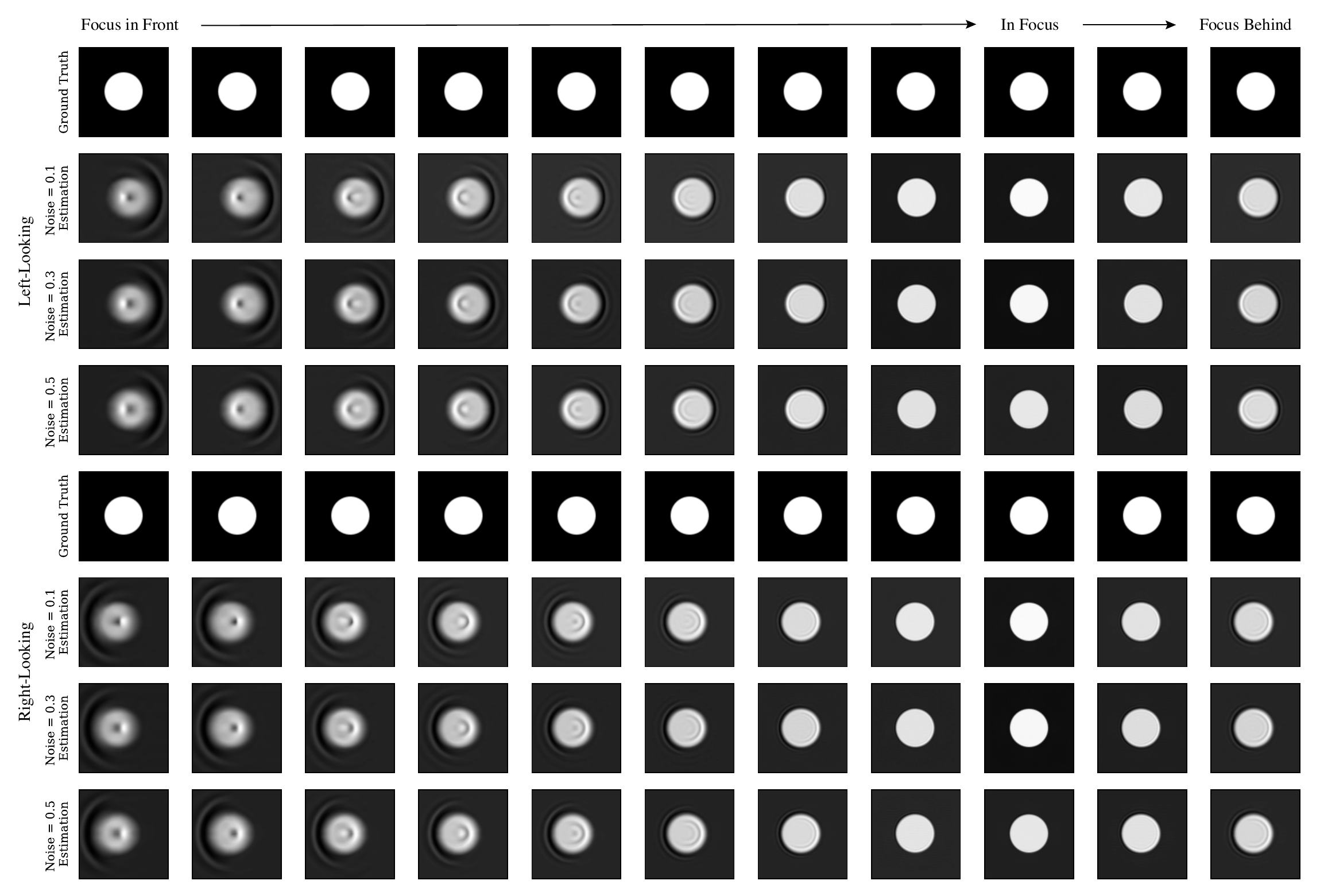}
\caption{\edit{Zhang et al.~\cite{zhang2017learning} sharp image restorations from the full centered dot pattern shown in~\cref{fig:mannan-solves}. Zhang et al.~\cite{zhang2017learning} provides 3 pretrained models on different noise levels, $0.1, 0.3, 0.5$. We present results for all three, on the training lens positions corresponding to Table I of the main paper. Qualitatively, all three methods do not successfully capture the ground truth, with a decrease in performance seen as a function of increasing blur size.}}
\label{fig:zhang-sharp-dot}
\end{figure*}

\begin{figure*}[!t]
\centering
\includegraphics[width=0.8\linewidth]{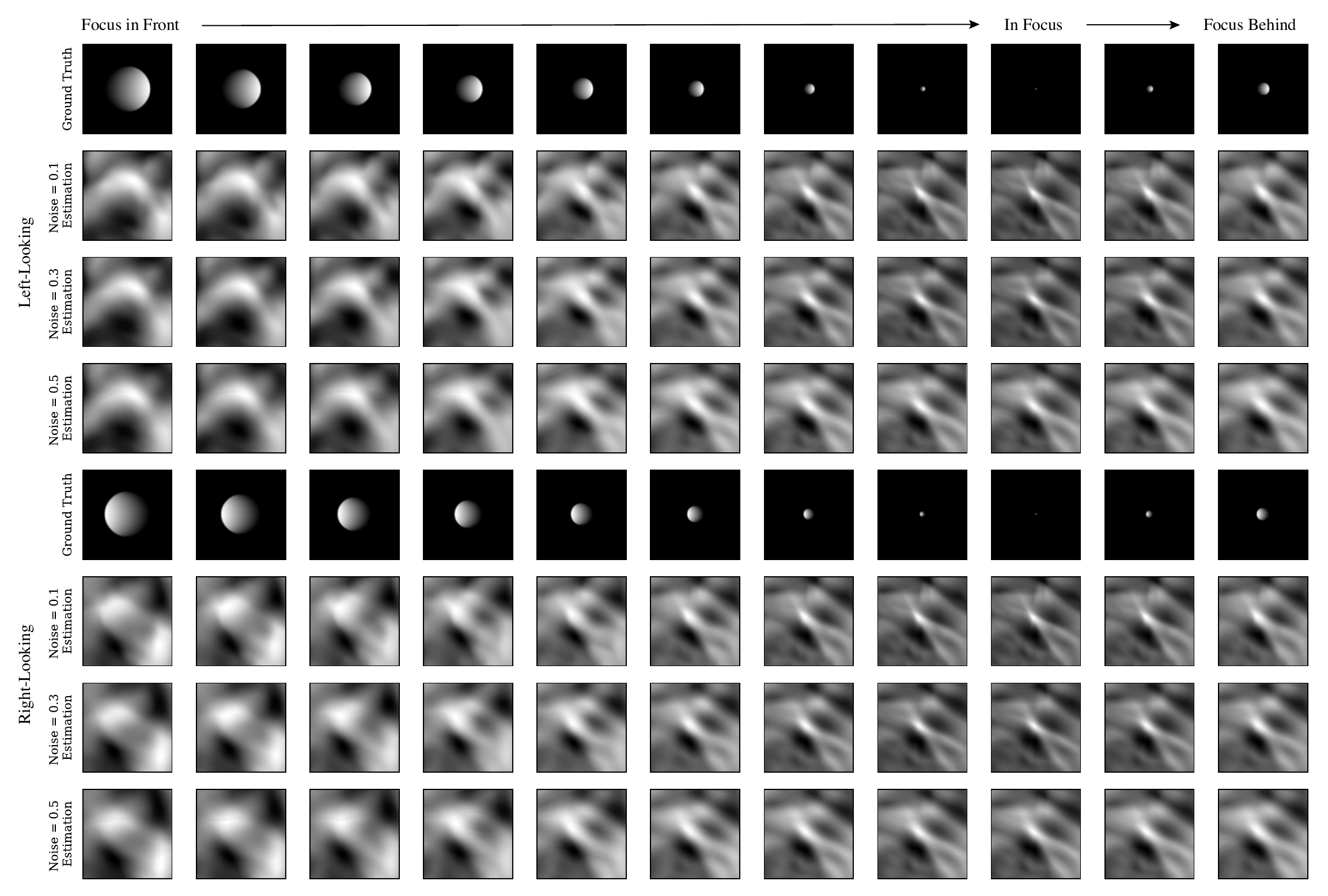}
\caption{\edit{Zhang et al.~\cite{zhang2017learning} PSF restorations from the lowest frequency random noise pattern shown in~\cref{fig:simulation-patterns}. Zhang et al.~\cite{zhang2017learning} provides 3 pretrained models on different noise levels, $0.1, 0.3, 0.5$. We present results for all three, on the training lens positions corresponding to Table I of the main paper.}}
\label{fig:zhang-psf-leopard}
\end{figure*}

\begin{figure*}[!t]
\centering
\includegraphics[width=0.8\linewidth]{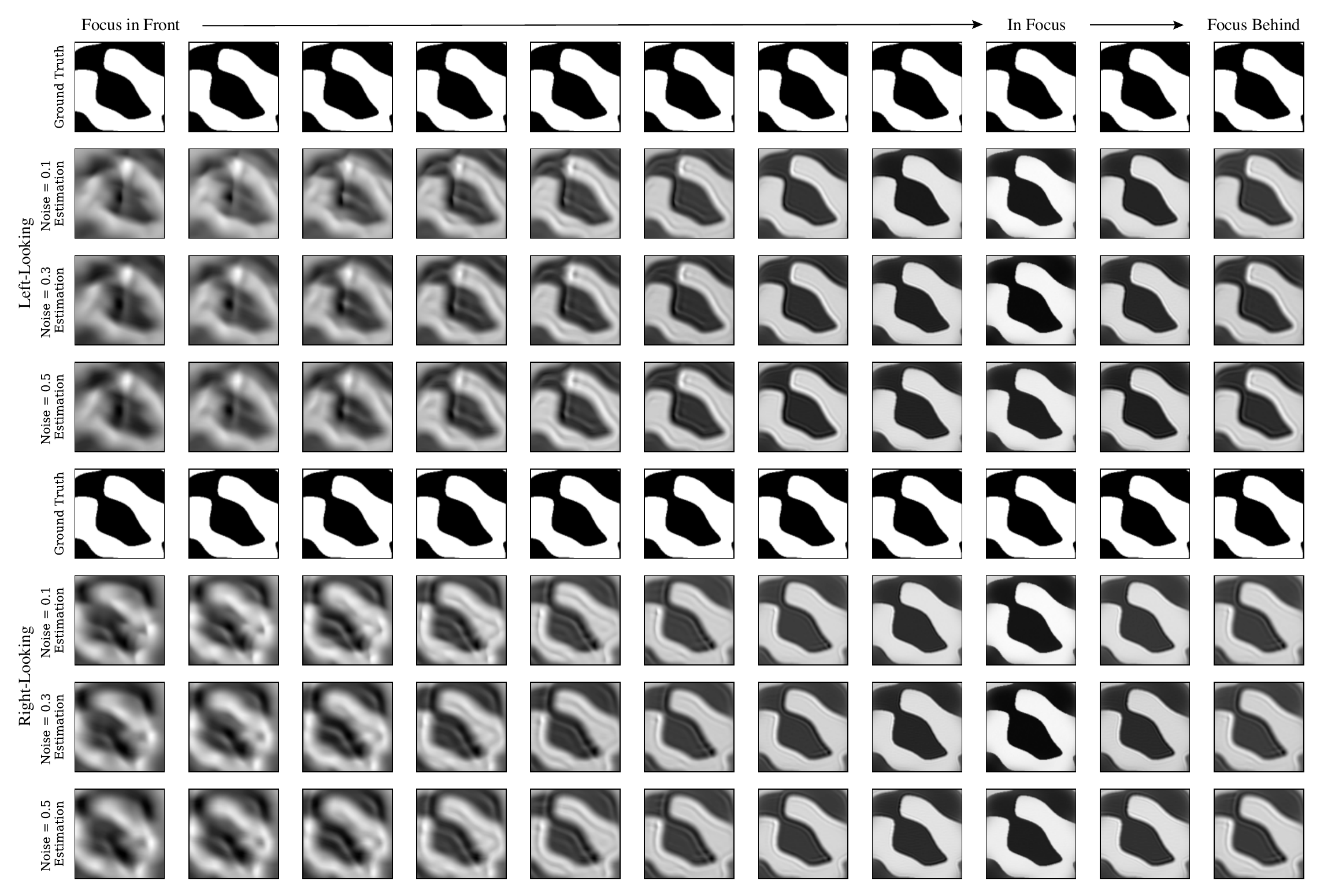}
\caption{\edit{Zhang et al.~\cite{zhang2017learning} sharp image restorations from the lowest frequency random noise pattern shown in~\cref{fig:simulation-patterns}. Zhang et al.~\cite{zhang2017learning} provides 3 pretrained models on different noise levels, $0.1, 0.3, 0.5$. We present results for all three, on the training lens positions corresponding to Table I of the main paper. Qualitatively, all three methods do not successfully capture the ground truth, with a decrease in performance seen as a function of increasing blur size.}}
\label{fig:zhang-sharp-leopard}
\end{figure*}

\edit{PSF and sharp image restoration results are shown in \cref{fig:zhang-psf-dot}, \cref{fig:zhang-sharp-dot}, \cref{fig:zhang-psf-leopard}, and \cref{fig:zhang-sharp-leopard}. We see that all settings (noise = $0.01, 0.03, 0.05$) in Zhang et al.'s~\cite{zhang2017learning} method perform well in recovering the sharp image when blur size is small. As blur size increases, performance degrades. On the other hand, Zhang et al.~\cite{zhang2017learning} does not perform as well at PSF restoration, with results from the dot pattern being better than those from the synthetic noise pattern. Image metrics comparing synthetic images rendered with the recovered PSFs to ground truth, following the process described in Sec. IV-A of the main paper, can be found in \cref{tab:neural-deconv}.}

\edit{In comparison, the proposed method does not rely on learned priors, supports a both PSF and sharp image restoration, and has a bigger PSF size allowance (80x80). Overall, we see that the proposed method outperforms Zhang et al.~\cite{zhang2017learning} in \cref{tab:zhang-deconv}.}

\begin{table*}
  \centering
  \resizebox{0.99\textwidth}{!}{
  \renewcommand{\arraystretch}{1.25}
  \begin{tabular}{lc@{\qquad}cccc@{\qquad}cccc}
    \toprule
    \multirow{2}{*}{\raisebox{-\heavyrulewidth}{}} && \multicolumn{4}{c}{Training Lens Positions} & \multicolumn{4}{c}{Validation Lens Positions} \\
    \cmidrule{3-10}
    && \edit{\makecell{Proposed }} & \edit{\makecell{Zheng et al.~\cite{zhang2017learning}\\ Noise=$0.01$}} & \edit{\makecell{Zheng et al.~\cite{zhang2017learning}\\ Noise=$0.03$}} & \edit{\makecell{Zheng et al.~\cite{zhang2017learning}\\ Noise=$0.05$}} 
    & \edit{\makecell{Proposed}} & \edit{\makecell{Zheng et al.~\cite{zhang2017learning}\\ Noise=$0.01$}} & \edit{\makecell{Zheng et al.~\cite{zhang2017learning}\\ Noise=$0.03$}} & \edit{\makecell{Zheng et al.~\cite{zhang2017learning}\\ Noise=$0.05$}} \\
    \midrule
    \multirow{3}{*}{\raisebox{-\heavyrulewidth}{\makecell{Training \\ Patterns}}}
                & PSNR $\uparrow$ & $ 32.109 \pm 1.201$ & \edit{$ 15.991 \pm 5.323$} & \edit{$ 16.027 \pm 5.300$} & \edit{$ 15.700 \pm 5.495$}
                                  & $ 30.462 \pm 3.850$ & \edit{$ 15.875 \pm 5.189$} & \edit{$ 15.913 \pm 5.165$} & \edit{$ 15.563 \pm 5.361$}\\
                & SSIM $\uparrow$   & $ 0.929 \pm 0.072$ & \edit{$ 0.299 \pm 0.187$} & \edit{$ 0.304 \pm 0.187$} & \edit{$ 0.294 \pm 0.186$}
                                    & $ 0.924 \pm 0.066$ & \edit{$ 0.298 \pm 0.188$} & \edit{$ 0.323 \pm 0.209$}  & \edit{$ 0.292 \pm 0.188$}\\
                & RMSE $\downarrow$ & $ 0.022 \pm 0.004$ & \edit{$ 0.176 \pm 0.112$} & \edit{$ 0.175 \pm 0.112$}  & \edit{$ 0.184 \pm 0.116 $}
                                    & $ 0.031 \pm 0.023$ & \edit{$ 0.177 \pm 0.107$} & \edit{$ 0.176 \pm 0.107$}  & \edit{$ 0.185 \pm 0.111$}\\
    \midrule
    \multirow{3}{*}{\raisebox{-\heavyrulewidth}{\makecell{Validation \\ Patterns}}}
                & PSNR $\uparrow$ & $ 32.051 \pm 1.179$ & \edit{$ 16.845 \pm 5.416$} & \edit{$ 16.909 \pm 5.390$} & \edit{$ 16.502 \pm 5.620$}
                                  & $ 30.396 \pm 3.840$ & \edit{$ 16.694 \pm 5.321$} & \edit{$ 16.759 \pm 5.294$} & \edit{$ 16.332 \pm 5.525$} \\
                & SSIM $\uparrow$   & $ 0.924 \pm 0.078$ & \edit{$ 0.311 \pm 0.205$} & \edit{$ 0.323 \pm 0.209$}  & \edit{$ 0.300 \pm 0.200$}
                                    & $ 0.919 \pm 0.072$ & \edit{$ 0.309 \pm 0.205$} & \edit{$ 0.321 \pm 0.209$}  & \edit{$ 0.298 \pm 0.199$}\\
                & RMSE $\downarrow$ & $ 0.022 \pm 0.004$ & \edit{$ 0.162 \pm 0.113$} & \edit{$ 0.160 \pm 0.112$}  & \edit{$ 0.170 \pm 0.117$}
                                    & $ 0.031 \pm 0.023$ & \edit{$ 0.163 \pm 0.108$} & \edit{$ 0.162 \pm 0.108$}  & \edit{$ 0.172 \pm 0.113$}\\
    \bottomrule
  \end{tabular}}
  \caption{\edit{Evaluation of reconstruction accuracy on synthetic noise patterns. The proposed approach quantitatively outperforms all three of Zhang et al.'s~\cite{zhang2017learning} neural deconvolution methods in terms of PSNR, SSIM, and RMSE. Results were obtained following the same method in Sect. IV-A of the main paper. }}
\label{tab:zhang-deconv}
\end{table*}

\subsection{\edit{Comparison to Ren et al.~\cite{ren_2020_cvpr}}}
\edit{SelfDeblur~\cite{ren_2020_cvpr}, presented by Ren et al.~\cite{ren_2020_cvpr} is a blind deconvolution that assume spatial invariance of the blur kernel to estimate a sharp image and PSF restoration given a blurry image. }

\begin{figure*}[!t]
\centering
\includegraphics[width=0.9\linewidth]{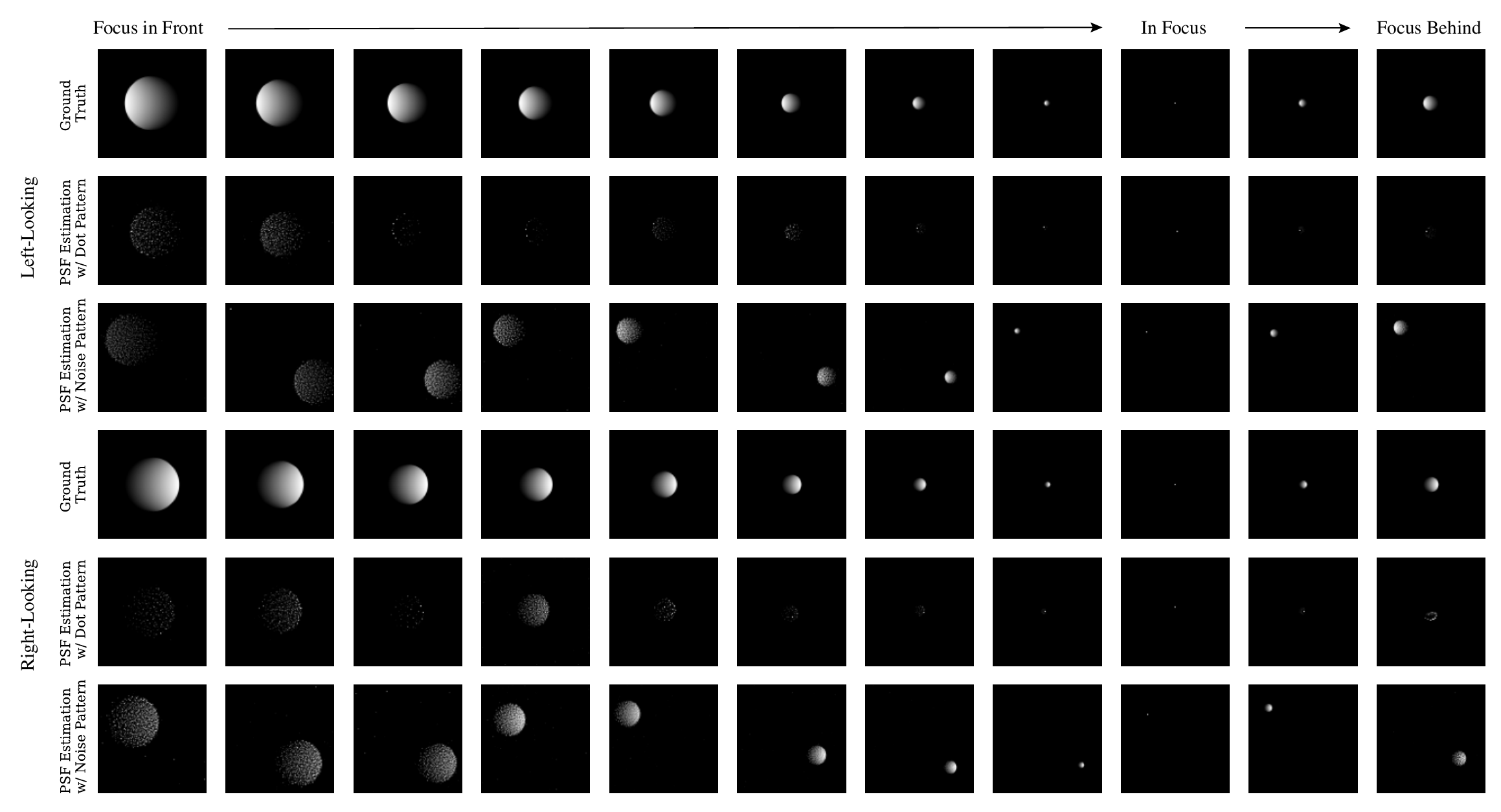}
\caption{\edit{Ren et al.~\cite{ren_2020_cvpr} PSF restorations from full centered dot pattern shown in~\cref{fig:mannan-solves} and the lowest frequency random noise pattern shown in~\cref{fig:simulation-patterns}. We present results corresponding training lens positions corresponding to Table I of the main paper. Qualitatively, neither PSF estimation from the dots pattern nor the synthetic noise pattern successfully captures the ground truth. When comparing synthetic images rendered with these PSFs with ground truth renders in~\cref{tab:neural-deconv}, the proposed method outperforms all three settings across PSNR, SSIM and RMSE.}}
\label{fig:ren-psf}
\end{figure*}

\begin{figure*}[!t]
\centering
\includegraphics[width=0.9\linewidth]{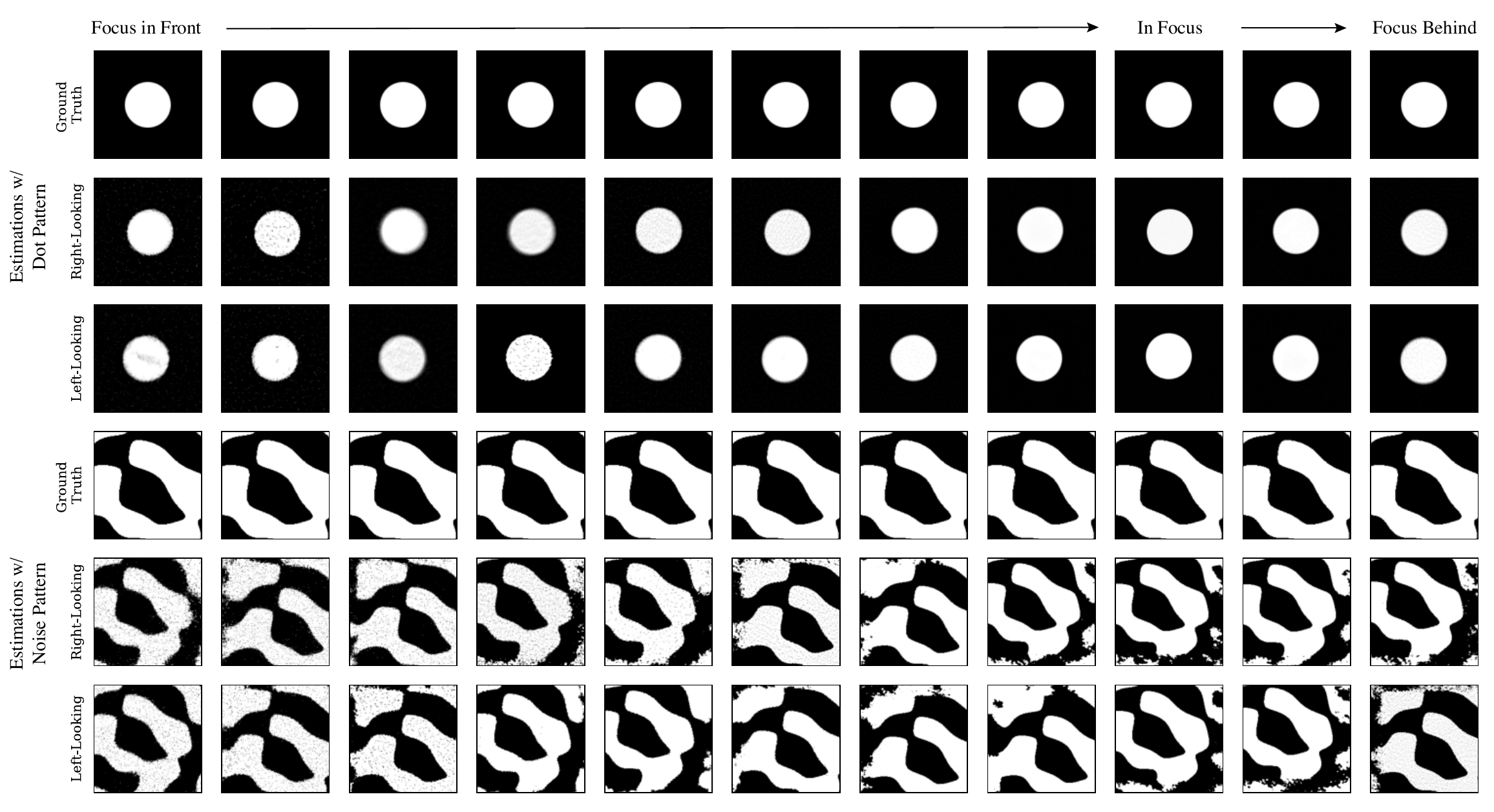}
\caption{\edit{Ren et al.~\cite{ren_2020_cvpr} sharp image restorations from full centered dot pattern shown in~\cref{fig:mannan-solves} and the full centered dot pattern shown in~\cref{fig:mannan-solves}. We present results corresponding to the training lens positions corresponding to Table I of the main paper.}}
\label{fig:ren-sharp}
\end{figure*}

\edit{PSF and sharp image restoration results are shown in ~\cref{fig:ren-psf} and ~\cref{fig:ren-sharp}. We see that Ren et al.~\cite{ren_2020_cvpr} performs well in recovering the sharp image regardless of the choice of pattern (dot vs synthetic noise). However, Ren et al.~\cite{ren_2020_cvpr} does not perform as well at PSF restoration. Image metrics when synthetic images rendered with the recovered PSFs compared to ground truth following the process described in Sec. IV-A of the main paper can be found in ~\cref{tab:neural-deconv}.}

\subsection{\edit{Comparison to Dong et al.~\cite{dong2020deep}}}
\edit{Dong et al.~\cite{dong2020deep} is a non-blind neural deconvolution work that proposes a deep Wiener deconvolution network for non-blind image deblurring, integrating classical Wiener deconvolution with learned deep features to perform deconvolution in feature space, followed by a multi-scale cascaded feature refinement module to recover fine details.}

\begin{figure*}[!t]
\centering
\includegraphics[width=0.9\linewidth]{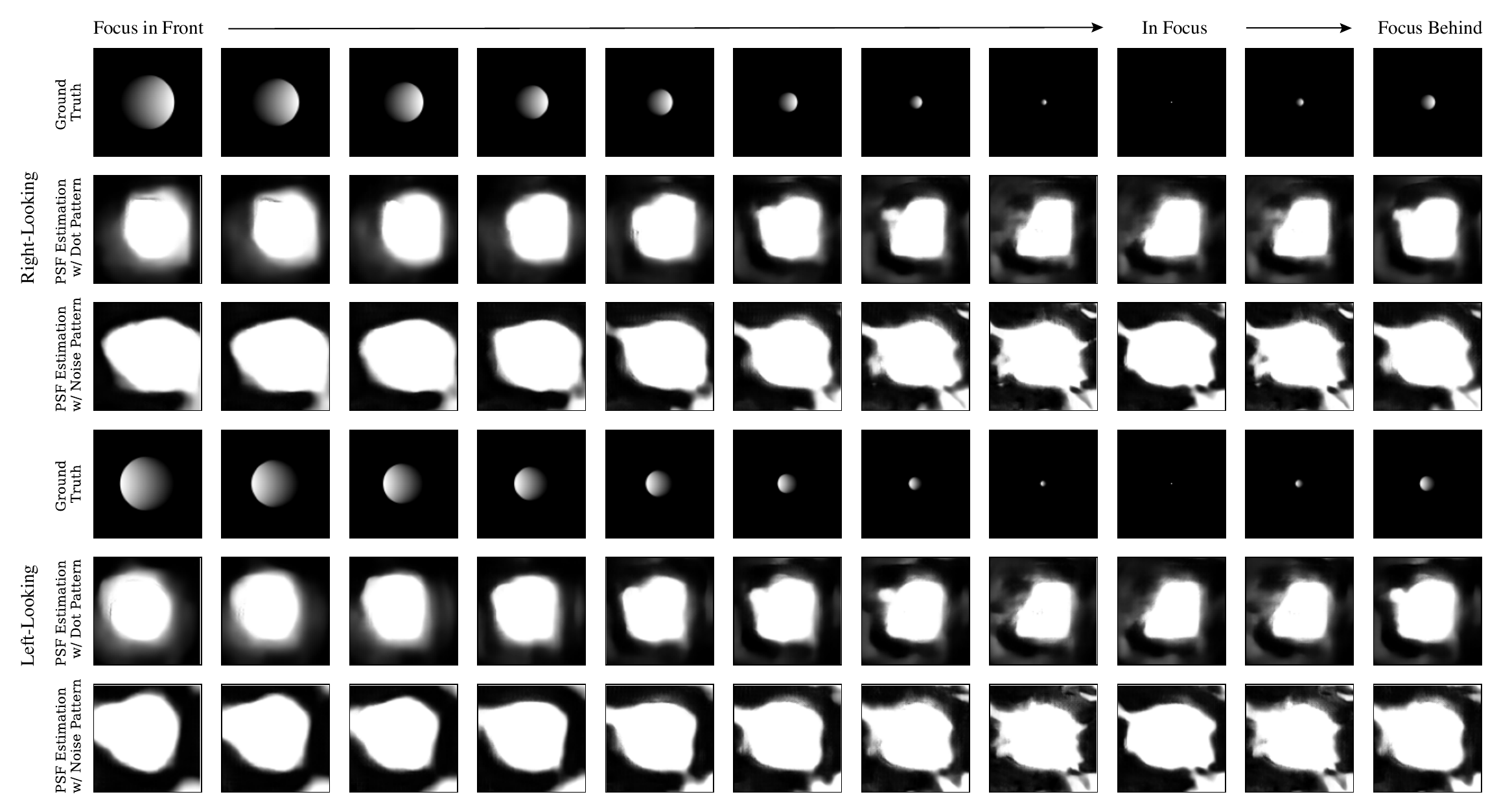}
\caption{\edit{Dong et al.~\cite{dong2020deep} PSF restorations from full centered dot pattern shown in~\cref{fig:mannan-solves} and the lowest frequency random noise pattern shown in~\cref{fig:simulation-patterns}. We present results corresponding training lens positions corresponding to Table I of the main paper. Qualitatively, neither PSF estimation from the dots pattern nor the synthetic noise pattern successfully captures the ground truth. When comparing synthetic images rendered with these PSFs with ground truth renders in~\cref{tab:neural-deconv}, the proposed method outperforms all three settings across PSNR, SSIM and RMSE.}}
\label{fig:dwdn-psf}
\end{figure*}

\begin{figure*}[!t]
\centering
\includegraphics[width=0.9\linewidth]{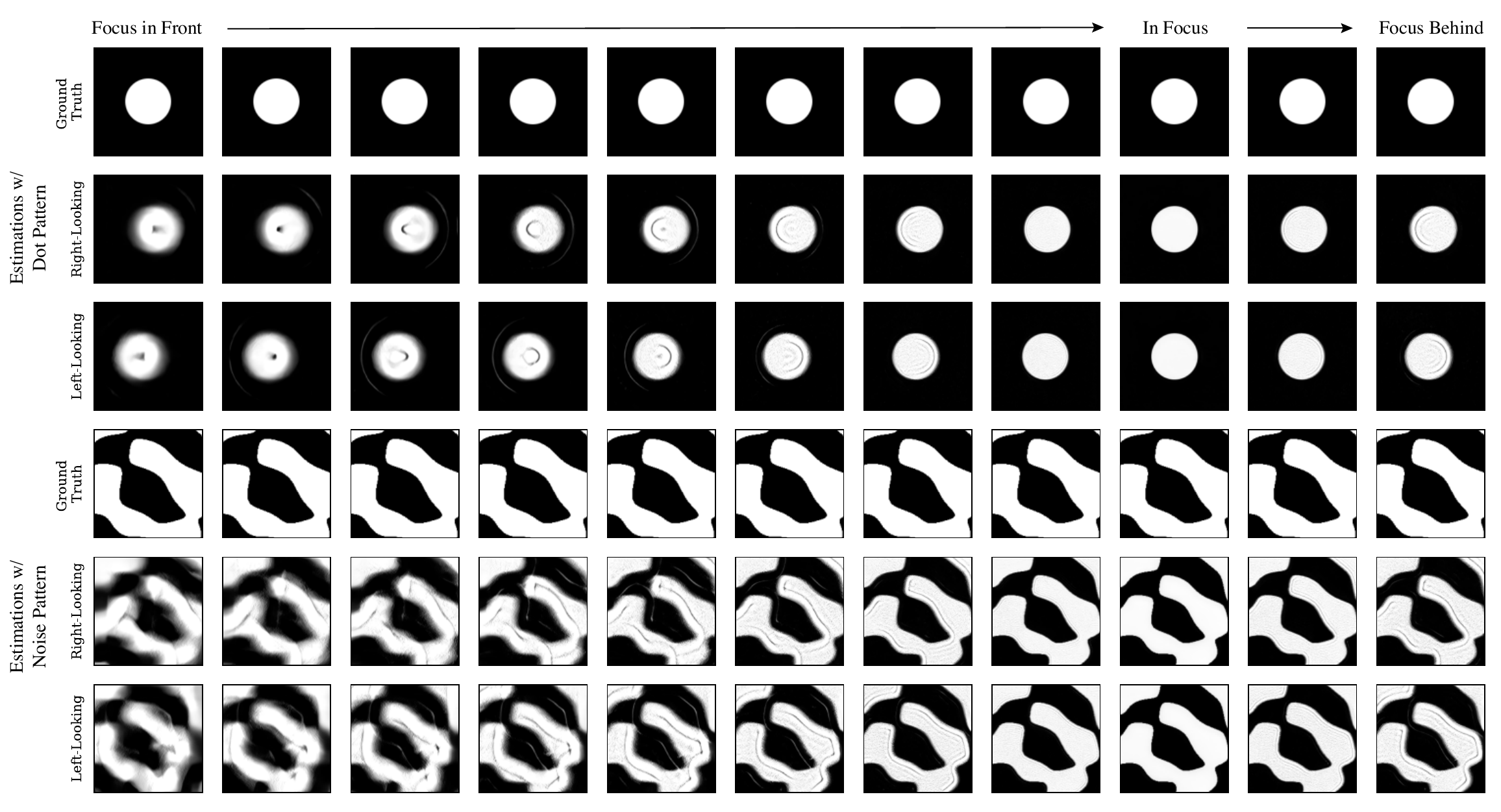}
\caption{\edit{Dong et al.~\cite{dong2020deep} sharp image restorations from full centered dot pattern shown in~\cref{fig:mannan-solves} and the full centered dot pattern shown in~\cref{fig:mannan-solves}. We present results corresponding to the training lens positions corresponding to Table I of the main paper. Qualitatively, Dong et al.~\cite{dong2020deep} sees a decrease in performance as a function of increasing blur size.}}
\label{fig:dwdn-sharp}
\end{figure*}

\edit{PSF and sharp image restoration results are shown in ~\cref{fig:dwdn-psf} and ~\cref{fig:dwdn-sharp}. We see that Dong et al.~\cite{dong2020deep} performs well in recovering the sharp image when blur size is small, but performance degrades as blur size increases. Additionally, Dong et al.~\cite{dong2020deep} performs better at recovering the sharp image than PSF restoration. Image metrics when synthetic images rendered with the recovered PSFs compared to ground truth following the process described in Sec. IV-A of the main paper can be found in ~\cref{tab:neural-deconv}.}

\bibliographystyle{IEEEtran}
\bibliography{IEEEabrv,references}

